\documentclass[showpacs,prd,twocolumn,floatfix,nofootinbib,superscriptaddress]{revtex4}

\usepackage{hyperref}

\hypersetup{
    colorlinks=true,       % false: boxed links; true: colored links
    linkcolor=blue,          % color of internal links
    citecolor=blue,        % color of links to bibliography
    filecolor=blue,      % color of file links
    urlcolor=blue           % color of external links
}

\usepackage{graphicx}
\usepackage{nicefrac}
\usepackage{lipsum}
\usepackage{amsmath,bm}
\usepackage{subfig}
\usepackage{multirow}
\usepackage{enumerate}

\newcommand\T{\rule{0pt}{2.6ex}}

\begin{document}

%-----------------------------------o
%  Front page stuff
%-----------------------------------o

\title{ $B_s \to K \ell \nu$ form factors from lattice QCD }

\author{C.M. Bouchard}
\thanks{bouchard.18@osu.edu}
\affiliation{Department of Physics,
The Ohio State University, Columbus, Ohio 43210, USA}

\author{{G.\ Peter} Lepage}
\affiliation{Laboratory of Elementary Particle Physics,
Cornell University, Ithaca, New York 14853, USA}

\author{Christopher Monahan}
\affiliation{Physics Department,
College of William and Mary, Williamsburg, Virginia 23187, USA}

\author{Heechang Na} 
\affiliation{Department Physics and Astronomy, 
University of Utah, Salt Lake City, Utah 84112, USA}

\author{Junko Shigemitsu}
\affiliation{Department of Physics,
The Ohio State University, Columbus, Ohio 43210, USA}

\collaboration{HPQCD Collaboration}
\noaffiliation
\date{\today}

%-----------------------------------o
%  Abstract
%-----------------------------------o

\begin{abstract}
We report the first lattice QCD calculation of the form factors for the standard model tree-level decay $B_s\to K \ell\nu$.  
In combination with future measurement, this calculation will provide an alternative exclusive semileptonic determination of $|V_{ub}|$.
We compare our results with previous model calculations, make predictions for differential decay rates and branching fractions, and predict the ratio of differential branching fractions between $B_s\to K\tau\nu$ and $B_s\to K\mu\nu$.
We also present standard model predictions for differential decay rate forward-backward asymmetries and polarization fractions and calculate potentially useful ratios of $B_s\to K$ form factors with those of the fictitious $B_s\to\eta_s$ decay.
Our lattice simulations utilize nonrelativistic QCD $b$ and highly improved staggered light quarks on a subset of the MILC Collaboration's $2+1$ asqtad gauge configurations, including two lattice spacings and a range of light quark masses.
\end{abstract}

\pacs{12.38.Gc,  % Lattice QCD
13.20.He, % bottom meson leptonic decays
14.40.Nd, % bottom mesons
14.40.Df} % strange mesons

\maketitle

%================================= BODY =================================

\section{Introduction}
\label{sec-Intro}
%%%%%%%%%%%%%%%%%%%%%%%%%%%%%%%%%%%%%%%%%%%%%%%%%%%%%%%
%%%%%%%%%%%%%%%%%%%%%%%%%%%%%%%%%%%%%%%%%%%%%%%%%%%%%%%
%%%%%%%%%%%%%%%%%%%%%%%%%%%%%%%%%%%%%%%%%%%%%%%%%%%%%%%

The decay $B_s \to K\ell\nu$ occurs at tree-level in the standard model via the flavor-changing charged-current $b \to u$ transition, making it an alternative to $B\to\pi\ell\nu$ in the determination of $|V_{ub}|$ from exclusive semileptonic decays.  
The difference in these processes, a spectator strange quark in $B_s\to K\ell\nu$ vs a spectator down quark in $B\to\pi\ell\nu$, is beneficial for lattice QCD simulations, because it improves the ratio of signal to noise.
Though this process has not yet been observed, its measurement is planned at LHCb and is possible during an $\Upsilon(5S)$ run at BelleII.  This provides a prediction opportunity for lattice QCD.

In addition to the calculation of form factors for $B_s\to~K$, we also calculate their ratios with form factors for the fictitious $B_s \to \eta_s$ decay.  Such ratios are essentially free of our largest systematic error, perturbative matching.  In combination with a future calculation of $B_s \to \eta_s$ using a highly improved staggered (HISQ) $b$ quark, these ratios would yield a non-perturbative evaluation of the matching factor for the $b \to u$ current with nonrelativistic QCD (NRQCD) $b$ quark.  This matching factor would be applicable to $B_s\to K \ell \nu$ and $B \to \pi \ell \nu$ simulations using NRQCD $b$ quarks.

To include correlations among the data for both decays, correlation function fits must include vast amounts of correlated data.  To make such fits feasible, we have developed a new technique, called chaining, discussed in Appendix~\ref{app-basics}.  
In addition, the use of marginalization techniques developed in Ref.~\cite{Hornbostel:2011} significantly reduces the time required for the fits.

The chiral, continuum, and kinematic extrapolations are performed simultaneously using the modified $z$~expansion~\cite{Na:2010, Na:2011} with the chiral logarithmic corrections fixed by the results of hard pion chiral perturbation theory (HPChPT)~\cite{Bijnens:2010, Bijnens:2011}.  
The factorization of chiral corrections and kinematics, as found at one-loop order by HPChPT, suggests the modified $z$~expansion is a natural choice for carrying out this simultaneous extrapolation.
We refer to the combination of HPChPT chiral logarithmic corrections and the modified $z$~expansion as the HPChPT $z$~expansion.

\section{Form factors and matrix elements}
\label{sec-FFsnMEs}
%%%%%%%%%%%%%%%%%%%%%%%%%%%%%%%%%%%%%%%%%%%%%%%%%%%%%%%
%%%%%%%%%%%%%%%%%%%%%%%%%%%%%%%%%%%%%%%%%%%%%%%%%%%%%%%
%%%%%%%%%%%%%%%%%%%%%%%%%%%%%%%%%%%%%%%%%%%%%%%%%%%%%%%
The vector hadronic matrix element is parametrized by the scalar and vector form factors $f_{0,+}$
\begin{eqnarray}
\langle K | V^\mu | B_s \rangle &=& f_+ \left( p_{B_s}^\mu +p_K^\mu - \frac{ M_{B_s}^2 - M_K^2 }{ q^2 }\,q^\mu \right) \nonumber \\
& &  +\ f_0\ \frac{ M_{B_s}^2 - M_K^2 }{ q^2 }\, q^\mu,
\end{eqnarray}
where $V^\mu = \bar{u} \gamma^\mu b$ and $q^\mu = p_{B_s}^\mu - p_K^\mu$.  At intermediate stages of the calculation we recast $f_{0,+}$ in terms of the more convenient form factors $f_{\parallel,\perp}$
\begin{equation}
\langle K | V^\mu | B_s \rangle = \sqrt{2M_{B_s}} \left( \frac{p_{B_s}^\mu}{M_{B_s}}\ f_\parallel + p_\perp^\mu\  f_\perp \right),
\end{equation}
where $p_\perp^\mu = p_K^\mu - p_{B_s}^\mu (p_K\cdot p_{B_s}) / M_{B_s}^2$.  In the $B_s$ meson rest frame, the form factors $f_{\parallel,\perp}$ are simply related to the temporal and spatial components of the hadronic vector matrix elements,
\begin{eqnarray}
\langle K | V^0 | B_s \rangle &=& \sqrt{2M_{B_s}}\ f_\parallel, \label{eq-fpardef} \\
\langle K | V^k | B_s \rangle &=& \sqrt{2M_{B_s}}\ p_K^k\ f_\perp. \label{eq-fperpdef}
\end{eqnarray}
The scalar and vector form factors are related to $f_{\parallel,\perp}$ by
\begin{eqnarray}
f_0 &=& \frac{\sqrt{2M_{B_s}}}{M_{B_s}^2-M_K^2} \left[ (M_{B_s}-E_K) f_{\parallel} + {\bf p}_K^2 f_{\perp} \right], \label{eq-f0def} \\
f_+ &=& \frac{1}{\sqrt{2M_{B_s}}}\left[ f_{\parallel} + (M_{B_s}-E_K) f_{\perp} \right], \label{eq-fplusdef}
\end{eqnarray}
where ${\bf p}_K$ is the kaon three-momentum.  This discussion generalizes in a straightforward way for the $B_s\to\eta_s$ matrix elements.

\section{Simulation}
\label{sec-Simulation}
%%%%%%%%%%%%%%%%%%%%%%%%%%%%%%%%%%%%%%%%%%%%%%%%%%%%%%%
%%%%%%%%%%%%%%%%%%%%%%%%%%%%%%%%%%%%%%%%%%%%%%%%%%%%%%%
%%%%%%%%%%%%%%%%%%%%%%%%%%%%%%%%%%%%%%%%%%%%%%%%%%%%%%%
\setlength{\tabcolsep}{0.1in}
\begin{table*}[t!]
\caption{Left to right:  labels for the ensembles used in this analysis; lattice volume; inverse lattice spacing in $r_1$-units; light/strange sea-quark masses; tadpole improvement factor $u_0 = \langle {\rm plaquette}\rangle^{\nicefrac{1}{4}}$; number of configurations; number of time sources; valence $u$-quark mass; valence $s$-quark mass; $b$-quark mass; and the spin-averaged $b\bar b$ ground state energies used to relate our $B_s$ meson simulation energies to their physical values.}
\begin{tabular}{lcccccccccc}
\hline\hline	
	\T Ensemble   	& $L^3\times N_t$	& $r_1/a$	& $au_0m_{\rm sea}$	& $u_0$ 	& $N_{\rm conf}$	& $N_{\rm tsrc}$	& $am_u$		& $am_s$ 	& $am_b$		& $aE_{b\bar b}^{\rm sim}$	\\ [0.5ex]
	\hline
	\T C1 	& $24^3 \times 64$	& 2.647(3)& 0.005/0.05			& 0.8678	& 1200			& 2				& 0.0070		& 0.0489		& 2.650		& 0.28356(15)	\\ [-0.2ex]
	\T C2	& $20^3 \times 64$	& 2.618(3)& 0.01/0.05			& 0.8677	& 1200			& 2				& 0.0123		& 0.0492		& 2.688 		& 0.28323(18)	\\ [-0.2ex]
	\T C3	& $20^3 \times 64$	& 2.644(3)& 0.02/0.05			& 0.8688	& 600			& 2				& 0.0246		& 0.0491		& 2.650		& 0.27897(20)	\\  [-0.2ex]
	\T F1		& $28^3 \times 96$	& 3.699(3)& 0.0062/0.031			& 0.8782	& 1200			& 4				& 0.00674		& 0.0337		& 1.832		& 0.25653(14)	\\ [-0.2ex]
	\T F2		& $28^3 \times 96$	& 3.712(4)& 0.0124/0.031			& 0.8788	& 600			& 4				& 0.01350		& 0.0336		& 1.826		& 0.25558(28)	\\ [0.5ex]
\hline\hline
\end{tabular}
\label{tab-ens}
\end{table*}

Ensemble averages are performed with the MILC Collaboration's $2+1$ asqtad gauge configurations~\cite{Bazavov:2010} listed in Table~\ref{tab-ens}.  Valence quarks in our simulation are nonrelativistic QCD (NRQCD)~\cite{Lepage:1992} $b$ quarks, tuned in Ref.~\cite{Na:2012}, and highly improved staggered (HISQ)~\cite{Follana:2007} light and $s$ quarks, the propagators for which were generated in Refs.~\cite{Na:2010, Na:2011}.  Valence quark masses for each ensemble used in the simulations are collected in Table~\ref{tab-ens} and correspond to pion masses ranging from, approximately, 260\,MeV to 500\,MeV.

\begin{figure}[t]
\vspace{0.0in}
\centering
%{\scalebox{0.6}{\includegraphics[angle=0,width=0.78\textwidth,bb = 175 330 445 480,clip]{/Users/cmb/Documents/writeups/BstoKlv/Diagram/diagram.pdf}}}
{\scalebox{0.6}{\includegraphics[angle=0,width=0.78\textwidth,bb = 175 330 445 480,clip]{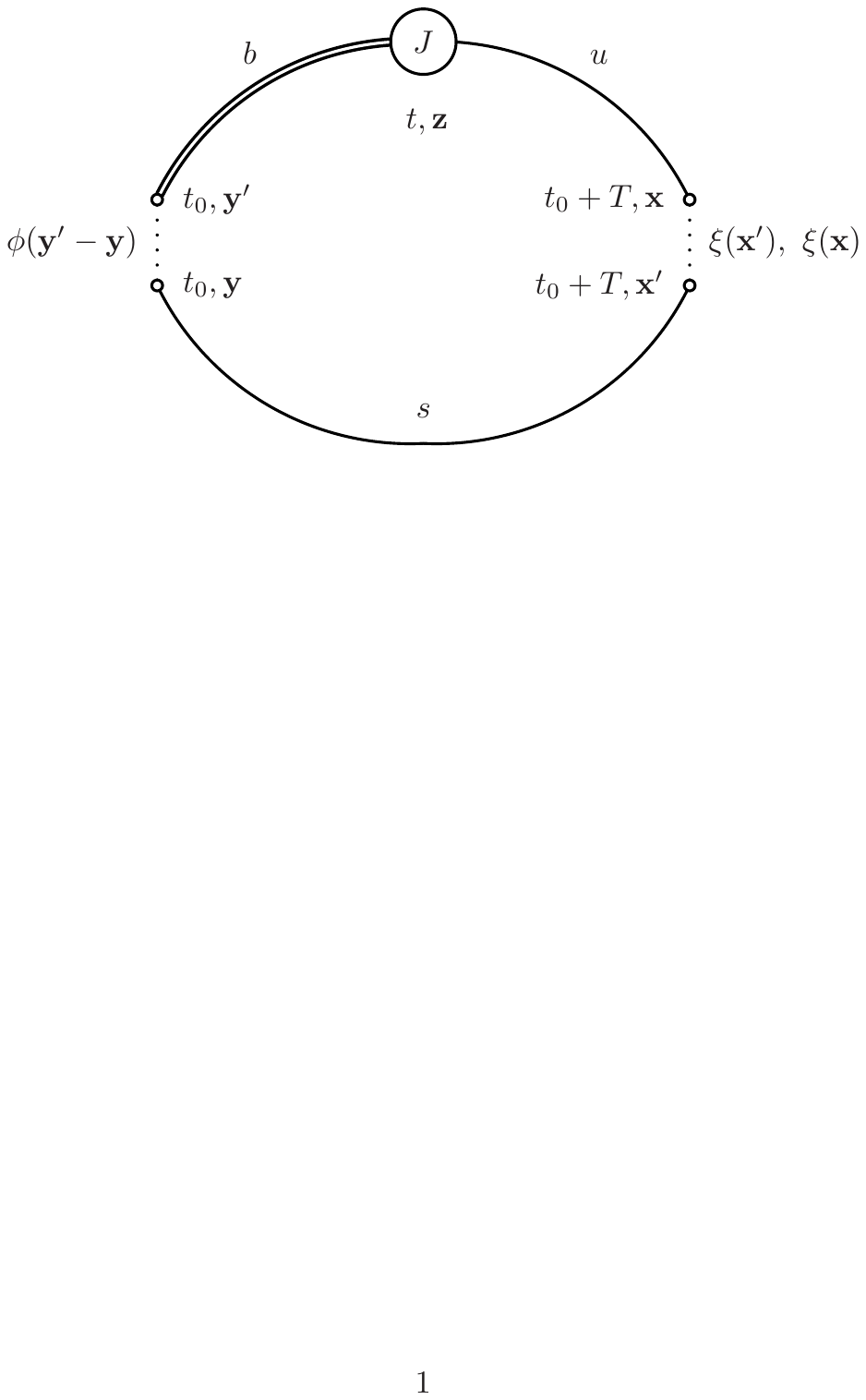}}}
\caption{Setup for three point correlator data generation.}
\vspace{0.0in}
\label{fig-feyndiag}
\end{figure}

Heavy-light $B_s$ meson bilinears $\Phi_{B_s}^\alpha$ are built from NRQCD $b$ and HISQ $s$ quarks (for details see Ref.~\cite{Na:2012}) and light-light kaon (and similarly for the $\eta_s$) bilinears $\Phi_K$ are built from HISQ light and $s$ quarks (for details see Ref.~\cite{Na:2010}).  From these bilinears we build two and three point correlation function data
\begin{eqnarray}
C^{\alpha\beta}_{B_s}(t_0,t) &=& \frac{1}{L^3} \sum_{{\bf x}, {\bf y}}  \langle \Phi^\beta_{B_s}(t,{\bf y})\ \Phi^{\alpha\dagger}_{B_s}(t_0,{\bf x}) \rangle, \label{eq-B2pt} \\
C_{K,{\bf p}}(t_0,t) &=& \frac{1}{L^3} \sum_{{\bf x}, {\bf y}}  e^{i\,{\bf p} \cdot ({\bf x} - {\bf y})}\langle \Phi_K(t,{\bf y})\ \Phi^\dagger_K(t_0,{\bf x}) \rangle, \nonumber \\
& & \label{eq-X2pt} \\
C^\alpha_{J,{\bf p}}(t_0,t,T) &=& \frac{1}{L^3} \sum_{{\bf x}, {\bf y}, {\bf z}} e^{i\,{\bf p}\cdot ({\bf z} - {\bf x})} \nonumber \\
& &\!\!\times \langle \Phi_K(t_0+T,{\bf x})\ J(t,{\bf z})\ \Phi^{\alpha\dagger}_{B_s}(t_0,{\bf y}) \rangle, \label{eq-3pt}
\end{eqnarray}
where indices $\alpha, \beta$ specify $b$ quark smearing.
We generate data for both a local and Gaussian smeared $b$ quark, with smearing function $\phi$ introduced via the replacement $\sum_{\bf y} \to \sum_{{\bf y}, {\bf y}'} \phi({\bf y}'-{\bf y})$ in Eqs.~(\ref{eq-B2pt}) and (\ref{eq-3pt}).
Three point and daughter meson two point correlation function data are generated at four daughter meson momenta, corresponding to ${\bf p}L\,\in\,2\pi\{(000), (100), (110), (111)\}$.  
In three point data, these momenta are inserted at ${\bf x}$ in Fig.~\ref{fig-feyndiag}.
The sum over ${\bf x}$ in Eqs.~(\ref{eq-X2pt}) and (\ref{eq-3pt}) is performed using random wall sources with U(1) phases $\xi$, i.e. $\sum_{\bf x} \to \sum_{{\bf x}, {\bf x}'} \xi({\bf x}) \xi({\bf x}')$.  
In the three point correlator a $B_s$ meson source is inserted at timeslice $t_0$, selected at random on each configuration to reduce autocorrelations.  
The current $J$ is inserted at timeslices $t$ such that $t_0\leq t\leq t_0+T$ and the daughter meson is annihilated at timeslice $t_0+T$.
Prior to performing the fits, all data are shifted to a common $t_0=0$.
This three point correlator setup is depicted in Fig.~\ref{fig-feyndiag}.
Additional details regarding the two and three point correlation function generation can be found in Ref.~\cite{Bouchard:2013a}.

The flavor-changing current $J$ is an effective lattice vector current $\mathcal{V}_\mu$ corrected through $\mathcal{O}(\alpha_s, \Lambda_{\rm QCD}/m_b, \alpha_s/(am_b))$.  The lattice currents that contribute through this order are
\begin{eqnarray}
\mathcal{V}_\mu^{(0)} &=& \overline{\Psi}_u\, \gamma_\mu\, \Psi_b, \\
\mathcal{V}_\mu^{(1)} &=&-\frac{1}{2 am_b} \overline{\Psi}_u\, \gamma_\mu\, {\boldsymbol \gamma} \cdot {\boldsymbol \nabla}\, \Psi_b.
\end{eqnarray}
Matrix elements of the continuum vector current $\langle V_\mu \rangle$ are matched to those of the lattice vector current according to
\begin{equation}
\langle V_\mu\rangle = (1+\alpha_s \rho_0^{(V_\mu)})\langle \mathcal{V}_\mu^{(0)}\rangle + \langle \mathcal{V}_\mu^{(1), {\rm sub}}\rangle ,
\label{eq-Vmatch}
\end{equation}
where
\begin{equation}
\langle \mathcal{V}_\mu^{(1), {\rm sub}}\rangle \equiv \langle \mathcal{V}_\mu^{(1)}\rangle - \alpha_s \zeta_{10}^{V_\mu} \langle \mathcal{V}_\mu^{(0)}\rangle.
\end{equation}
The matching calculation is done to one loop using massless HISQ lattice perturbation theory~\cite{Monahan:2013}.  
In implementing the matching, we omit $\mathcal{O}\left( \alpha_s\Lambda_{\rm QCD}/m_b \right)$ contributions.
Ref.~\cite{Gulez:2007}, which used asqtad valence quarks, found contributions of this order to be negligible.  
In Ref.~\cite{Bouchard:2013a}, which used HISQ valence quarks, these contributions to the temporal component of the vector current were studied and again were found to be negligible.  We also omit $\mathcal{O}(\Lambda_{\rm QCD}/m_b)^2$ relativistic matching corrections.  These, and higher order, omitted contributions to the matching result in our leading systematic error.  An estimate of this error, and its incorporation in our fit results, is discussed in the following section.

\section{Correlation function fits}
\label{sec-CorrFits}
%%%%%%%%%%%%%%%%%%%%%%%%%%%%%%%%%%%%%%%%%%%%%%%%%%%%%%%
%%%%%%%%%%%%%%%%%%%%%%%%%%%%%%%%%%%%%%%%%%%%%%%%%%%%%%%
%%%%%%%%%%%%%%%%%%%%%%%%%%%%%%%%%%%%%%%%%%%%%%%%%%%%%%%

Two and three point correlation function fit Ans\"atze, and the selection of priors, closely follows the methods of Ref.~\cite{Bouchard:2013a}.  Two point $B_s$ data are fit to
\begin{multline}
C^{\alpha \beta}_{B_s}(t) = \sum_{n=0}^{N-1} b^{\alpha(n)}b^{\beta(n)\dagger} e^{-E^{{\rm sim} (n)}_{B_s}t} \\
+ \sum_{m=0}^{\tilde{N}-1} \tilde{b}^{\alpha(m)}\tilde{b}^{\beta(m)\dagger}(-1)^te^{-\tilde{E}^{{\rm sim} (m)}_{B_s}t}, \label{eq-B2ptfit}
\end{multline}
where tildes denote oscillating state contributions and $E^{\rm sim}_{B_s}$ is the simulated $B_s$ energy.  The physical ground state $B_s$ mass is related to the simulation ground state energy by
\begin{equation}
E^{(0)}_{B_s} = E_{B_s}^{{\rm sim}(0)} + \frac{1}{2} (M_{b\bar b}^{\rm expt} - E_{b \bar b}^{\rm sim})
\end{equation}
where $M_{b\bar b}^{\rm expt} = 9.450(4)$ GeV~\cite{Gregory:2011} is adjusted from experiment to remove electromagnetic, $\eta_b$ annihilation, and charmed sea effects not present in our simulations, and $E^{\rm sim}_{b\bar b}$ is the spin-averaged energy of $b\bar b$ states calculated on the ensembles used in the simulation and listed in Table~\ref{tab-ens}. 
The $b$ quark smearing is indicated by indices $\alpha, \beta$.  
Kaon and $\eta_s$ two point correlator data are fit to an expression of the form\footnote{The zero momentum $\eta_s$ has no oscillating state contributions due to mass degeneracy of its valence quarks.} 
\begin{multline}
C_{{\bf p}}(t) = \sum_{n=0}^{N-1} |d_{{\bf p}}^{(n)}|^2 \big( e^{-E^{(n)}t} + e^{-E^{(n)}(N_t-t)} \big) \\
+ \sum_{m=0}^{\tilde{N}-1} | \tilde{d}_{{\bf p}}^{(m)}|^2 (-1)^t \big( e^{-\tilde{E}^{(m)}t} + e^{-\tilde{E}^{(m)}(N_t-t)} \big). \label{eq-X2ptfit}
\end{multline}
Results of two point fits satisfy the dispersion relation and are stable with respect to variations in $(N,\tilde{N})$ and the range of timeslices included in the fits, as demonstrated for kaon two point data in Ref.~\cite{Bouchard:2013a}.

Three point correlation function data are described by
%\begin{multline}
%C^{\alpha}_{J, {\bf p}}(t,T) = \sum_{n,m=0}^{N-1} d^{(n)}_{{\bf p}} A_{J, {\bf p}}^{(n,m)} b^{\alpha(m)\dagger}   \\
% \times\ e^{-E^{(n)}(T-t)} e^{-E^{{\rm sim} (m)}_{B_s}t} \\
% +  \sum_{n=0}^{N-1} \sum_{m=0}^{\tilde{N}-1} d^{(n)}_{{\bf p}} B_{J, {\bf p}}^{(n,m)} \tilde{b}^{\alpha(m)\dagger} (-1)^t \\
% \times\ e^{-E^{(n)}(T-t)} e^{-\tilde{E}^{{\rm sim} (m)}_{B_s}t} \\
% + \sum_{n=0}^{\tilde{N}-1} \sum_{m=0}^{N-1} \tilde{d}^{(n)}_{{\bf p}} C_{J, {\bf p}}^{(n,m)} b^{\alpha(m)\dagger} (-1)^{T-t} \\
% \times\ e^{-\tilde{E}^{(n)}(T-t)} e^{-E^{{\rm sim} (m)}_{B_s}t} \\
% + \sum_{n,m=0}^{\tilde{N}-1} \tilde{d}^{(n)}_{{\bf p}} D_{J, {\bf p}}^{(n,m)} \tilde{b}^{\alpha(m)\dagger} (-1)^T \\
% \times\ e^{-\tilde{E}^{(n)}(T-t)} e^{-\tilde{E}^{{\rm sim} (m)}_{B_s}t},
% \label{eq-3ptfit}
%\end{multline}
\begin{multline}
C^{\alpha}_{J, {\bf p}}(t,T) \\
= \sum_{n,m=0}^{N-1} d^{(n)}_{{\bf p}} A_{J, {\bf p}}^{(n,m)} b^{\alpha(m)\dagger} e^{-E^{(n)}(T-t)} e^{-E^{{\rm sim} (m)}_{B_s}t} \\
 +  \sum_{n=0}^{N-1} \sum_{m=0}^{\tilde{N}-1} d^{(n)}_{{\bf p}} B_{J, {\bf p}}^{(n,m)} \tilde{b}^{\alpha(m)\dagger} (-1)^t e^{-E^{(n)}(T-t)} e^{-\tilde{E}^{{\rm sim} (m)}_{B_s}t} \\
 + \sum_{n=0}^{\tilde{N}-1} \sum_{m=0}^{N-1} \tilde{d}^{(n)}_{{\bf p}} C_{J, {\bf p}}^{(n,m)} b^{\alpha(m)\dagger} (-1)^{T-t} e^{-\tilde{E}^{(n)}(T-t)} e^{-E^{{\rm sim} (m)}_{B_s}t} \\
 + \sum_{n,m=0}^{\tilde{N}-1} \tilde{d}^{(n)}_{{\bf p}} D_{J, {\bf p}}^{(n,m)} \tilde{b}^{\alpha(m)\dagger} (-1)^T e^{-\tilde{E}^{(n)}(T-t)} e^{-\tilde{E}^{{\rm sim} (m)}_{B_s}t},
 \label{eq-3ptfit}
\end{multline}
where the three point amplitudes $A$, $B$, $C$, and $D$ are proportional to the hadronic matrix elements.  
The ground state hadronic matrix element is obtained from $A^{(0,0)}$
\begin{equation}
\frac{4}{\sqrt{2}}A_{J,\bf p}^{(0,0)} = \frac{a^3 \langle K_{\bf p}^{(0)} |J| B_s^{(0)} \rangle }{\sqrt{2a^3E_K^{(0)}} \sqrt{2a^3 E^{(0)}_{B_s}}},
\end{equation}
where the factor of $4/\sqrt{2}$ accounts for numerical factors introduced in the simulation and associated with taste averaging and HISQ inversion.
In the correlator fits we include data for several temporal separations $T$ between the mother and daughter mesons.  On the coarse ensembles we include data for $T=13,14,15$  while for the fine ensembles we include $T=23, 24$ data.

On each ensemble we perform a simultaneous fit to two and three point correlation function data for the $B_s\to K$ and $B_s\to \eta_s$ decays, at all simulated momenta, including both spatial and temporal currents, and for the temporal separations listed above.
This ensures correlations among these data are accounted for in the analysis.
However, fits to such large data sets produce unwieldy data covariance matrices and are typically not convergent, or require a prohibitively large number of iterations.  This can be partially addressed by thinning the data, e.g. by the use of singular value decomposition (SVD) cuts, but this reduces the accuracy of the fits.

\begin{figure}[t!]
%{\scalebox{1.1}{\includegraphics[angle=-90,width=0.45\textwidth]{/Users/cmb/fitting/09July2013/fitscripts/plots/Marginalization/BsKVt100_v_Ninc-eps-converted-to.pdf}}}
{\scalebox{1.1}{\includegraphics[angle=-90,width=0.45\textwidth]{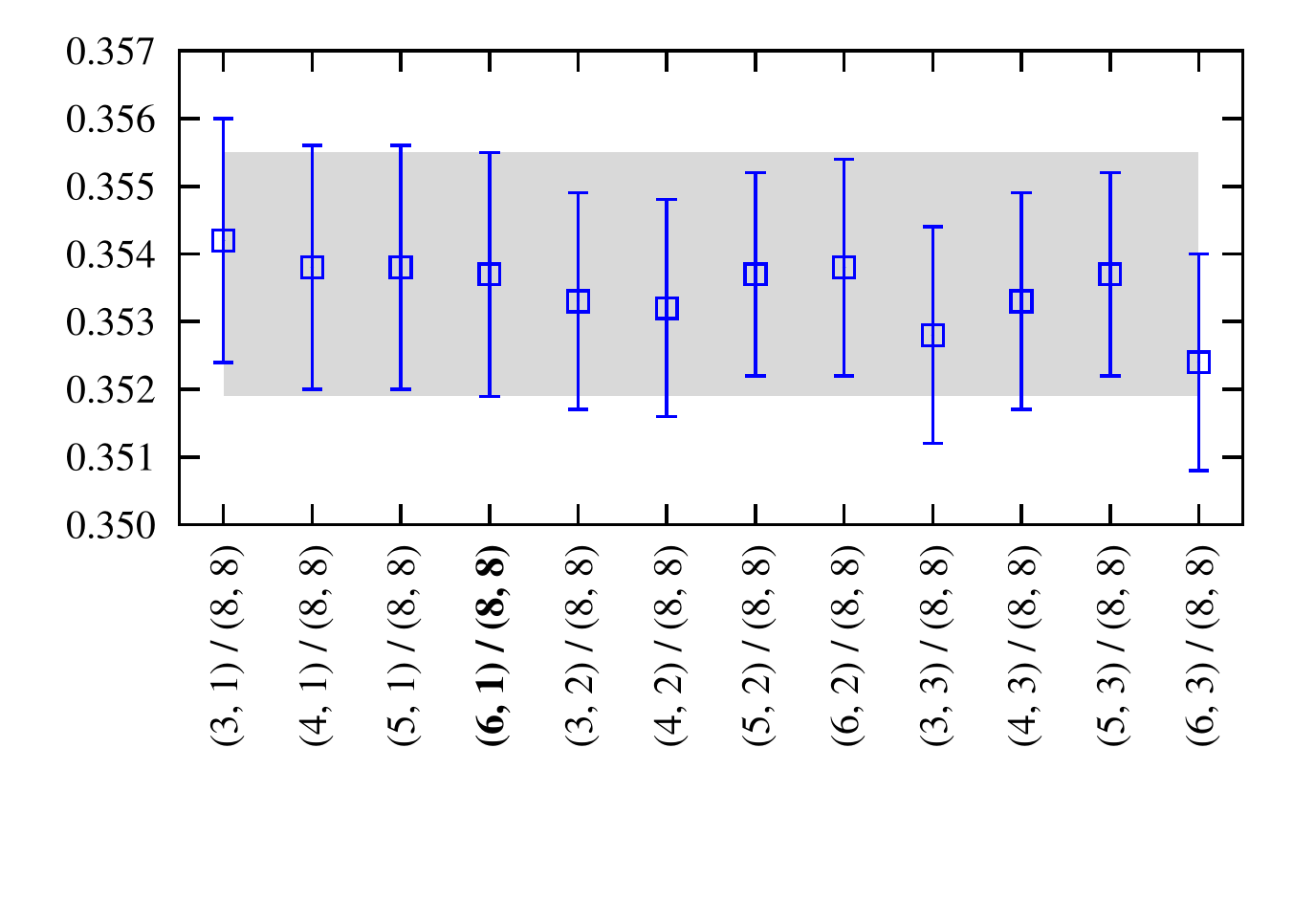}}}
\vspace{-0.1in}
%
%{\scalebox{1.0}{\includegraphics[angle=-90,width=0.5\textwidth]{/Users/cmb/fitting/09July2013/fitscripts/plots/Marginalization/BsKVt100_v_Ntot-eps-converted-to.pdf}}}
{\scalebox{1.0}{\includegraphics[angle=-90,width=0.5\textwidth]{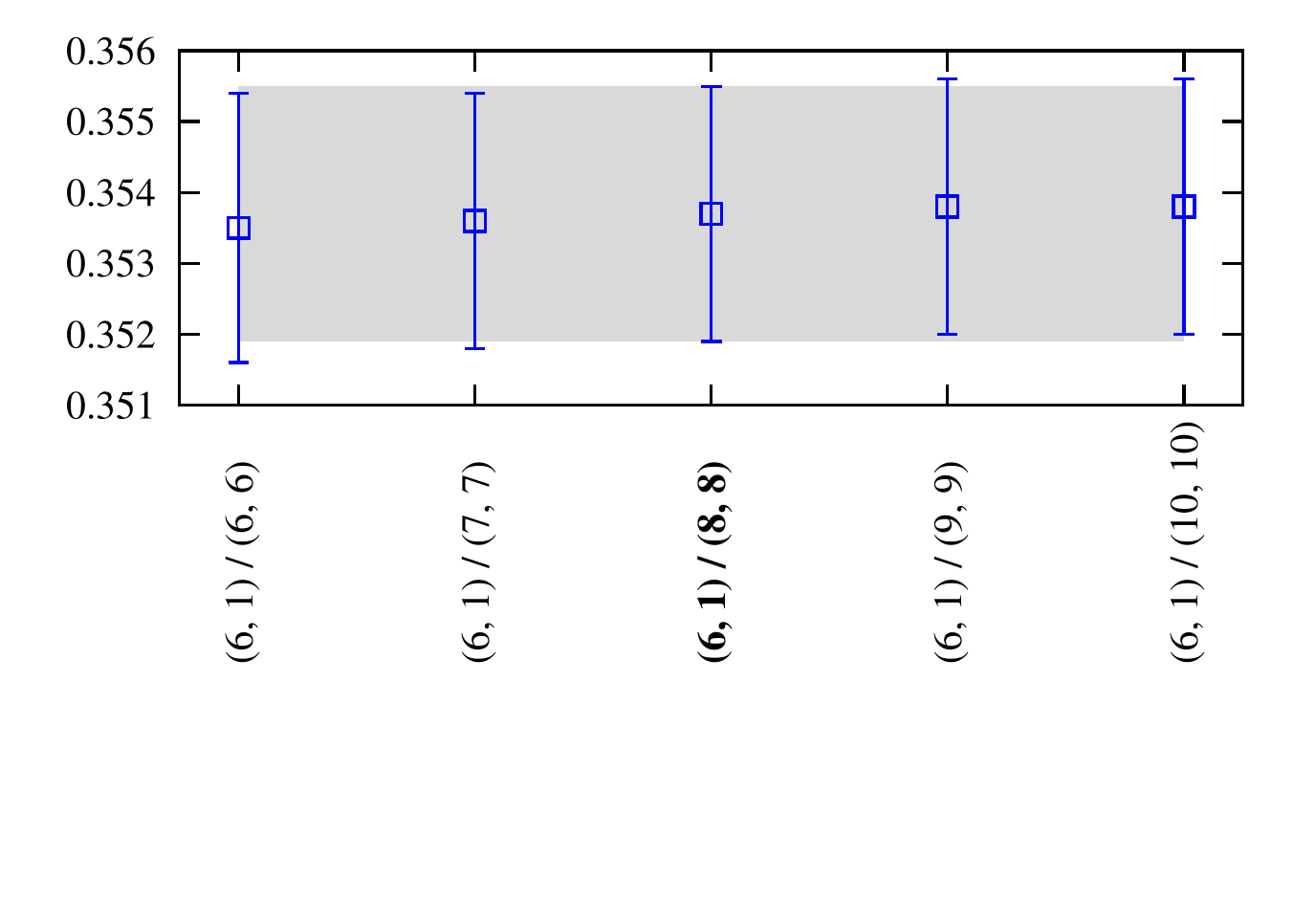}}}
\vspace{-0.45in}
\caption{(color online).  Chained and marginalized fit results for the ground state amplitude $A^{(0,0)}_{V_t,(1,0,0)}$ of the $B_s\to K$ decay on ensemble F2.  Fit results are shown as a function of the number of ({\it top}) states explicitly included in the fit and ({\it bottom}) total states accounted for in the fit.  Final results are taken from (6,1)/(8,8) fits, represented by gray bands.}
\label{fig-marg}
\end{figure}
To address this problem we introduce a technique, which we refer to as chaining, to simplify fits to very large data sets.  
Consider a data set consisting of $N$ correlators, ${\rm data}=({\rm correlator}_1, {\rm correlator}_2, \dots, {\rm correlator}_N)$.  
Before the fit, all fit parameters are assigned priors.  
Chaining first fits ${\rm correlator}_1$ then uses the best fit mean values and covariances to replace the corresponding priors in subsequent fits.  
The updated set of priors is then used in the fit to ${\rm correlator}_2$.  
In this and all subsequent fits, correlations are accounted for between the data being fit and those priors which are best fit results from previous fits\,---\,this is an important step as it prevents ``double counting" data.
After this second fit, the priors are again updated according to the best fit mean values and covariances.  This process is repeated for all correlators.  The collection of best fit mean values and covariances following the fit to ${\rm correlator}_N$ are the final fit results.  Chaining is described in greater detail in Appendix~\ref{app-basics}.  

We combine the use of Bayesian~\cite{Lepage:2002}, marginalized~\cite{Hornbostel:2011}, and chained fitting techniques.  
Our final fit results use marginalization with a total of $(N, \tilde{N})=(8,8)$ states accounted for, of which $(6,1)$ are explicitly fit.  We refer to such fits with the shorthand notation, $(6,1)/(8,8)$.  States accounted for but not explicitly fit are marginalized in that their contributions are subtracted from the data prior to the fit.  This technique reduces significantly the time required to perform the fits.
In Fig.~\ref{fig-marg} we show the stability of the fits under variations in the numbers of states explicitly included and the total number of states accounted for in the fit.
\setlength{\tabcolsep}{0.045in}
\begin{table}[t]
\caption{Fit results for the scalar and vector $B_s\to K$ form factors on each ensemble and for each simulated momentum.}
\begin{tabular}{lcccc}
\hline\hline
	\T Ensemble   	& $f^{B_s K}_0(000)$	& $f^{B_s K}_0(100)$	& $f^{B_s K}_0(110)$	& $f^{B_s K}_0(111)$ \\ [0.5ex]
	\hline
	\T C1 	& 0.8244(23)			& 0.7081(27)			& 0.6383(30)			& 0.5938(41)	\\  [-0.2ex]
	\T C2	& 0.8427(25)			& 0.6927(35)			& 0.6036(49)			& 0.536(12)	\\ [-0.2ex]
	\T C3	& 0.8313(29)			& 0.6953(33)			& 0.6309(30)			& 0.5844(46)	\\ [-0.2ex]
	\T F1		& 0.8322(25)			& 0.6844(35)			& 0.5994(43)			& 0.5551(56)	\\ [-0.2ex]
	\T F2		& 0.8316(27)			& 0.6915(38)			& 0.6119(43)			& 0.5563(61)	\\ [0.5ex]
\\ [-2.5ex]
	\T Ensemble   	&& $f^{B_s K}_+(100)$& $f^{B_s K}_+(110)$	& $f^{B_s K}_+(111)$ \\ [0.5ex]
	\hline
	\T C1 	&& 2.087(16)	& 1.657(14)	& 1.378(13)	\\ [-0.2ex]
	\T C2	&& 1.880(12)	& 1.412(16)	& 1.142(33)	\\ [-0.2ex]
	\T C3	&& 1.773(11)	& 1.4212(84)	& 1.184(10)	\\ [-0.2ex]
	\T F1		&& 1.878(13)	& 1.385(12)	& 1.158(13)	\\ [-0.2ex]
	\T F2		&& 1.834(14)	& 1.396(10)	& 1.163(14)	\\ [0.5ex]
\hline\hline
\end{tabular}
\label{tab-BsKcorrfits}
\end{table}
The $B_s\to K$ form factor results from the correlation function fits are tabulated in Table~\ref{tab-BsKcorrfits} and additional details are given in Appendix~\ref{sec-corrfit_results}.

\begin{figure}[t!]
%{\scalebox{1.1}{\includegraphics[angle=0,width=0.5\textwidth]{/Users/cmb/fitting/09July2013/fitscripts/modz/BstoK_Etas/plots/correlations/v9a2/C1_corr.pdf}}}
{\scalebox{1.1}{\includegraphics[angle=0,width=0.5\textwidth]{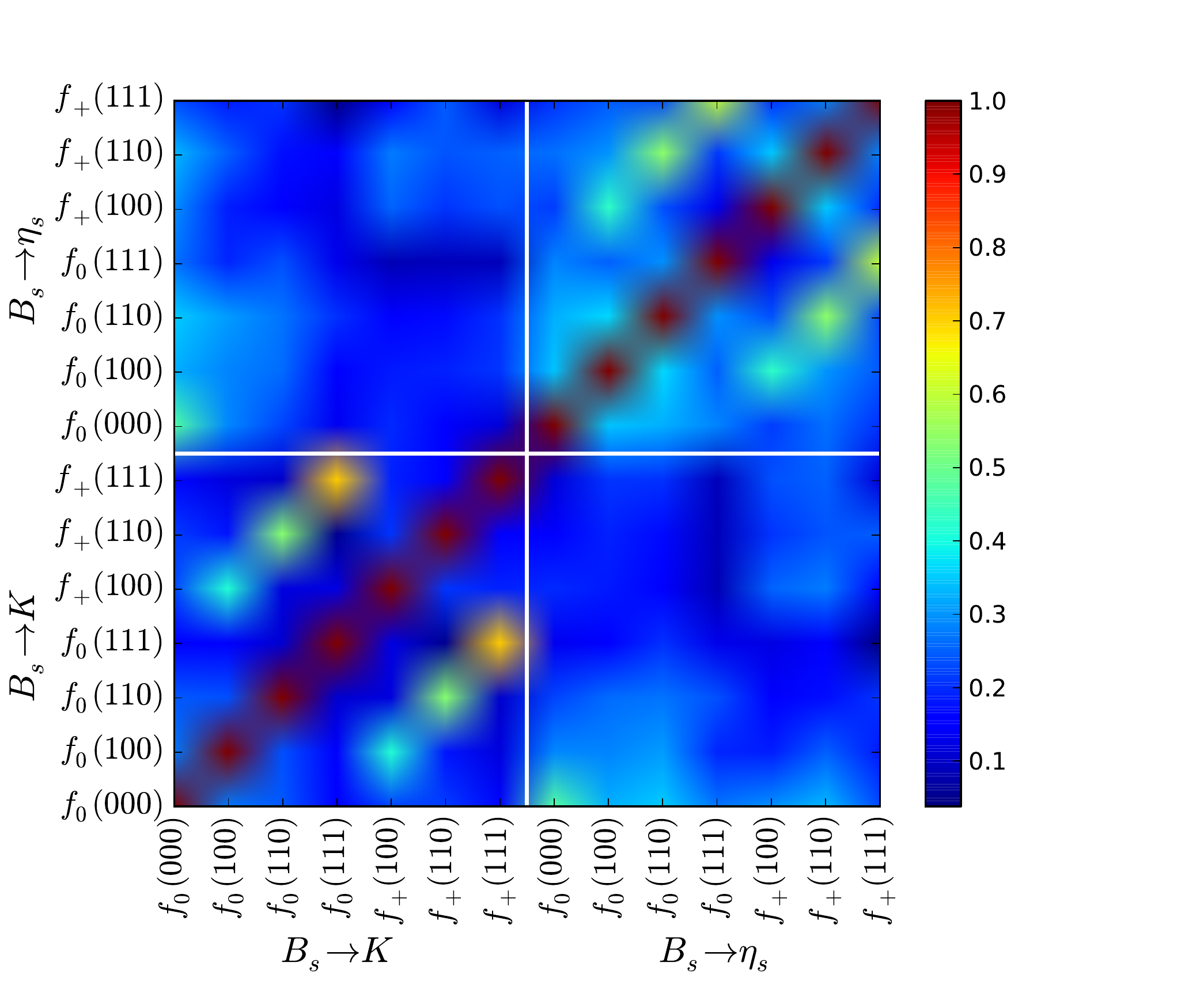}}}
%
%{\scalebox{1.0}{\includegraphics[angle=0,width=0.5\textwidth]{/Users/cmb/fitting/09July2013/fitscripts/modz/BstoK_Etas/plots/correlations/v9a2/corr_histo.pdf}}}
{\scalebox{1.0}{\includegraphics[angle=0,width=0.5\textwidth]{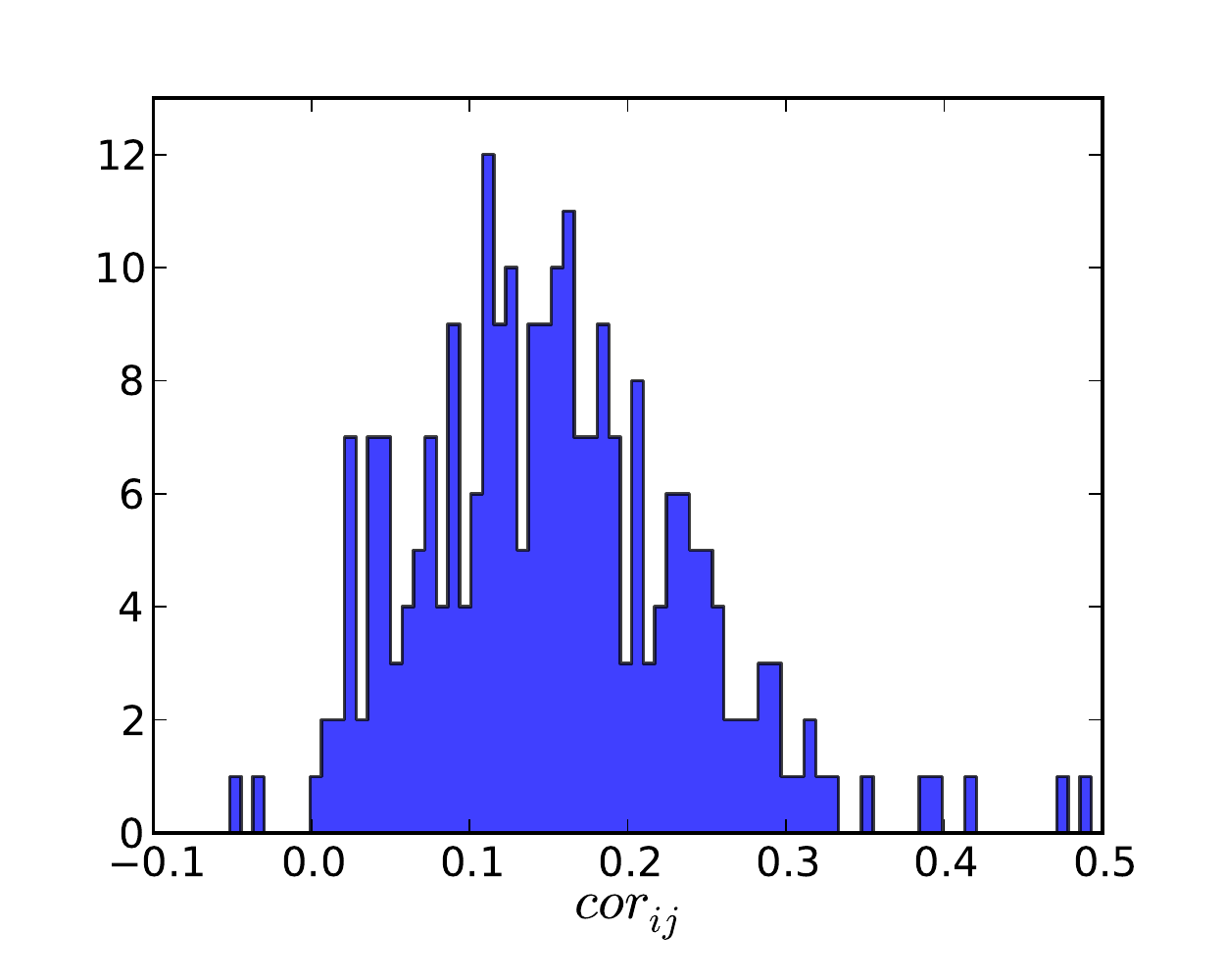}}}
\caption{(color online). ({\it top}) Heat map of the correlation matrix for ensemble C1.  ({\it bottom}) Distribution of correlations among the form factors for $B_s\to K$ and $B_s\to\eta_s$ for all ensembles.}
\label{fig-correlations}
\end{figure}
The form factors obtained from these fits preserve correlations resulting from shared gauge field configurations and quark propagators used in data generation.  The preservation of correlations is demonstrated in the top panel of Fig.~\ref{fig-correlations} where, e.g., significant correlations among the $B_s\to K$ form factor fit results are seen at common momenta and nonzero correlations among form factors for the two decays is suggested.  The bottom panel of Fig.~\ref{fig-correlations} shows the distribution over all ensembles of correlations among form factors for the two decays.  Accounting for these correlations is useful in our determination of the ratio of form factors for the two decays.
Fit results for $B_s\to\eta_s$, and the resulting form factor ratios, are presented in Appendix~\ref{sec-Ratio}.

\section{Chiral, continuum, and kinematic extrapolation}
\label{sec-Extrap}
%%%%%%%%%%%%%%%%%%%%%%%%%%%%%%%%%%%%%%%%%%%%%%%%%%%%%%%
%%%%%%%%%%%%%%%%%%%%%%%%%%%%%%%%%%%%%%%%%%%%%%%%%%%%%%%
%%%%%%%%%%%%%%%%%%%%%%%%%%%%%%%%%%%%%%%%%%%%%%%%%%%%%%%

The results of HPChPT~\cite{Bijnens:2010, Bijnens:2011} suggest a factorization, to at least one-loop order, of the soft physics of logarithmic chiral corrections and the physics associated with kinematics in the form factors describing semileptonic decays of heavy mesons,
\begin{equation}
f_{\parallel, \perp}(E) = (1 + [\text{logs}])\ \mathcal{K}_{\parallel, \perp}(E).
\label{eq-fact}
\end{equation}
The logarithmic chiral corrections, calculated in Ref.~\cite{Bijnens:2011} for several $B_{(s)}$ decays, are independent of $E$.  
An unspecified function $\mathcal{K}$ characterizes the kinematics.  

To obtain results over the full kinematic range one must include lattice simulation data over a range of energies.
However, for any relevant physical scale $\Lambda$ (e.g. $\Lambda_{QCD}$, $1/r_1,\ \Lambda_{\rm ChPT},\ \dots$), $E \gtrsim \Lambda$ at nominal lattice momenta and there is no convergent expansion of the unknown function $\mathcal{K}(E)$ in powers of $E/\Lambda$.  
This is an inherent limitation of characterizing the kinematics in terms of energy.  The energy of the daughter meson is a poor variable with which to describe the kinematics.

In contrast, the $z$~expansion~\cite{Boyd:1996, Arnesen:2005, Bourrely:2010} provides a convergent, model-independent characterization of the kinematics over the entire kinematically accessible range.
Combining a $z$~expansion on each ensemble\footnote{This assumes the general arguments on which the $z$~expansion is based hold for heavier than physical quark masses and at finite lattice spacing.} with the HPChPT inspired factorization of Eq.~(\ref{eq-fact}) allows a simultaneous chiral, continuum, and kinematic extrapolation of lattice data at arbitrary energies.
Because the chiral logs are the same for $f_{\parallel}$ and $f_{\perp}$, linear combinations (i.e. $f_{0}$ and $f_+$) factorize in the same way and have the same chiral logs.
Motivated by these observations, we construct a HPChPT-motivated modified $z$~expansion, which we call the ``HPChPT $z$~expansion", and 
 fit the lattice data of Tables~\ref{tab-BsKcorrfits} and~\ref{tab-BsEtascorrfits}, with accompanying covariance matrix, to fit functions of the form
\begin{eqnarray} \label{eq-genform}
P_{0,+}(q^2) f_{0,+}(q^2) &=& (1 + [\text{logs}]) \nonumber \\
&\times& \sum_{k=0}^{K} a_k^{(0,+)} D_k^{(0,+)} z(q^2)^k,
\label{eq-basicz}
\end{eqnarray}
where [logs] are the continuum HPChPT logs of Ref.~\cite{Bijnens:2011},  
and generic analytic chiral and discretization effects are accounted for by $D_k$. 
Resonances above $q^2_{\rm max}$ but below the $B_sK$ production threshold, i.e. those in the range $q^2_{\rm max} < q^2 < (M_{B_s}+M_K)^2$, are accounted for via the Blaschke factor, $P=1-q^2/M^2_{\rm res}$.
Though not observed, we allow for the possibility of a $J^P=0^+$ state in $P_0$, with choice of mass guided by Ref.~\cite{Gregory:2011}.
Our fit results are insensitive to the presence of this state.
The factorization suggested by HPChPT may not hold at higher order~\cite{Colangelo:2012} so we allow chiral analytic terms, which help parametrize effects from omitted higher order chiral logs, to have energy dependence (i.e. to vary with $k$).  

We note that Eq.~(\ref{eq-genform}) is the modified $z$~expansion introduced in Refs.~\cite{Na:2010, Na:2011}, with the coefficients of the chiral logarithmic corrections fixed by the results of HPChPT.
In the chiral and continuum limits
\begin{equation}
\lim_{\substack{
            m\to m_{\rm physical} \\
            a\to 0
            }}
(1+[\text{logs}])\, a_k D_k = b_k\ \text{of Ref.~\cite{Bourrely:2010}},
\label{eq-limit}
\end{equation}
and Eq.~(\ref{eq-genform}) is equivalent to the Bourrely-Caprini-Lellouch parametrization~\cite{Bourrely:2010} of the form factors.

Following Ref.~\cite{Bourrely:2010} we impose a constraint on $a_K^{(+)}$ from the expected scaling behavior of $f_+(q^2)$ in the neighborhood of $q^2_{\rm max}$.  The resulting fit function for $f_+$ is
%\begin{multline}
%P_+(q^2) f_+(q^2,a) = ( 1+[\text{logs}] ) \\
%\times \sum_{k=0}^{K-1} a_k^{(+)} D_k^{(+)}(a) \big[ z(q^2)^k - (-1)^{k-K} \frac{k}{K} z(q^2)^K \big].
%\label{eq-P+}
%\end{multline}
\begin{multline}
P_+(q^2) f_+(q^2,a) \\
= ( 1+[\text{logs}] ) \sum_{k=0}^{K-1} a_k^{(+)} D_k^{(+)}(a) \big[ z(q^2)^k - (-1)^{k-K} \frac{k}{K} z(q^2)^K \big].
\label{eq-P+}
\end{multline}
We write $f(q^2,a)$, $z(q^2)$, and $D_k(a)$, explicitly exposing the dependence on $q^2$ and $a$.  
This is useful in explaining the implementation of a second kinematic constraint we impose on the form factors.  
At the kinematic endpoint $q^2=0$, the continuum extrapolated form factors $f_0$ and $f_+$ are equal, i.e. $f_0(0,0) = f_+(0,0)$.  
We impose this constraint by fixing the coefficient $a_0^{(0)}$, 
\begin{multline}
a_0^{(0)} D_0^{(0)}(0) = - \sum_{k=1}^K a_k^{(0)} D_k^{(0)}(0) z(0)^k \\
 + \sum_{k=0}^{K-1} a_k^{(+)}D_k^{(+)}(0) \big[ z(0)^k - (-1)^{k-K} \frac{k}{K} z(0)^K \big].
\end{multline}
Imposing this constraint results in the fit function for $f_0$:
\begin{multline}
P_0(q^2) f_0(q^2,a) = ( 1+[\text{logs}] ) \\
 \times \bigg\{ \sum_{k=1}^{K} a_k^{(0)} \Big[ D_k^{(0)}(a) z(q^2)^k - \frac{ D_0^{(0)}(a) }{ D_0^{(0)}(0) } D_k^{(0)}(0) z(0)^k \Big]  \\
+ \frac{ D_0^{(0)}(a) }{ D_0^{(0)}(0) } \sum_{k=0}^{K-1} a_k^{(+)} D_k^{(+)}(0) \big[ z(0)^k - (-1)^{k-K} \frac{k}{K} z(0)^K \big] \bigg\}.
\label{eq-P0}
\end{multline}
In the fit functions for $f_0$ and $f_+$, Eqs.~(\ref{eq-P+}) and (\ref{eq-P0}), $D_k$ and [logs] are given by, 
\begin{eqnarray}
D_k &=& 1 + c^{(k)}_1 x_\pi + c^{(k)}_2 \big( \frac{1}{2}\delta x_\pi + \delta x_K \big)  \nonumber \\
&+& c^{(k)}_3 \delta x_{\eta_s} + d^{(k)}_1 (a/r_1)^2 + d^{(k)}_2 (a/r_1)^4 \nonumber \\
&+& e^{(k)}_1 (aE_K)^2 + e^{(k)}_2 (aE_K)^4, \label{eq-Dk} \\{}
[\text{logs}] &=& -\frac{3}{8} x_\pi ( \log x_\pi + \delta_{FV}) - \frac{1+6g^2}{4} x_K \log x_K \nonumber \\
&-& \frac{1+12g^2}{24}x_{\eta} \log x_{\eta}, \label{eq-logs}
\end{eqnarray}
with implicit indices in Eq.~(\ref{eq-Dk}) specifying the scalar or vector form factor.
We account for momentum-independent and momentum-dependent discretization effects in $D_k$.  
The values of $aE_K$ that enter the fit are the values from the simulation and are, of course, small.
Finite volume effects in the simulation are included via a shift $\delta_{FV}$ in the pion log~\cite{Bernard:2002}.  
The infinite volume limit is taken by setting this shift to zero.  
Eq.~(\ref{eq-logs}) gives the HPChPT~\cite{Bijnens:2011} result for the chiral logarithmic correction to $B_s\to K$ form factors.
These expressions make use of the dimensionless quantities
\begin{eqnarray}
x_{\pi, K, \eta} &=& \frac{M_{\pi, K, \eta}^2}{(4\pi f_\pi)^2} , \label{eq-x} \\
\delta x_{\pi, K} &=& \frac{(M^{\rm asqtad}_{\pi, K})^2 - (M^{\rm HISQ}_{\pi, K})^2}{(4\pi f_\pi)^2}, \label{eq-mixed} \\
\delta x_{\eta_s} &=& \frac{(M^{\rm HISQ}_{\eta_s})^2 - (M^{\rm physical}_{\eta_s})^2}{(4\pi f_\pi)^2},\label{eq-strange}
\end{eqnarray}
where $M_\eta^2 = (M^2_\pi + 2M^2_{\eta_s})/3$.
We determine $q^2$ and $z$ on each ensemble using correlator fit results for meson masses and simulation momenta.  Light and heavy quark discretization effects are accommodated for by making the $d_i^{(k)}$ mild functions of the masses, accomplished by the replacements
\begin{eqnarray}
d_1^{(k)} &\to& d_1^{(k)} (1 + l^{(k)} _1 x_\pi + l^{(k)} _2 x^2_\pi)  (1+ h^{(k)} _1 \delta x_b + h^{(k)} _2 \delta x_b^2 ), \nonumber \\
d_2^{(k)} &\to& d_2^{(k)} (1 + l^{(k)} _3 x_\pi + l^{(k)} _4 x^2_\pi)  (1+ h^{(k)} _3 \delta x_b + h^{(k)} _4 \delta x_b^2 ), \nonumber \\
\label{eq-disc}
\end{eqnarray}
where $\delta x_b = am_b - 2.26$ is chosen so that as $am_b$ varies over the coarse and fine ensembles $-0.4 \lesssim \delta x_b \lesssim 0.4$.

Lastly, we account for uncertainty associated with the perturbative matching of Sec.~\ref{sec-Simulation}.
With the matching coefficients calculated in Ref.~\cite{Monahan:2013}, we find $\mathcal{O}(\alpha_s, \Lambda_{\rm QCD}/m_b, \alpha_s/(am_b))$ contributions to be $\sim\!4\%$ of the total contribution to $\langle V_0\rangle$.  Of this $4\%$ the majority, $\sim\!3.5\%$, comes from the one loop $\mathcal{O}(\alpha_s)$ correction and $<\!1\%$ from the NRQCD matching via $\langle J_0^{(1),{\rm sub}}\rangle$.  For $\langle V_k\rangle$ we find contributions at this order to be $\sim\!2\%$, with $\sim\!1\%$ coming from the $\mathcal{O}(\alpha_s)$ correction and $<\!1\%$ from the NRQCD matching.  The matching error results from omitted higher order corrections, the size of which we estimate from observed leading order effects, where we conservatively use the larger 4\%.  
Following the arguments outlined in Ref.~\cite{Bouchard:2013a} we estimate the matching error to be the same size as the observed $\mathcal{O}(\alpha_s, \Lambda_{\rm QCD}/m_b, \alpha_s/(am_b))$ contributions and take the matching error to be 4\%.  
This is equivalent to taking the $\mathcal{O}(\alpha_s^2)$ matching coefficient to be four times larger than the $\mathcal{O}(\alpha_s)$ matching coefficient $\rho_0^{(V_0)}$ (13 times larger than $\rho_0^{(V_k)}$).  
This uncertainty is associated with the hadronic matrix elements and therefore, by Eqs.~(\ref{eq-fpardef}) and (\ref{eq-fperpdef}), with $f_\parallel$ and $f_\perp$.  To correctly incorporate it in the results for $f_0$ and $f_+$ we convert our fit functions for $f_{0,+}$ into $f_{\parallel, \perp}$, multiply by $(1+ m_{\parallel, \perp})$, where $m_{\parallel,\perp}$ is a coefficient representing the matching error with a prior central value of zero and width 0.04, then convert back to $f_{0,+}$ before performing the fit.  Schematically, we modify the fit functions, defined in Eqs.~(\ref{eq-P+}) and (\ref{eq-P0}), by
\begin{eqnarray}
&&\hspace{-0.3in} f_0, f_+ \to f_\parallel, f_\perp \\
&&\hspace{-0.3in} f_\parallel, f_\perp \to (1+m_\parallel)f_\parallel, (1+m_\perp)f_\perp \label{eq-matcherr} \\
&&\hspace{-0.3in} (1+m_\parallel)f_\parallel, (1+m_\perp)f_\perp \to f^{\rm corrected}_0, f^{\rm corrected}_+\!,
\end{eqnarray}
then we use $f^{\rm corrected}_{0,+}$ to fit the results of the correlation function fits of Sec.~\ref{sec-CorrFits}.
Conversions between the form factors $f_{0,+}$ and $f_{\parallel,\perp}$ are performed using Eqs.~(\ref{eq-f0def}) and (\ref{eq-fplusdef}).

\begin{figure}[t!]
%{\scalebox{1.0}{\includegraphics[angle=270,width=0.5\textwidth]{/Users/cmb/fitting/09July2013/fitscripts/modz/BstoK_Etas/plots/v9a2/f0p_BsK_vs_z_v9a2_k3_coarse_ensembles-eps-converted-to.pdf}}}
{\scalebox{1.0}{\includegraphics[angle=270,width=0.5\textwidth]{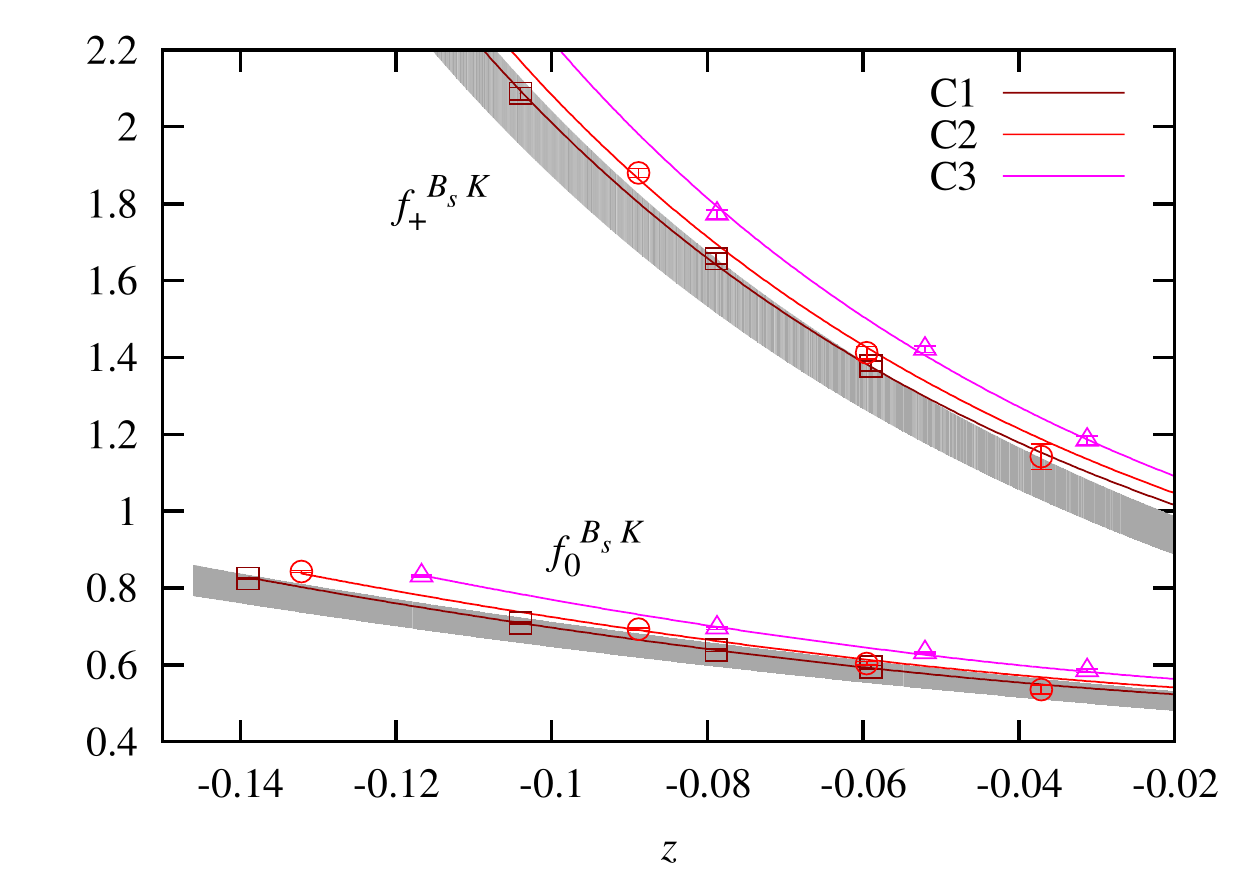}}}
%
%{\scalebox{1.0}{\includegraphics[angle=270,width=0.5\textwidth]{/Users/cmb/fitting/09July2013/fitscripts/modz/BstoK_Etas/plots/v9a2/f0p_BsK_vs_z_v9a2_k3_fine_ensembles-eps-converted-to.pdf}}}
{\scalebox{1.0}{\includegraphics[angle=270,width=0.5\textwidth]{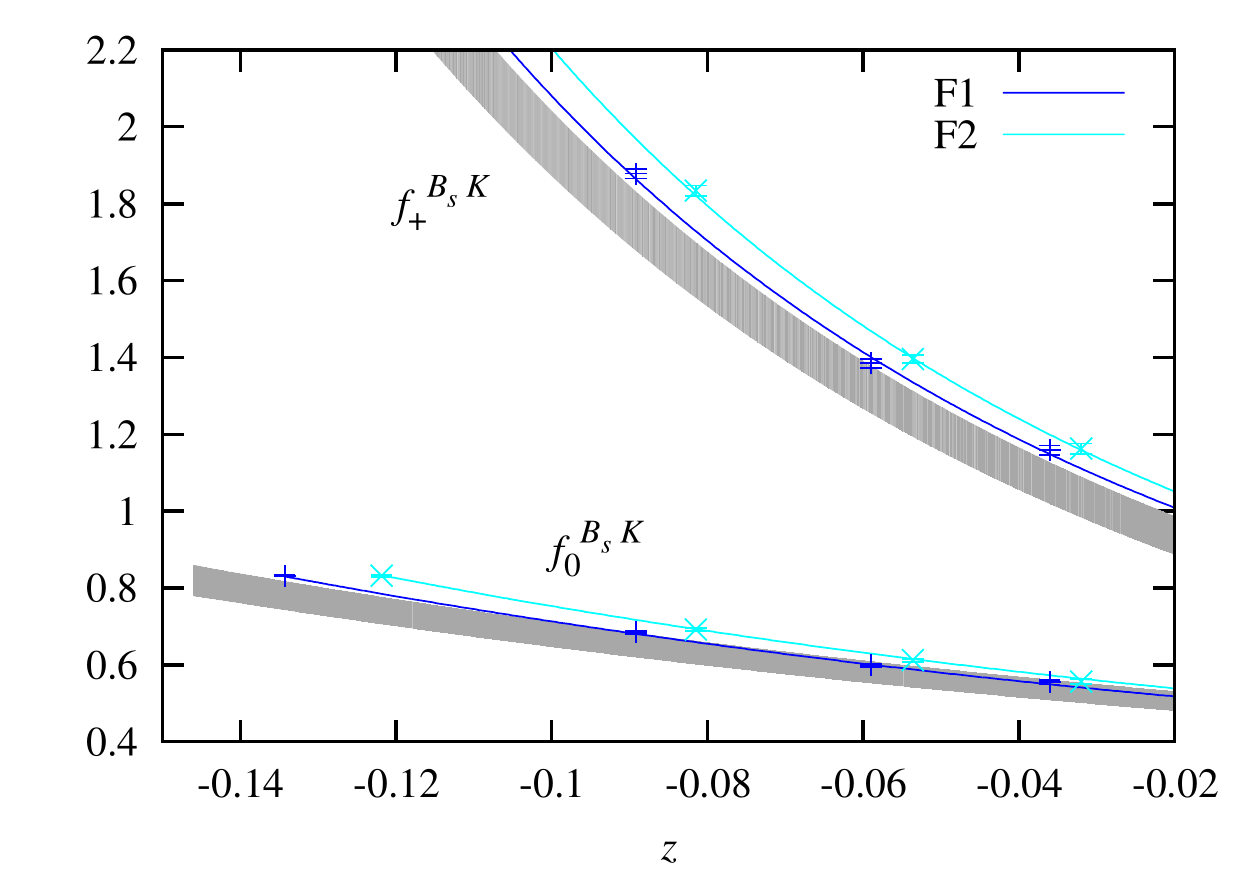}}}
\caption{(color online). $B_s\to K$ form factor results from a simultaneous chiral, continuum, and kinematic extrapolation via the HPChPT $z$~expansion are shown ({\it top}) relative to coarse ensemble data (C1, C2, and C3) and ({\it bottom}) relative to fine ensemble data (F1 and F2).}
\label{fig-HPChPTfit_BsK_ens}
\end{figure}
The results of a simultaneous fit to the data for $f_{0,+}^{B_sK}$ and $f_{0,+}^{B_s\eta_s}$, in which the maximum order of $z$ [specified by $K$ in Eqs.~(\ref{eq-P+}) and (\ref{eq-P0})] is 3 and $\chi^2/{\rm d.o.f.} = 84.0/70$, are shown relative to the data in Fig.~\ref{fig-HPChPTfit_BsK_ens} for $B_s\to K$.
Details of prior choices and fit results are given in Appendix~\ref{sec-HPChPTz_results}.

\begin{figure}[t!]
%{\scalebox{1.0}{\includegraphics[angle=270,width=0.5\textwidth]{/Users/cmb/fitting/09July2013/fitscripts/modz/BstoK_Etas/plots/stability_v9a2/BsK_Etas_v9a2_stab_chi2-eps-converted-to.pdf}}}
{\scalebox{1.0}{\includegraphics[angle=270,width=0.5\textwidth]{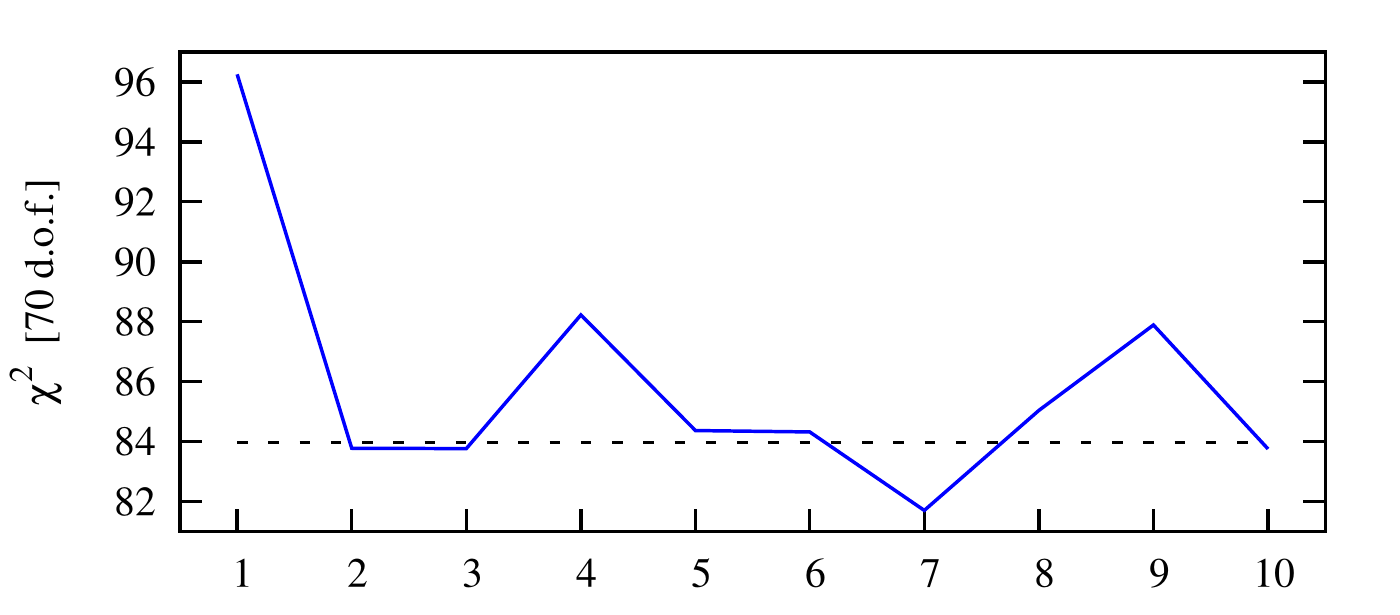}}}
\\
%{\scalebox{1.0}{\includegraphics[angle=270,width=0.505\textwidth]{/Users/cmb/fitting/09July2013/fitscripts/modz/BstoK_Etas/plots/stability_v9a2/BsK_Etas_v9a2_stab_BsK-eps-converted-to.pdf}}}
{\scalebox{1.0}{\includegraphics[angle=270,width=0.505\textwidth]{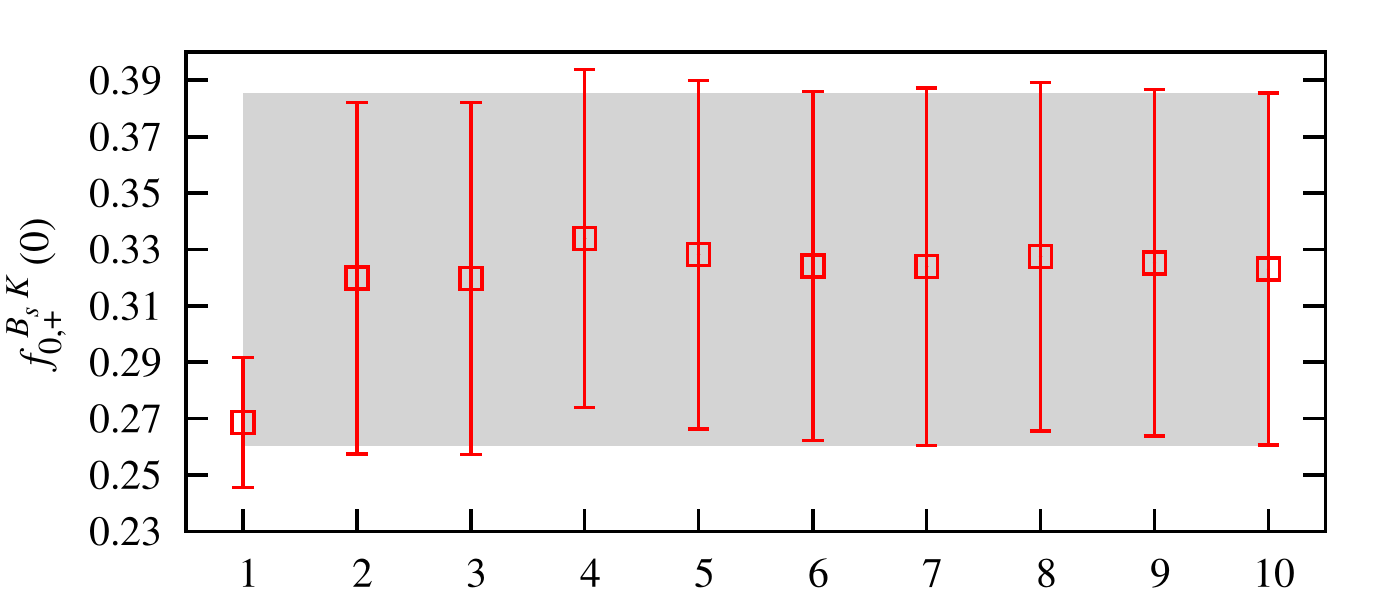}}}
\caption{(color online).  The stability of the HPChPT $z$~expansion is demonstrated by studying the fit results under various modifications, discussed in the text.  The top panel shows $\chi^2$ with 70 degrees of freedom (d.o.f.) for each test fit and the bottom panel shows form factors extrapolated to $q^2=0$.}
\label{fig-BsKstability}
\end{figure}

We test the stability of this fit to the following modifications of the fit Ans\"atze:
\begin{enumerate}[ {(}1{)} ] \itemsep-0.3em
\item Truncate the $z$~expansion at $\mathcal{O}(z^2)$.
\item Truncate the $z$~expansion at $\mathcal{O}(z^4)$.
\item Truncate the $z$~expansion at $\mathcal{O}(z^5)$.
\item Drop $\mathcal{O}(aE_K)^4$ momentum-dependent and $\mathcal{O}(a^4)$ momentum-independent discretization terms in Eq.~(\ref{eq-Dk}).
\item Drop the $am_b$-dependent discretization terms in Eq.~(\ref{eq-disc}).
\item Drop the light-quark mass-dependent discretization terms in Eq.~(\ref{eq-disc}).
\item Add the following next-to-next-to-leading-order (NNLO) chiral analytic terms to $D_k$ as defined in Eq.~(\ref{eq-Dk}):
\begin{eqnarray}
&& c_4^{(k)} x_\pi^2 +  c_5^{(k)}\big( \frac{1}{2}\delta x_\pi + \delta x_K \big)^2 + c_6^{(k)} \delta x^2_{\eta_s} \nonumber   \\
&&  +\ c_7^{(k)} x_\pi \big( \frac{1}{2}\delta x_\pi + \delta x_K \big) + c_8^{(k)} x_\pi \delta x_{\eta_s} \label{eq-NNLO}\\
&& +\ c_9^{(k)}\big( \frac{1}{2}\delta x_\pi + \delta x_K \big) \delta x_{\eta_s} + c_{10}^{(k)} x_\pi (a/r_1)^2 \nonumber \\
&& +\ c_{11}^{(k)}\big( \frac{1}{2}\delta x_\pi + \delta x_K \big)(a/r_1)^2 + c_{12}^{(k)} \delta x_{\eta_s} (a/r_1)^2 . \nonumber
\end{eqnarray}
\item Drop the sea- and valence-quark mass difference term $\big( \frac{1}{2}\delta x_\pi + \delta x_K \big)$ from Eq.~(\ref{eq-Dk}).
\item Drop the strange quark mistuning term $\delta x_{\eta_s}$ from Eq.~(\ref{eq-Dk}).
\item Drop finite volume effects, i.e. set $\delta_{FV}=0$ in Eq.~(\ref{eq-logs}).
\end{enumerate}
The stability of the $B_s\to K$ fit results to these modifications is shown in Fig.~\ref{fig-BsKstability}, where results are shown at the extrapolated $q^2=0$ point.  This point is furthest from the data region where simulations are performed and therefore is particularly sensitive to changes in the fit function.
In Fig.~\ref{fig-BsKstability} our final fit result, as defined by Eqs.~(\ref{eq-P+}) and (\ref{eq-P0}) with $K=3$ and by Eqs.~(\ref{eq-Dk})--(\ref{eq-disc}), is indicated by the dashed line and gray band.

Modifications 1, 2, and 3 vary the order of the truncation in $z$ and demonstrate that by $\mathcal{O}(z^3)$ fit results have stabilized and errors have saturated.  We therefore conclude that the error of the $\mathcal{O}(z^3)$ fit adequately accounts for the systematic error due to truncating the $z$~expansion.

Momentum-dependent and momentum-independent discretization effects proportional to $a^4$ are removed in modification 4.  This results in a modest increase in $\chi^2$ and a negligible shift in the fit result.  This suggests our final fit, which includes the $a^4$ effects, adequately accounts for all discretization effects observed in the data.

In modifications 5 and 6 we remove heavy- and light-quark mass-dependent discretization effects with essentially no impact on the fit.  That our results are independent of light-quark mass dependent discretization effects suggests that staggered taste violating effects are accommodated for by a generic $a^2$ dependence.  

Modification 7 tests the truncation of chiral analytic terms after next-to-leading-order (NLO) by adding the NNLO terms listed in Eq.~(\ref{eq-NNLO}).  This results in a slight decrease in $\chi^2$ but has no noticeable effect on the fit central value or error.  From this we conclude that errors associated with omitted higher order chiral terms are negligible.

Differences in sea and valence quark masses, due in part to our use of HISQ valence- and asqtad sea-quarks, are neglected in modification 8.  This results in a small increase in $\chi^2$ and negligible change in the fit results.  We account for these small mass differences in our final fit, though this test suggests they are unimportant in the fit.

Effects due to strange quark mass mistuning on the ensembles are omitted in modification 9, resulting in a modest increase in $\chi^2$ and no change in the fit central value and error.  We include these effects in our final fit.

Modification 10 results in nearly identical fit results, suggesting that finite volume effects are negligible in our data.  We include these effects in our final fit results.

\section{Form Factor Results}
\label{sec-FF_results}
%%%%%%%%%%%%%%%%%%%%%%%%%%%%%%%%%%%%%%%%%%%%%%%%%%%%%%%%%%%%%%%%%%%
%%%%%%%%%%%%%%%%%%%%%%%%%%%%%%%%%%%%%%%%%%%%%%%%%%%%%%%%%%%%%%%%%%%
\begin{figure}[t!]
%{\scalebox{1.0}{\includegraphics[angle=270,width=0.5\textwidth]{/Users/cmb/analysis/Observables/BstoK_Etas/FFs/BstoK_v9a2_f0fp_v_q2-eps-converted-to.pdf}}}
{\scalebox{1.0}{\includegraphics[angle=270,width=0.5\textwidth]{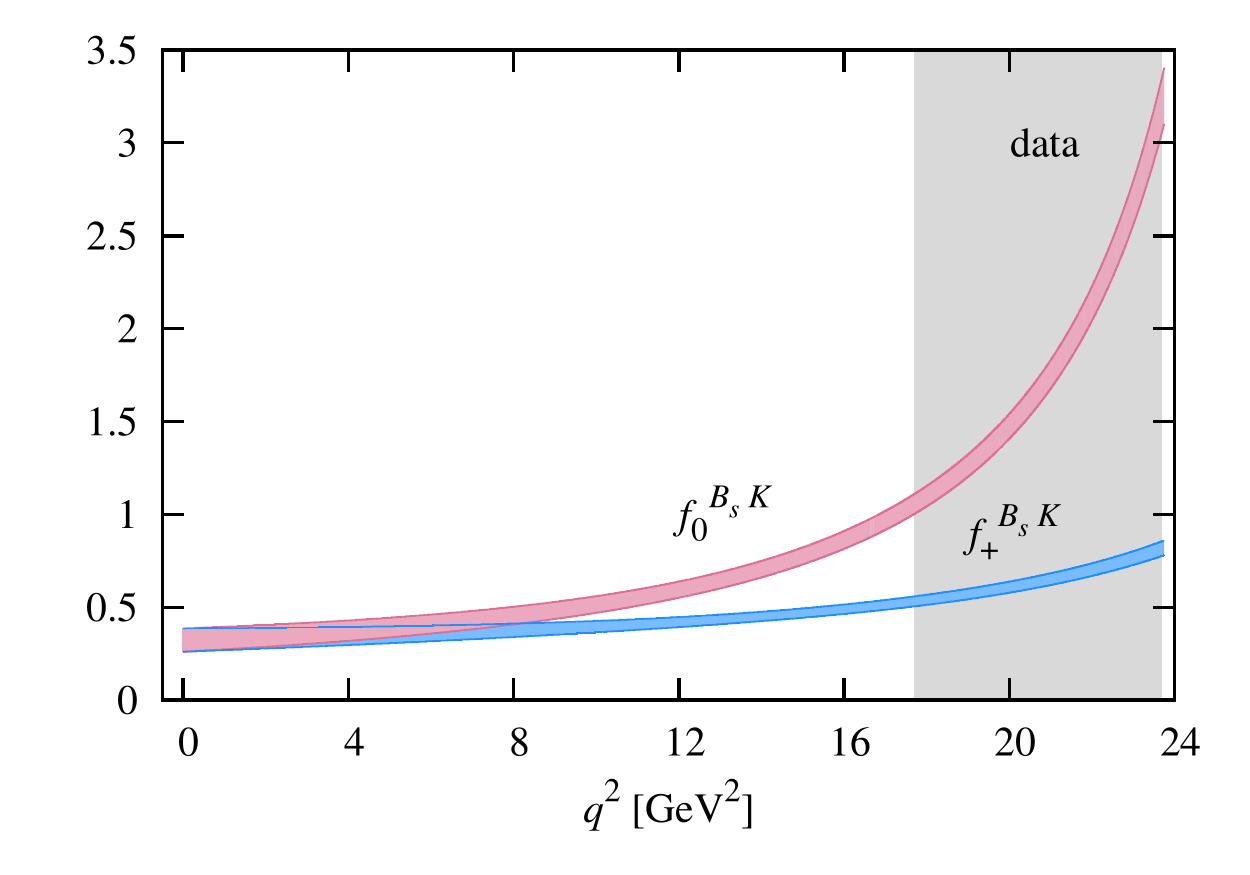}}}
\caption{(color online).  $B_s\to K$ form factor results from a simultaneous chiral, continuum, and kinematic extrapolation via the HPChPT $z$~expansion.  The $q^2$ region for which lattice simulation data exist is indicated by the shaded region.}
\label{fig-HPChPTfit_BsK_phys}
\end{figure}

In this section we present final results, with a complete error budget, for the $B_s\to K$ form factors.  We provide the needed information to reconstruct the form factors and compare our results with previous model calculations.

Fig.~\ref{fig-HPChPTfit_BsK_phys} shows the results of the chiral, continuum, and kinematic extrapolation of Sec.~\ref{sec-Extrap}, plotted over the entire kinematic range of $q^2$.  The form factors, extrapolated to $q^2=0$, have the value $f_{0,+}^{B_sK}(0)=0.323(63)$.

\subsection{Fit errors for the HPChPT $z$~expansion}
\label{sec-errbrkdwn}
\begin{figure}[h!]
\vspace{-0.03in}
%{\scalebox{1.0}{\includegraphics[angle=270,width=0.5\textwidth]{/Users/cmb/fitting/09July2013/fitscripts/modz/BstoK_Etas/plots/v9a2/BstoK_Etas_f0BsK_error_breakdown_v9a2_lines-eps-converted-to.pdf}}}
{\scalebox{1.0}{\includegraphics[angle=270,width=0.5\textwidth]{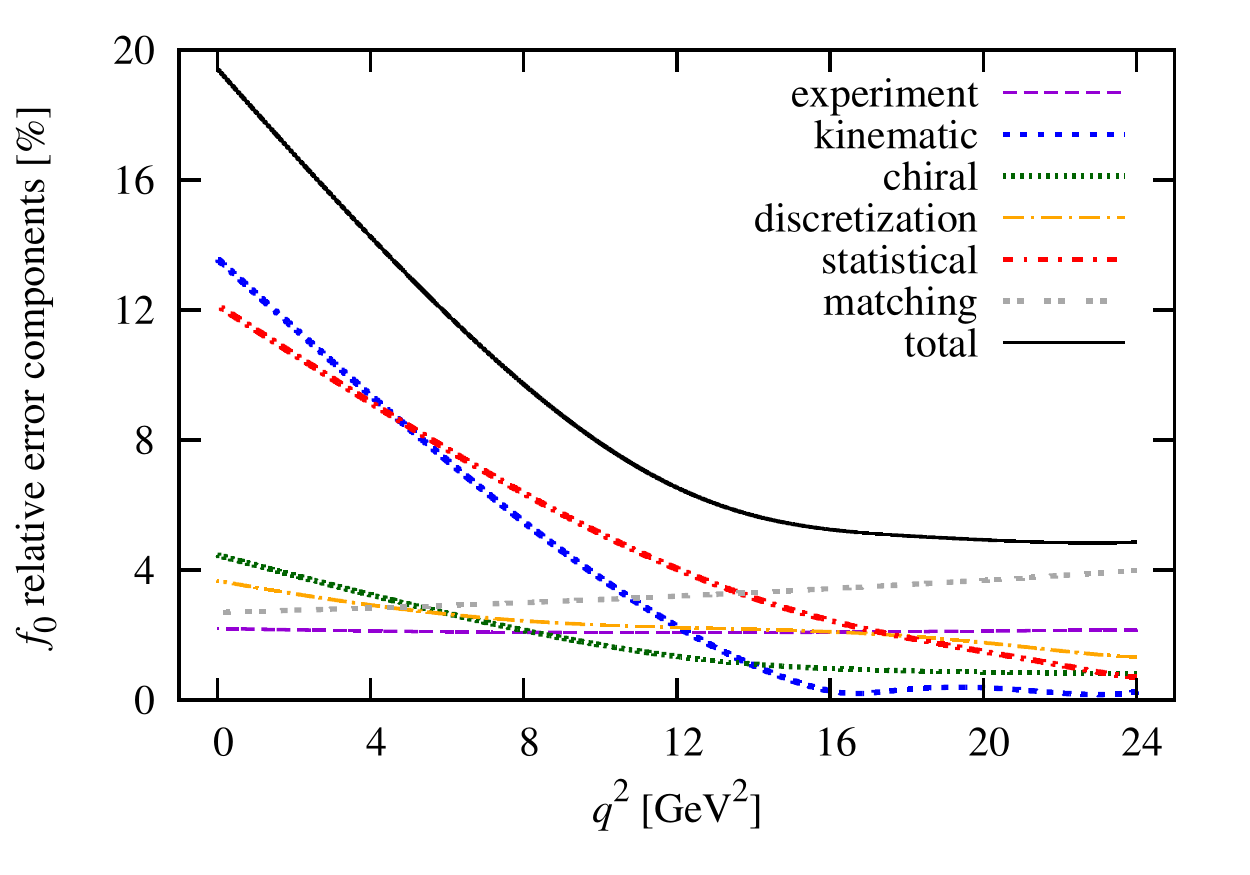}}}
\\
%{\scalebox{1.0}{\includegraphics[angle=270,width=0.5\textwidth]{/Users/cmb/fitting/09July2013/fitscripts/modz/BstoK_Etas/plots/v9a2/BstoK_Etas_fpBsK_error_breakdown_v9a2_lines-eps-converted-to.pdf}}}
{\scalebox{1.0}{\includegraphics[angle=270,width=0.5\textwidth]{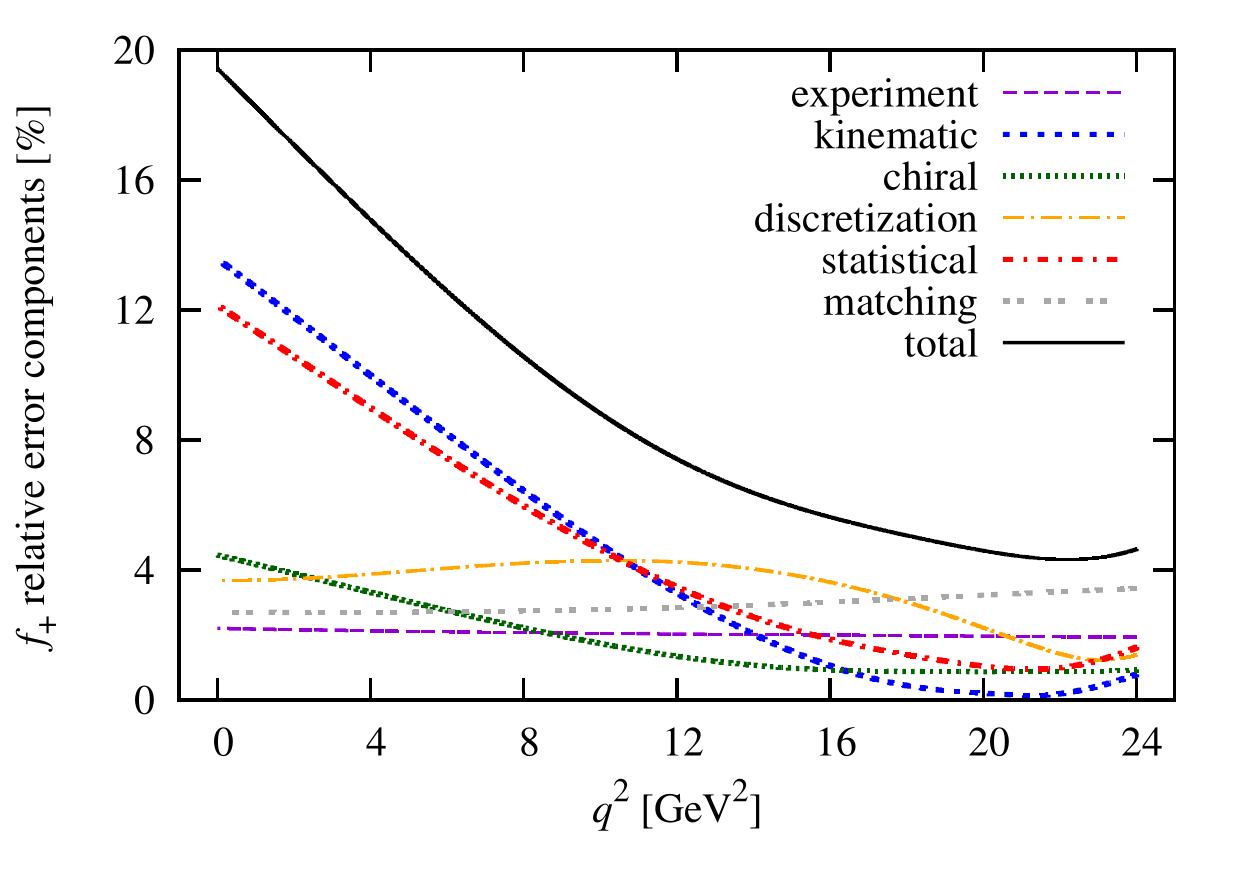}}}
\caption{\label{fig-modzerr} (color online). $B_s\to K$ ({\it top}) $f_0$ and ({\it bottom}) $f_+$ relative error components.  The total error (solid line) is the sum in quadrature of the components.}
\vspace{-0.03in}
\end{figure}

%%%%%%%%%%%%%%%%%%%%%%%%%%%%%%%%%%%%%%%%%%%%%%%%%%%%%%%%%%%%%%%%%%
%%%%%%%%%%%%%%%%%%%%%%%%%%%%%%%%%%%%%%%%%%%%%%%%%%%%%%%%%%%%%%%%%%
The inputs in our chiral, continuum, and kinematic extrapolation fits are data (the correlator fit results for $f_0$ and $f_+$ in Tables~\ref{tab-BsKcorrfits} and~\ref{tab-BsEtascorrfits} with the accompanying covariance matrix) and priors.  
The total hessian error of the fit can be described in terms of contributions from these inputs, as described in detail in Appendix~\ref{app-basics}.  
We group priors in a meaningful, though not unique, way and discuss the error associated with the chiral, continuum, and kinematic extrapolation based on these groupings.
As the priors are, by construction, uncorrelated with one another, we can group them together in any way we find meaningful.  The resulting error groupings are uncorrelated and add in quadrature to the total error.
In Fig.~\ref{fig-modzerr} we plot the following relative error components as functions of $q^2$:

\begin{enumerate}[ {(}i{)} ] \itemsep-0.3em

\item {\it experiment:} 
This is the error in the fit due to uncertainty of experimentally determined, and other, input parameters.  It is the sum in quadrature of the errors due to priors for the ``Group I" fit parameters listed in Table~\ref{tab-modzpriorsI}.  
This error is independent of $q^2$ and subdominant. 

\item {\it kinematic:}  
This error component is due to the priors for the coefficients $a_k^{(0,+)}$ in Eqs.~(\ref{eq-P+}) and (\ref{eq-P0}).  
A comparison of the fit results from modifications 1, 2, and 3 in Fig.~\ref{fig-BsKstability} shows that by $\mathcal{O}(z^3)$ the fit results have stabilized and errors have saturated.  
The kinematic error therefore includes the error associated with truncating the $z$~expansion.
The extrapolation to values of $q^2$ for which we have no simulation data is controlled by the $z$~expansion.  As a result, the growth in form factor errors away from the simulation region is due almost entirely to kinematic and statistical errors.

\item {\it chiral:} 
This error component is the sum in quadrature of errors associated with priors for $c_i^{(k)}$ in Eq.~(\ref{eq-Dk}).  These terms are responsible for extrapolating to the physical light quark mass and for accommodating for the slight strange quark mistuning and the small mismatch in sea and valence quark masses due to the mixed action used in the simulation.
As shown in Fig.~\ref{fig-modzerr}, these errors are subdominant and do not vary significantly with $q^2$.  

\item {\it discretization:}  
We account for momentum-dependent discretization effects via the $e_i^{(k)}$, and momentum-independent discretization effects via the $d_i^{(k)}$, terms of Eq.~(\ref{eq-Dk}).  In addition we allow for heavy- and light-quark mass-dependent discretization effects via the $h_i^{(k)}$ and $l_i^{(k)}$ terms in Eq.~(\ref{eq-disc}).
The discretization error component, which is essentially independent of $q^2$, is the sum in quadrature of the error due to the priors for these fit parameters.

\item {\it statistical:} 
The statistical component of the error is due to uncertainty in the data, i.e. the errors from form factor fit results of Table~\ref{tab-BsKcorrfits}.  Simulation data exist for $q^2 \gtrsim 17\ {\rm GeV}^2$ for $f_0$ and over the range $17\ {\rm GeV}^2 \lesssim q^2 \lesssim 22\ {\rm GeV}^2$ for $f_+$.  Extrapolation beyond these regions leads to increasing errors.

\item{\it matching:}
The matching error is due to the uncertainty associated with the priors for $m_{\parallel, \perp}$ introduced in Eq.~(\ref{eq-matcherr}) and discussed in the surrounding text.  

\end{enumerate}

In addition to the largest sources of error, which we account for directly in the fit, there are remaining systematic uncertainties.

We simulate with degenerate light quarks and neglect electromagnetism.  
By adjusting the physical kaon mass ($M_{K^\pm} \to M_{K^0}$) used in the chiral, continuum and kinematic extrapolation, we estimate the ``kinematic" effects of omitting electromagnetic and isospin symmetry breaking in our simulation to be $\lesssim 0.1\%$. 
It is more difficult to determine the size of the full effects.  However, in general electromagnetic and isospin effects are expected to be sub-percent.  
We assume the error in our form factor calculation due to these effects is negligible relative to other sources of uncertainty.

Our simulations include up, down, and strange sea quarks and we assume omitted charm sea quark effects are negligible.  This has been the case for processes in which it has been possible and appropriate to perturbatively estimate effects of charm quarks in the sea~\cite{Davies:2010}.

\begin{figure}[t!]
%{\scalebox{1.0}{\includegraphics[angle=270,width=0.5\textwidth]{/Users/cmb/analysis/Observables/BstoK_Etas/FFs/BstoK_v9a2_f0_v_q2_comparison-eps-converted-to.pdf}}}
{\scalebox{1.0}{\includegraphics[angle=270,width=0.5\textwidth]{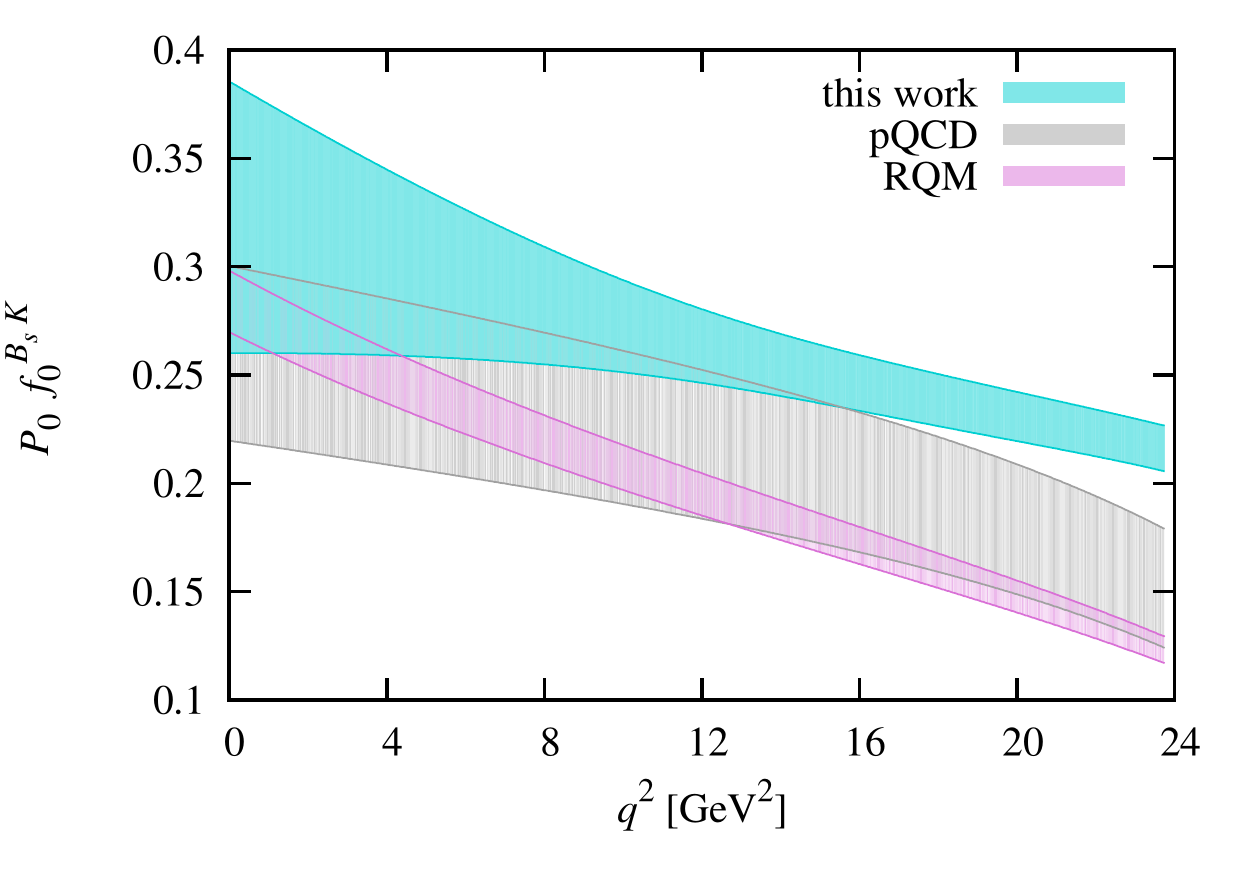}}}
%
%{\scalebox{1.0}{\includegraphics[angle=270,width=0.5\textwidth]{/Users/cmb/analysis/Observables/BstoK_Etas/FFs/BstoK_v9a2_fp_v_q2_comparison-eps-converted-to.pdf}}}
{\scalebox{1.0}{\includegraphics[angle=270,width=0.5\textwidth]{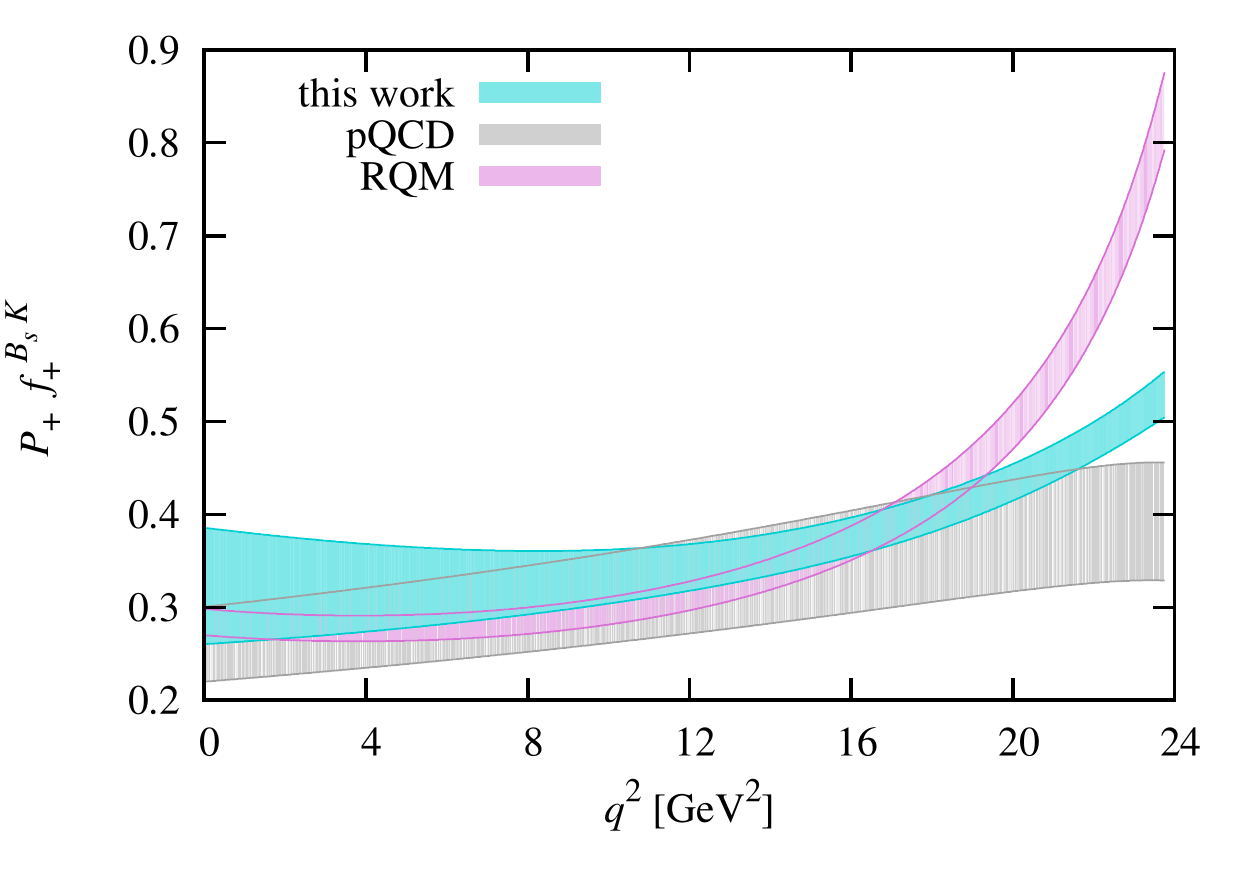}}}
\caption{(color online). Comparison of our $B_s\to K$ ({\it top}) $f_0$ and ({\it bottom}) $f_+$ form factors with those from a pQCD model~\cite{Wang:2012} and the RQM~\cite{Faustov:2013}.  To ease comparison the vertical scale is reduced by multiplying the form factors by the Blaschke factor $P_{0,+}$.}
\label{fig-FF_compare}
\end{figure}

Our final form factor results, multiplied by the Blaschke factor $P_{0,+}$, are shown in Fig.~\ref{fig-FF_compare} where they are compared with results from a model calculation using perturbative QCD (pQCD)~\cite{Wang:2012} and a relativistic quark model (RQM)~\cite{Faustov:2013}.  
Our results provide significant clarification on the form factors at large $q^2$.

\subsection{Reconstructing $B_s\to K \ell \nu$ Form Factors}
\label{sec-recon}
\setlength{\tabcolsep}{0.15in}
\begin{table*}[!t]
        \caption{({\it Top}) Physical extrapolated coefficients of the HPChPT $z$~expansion for the $B_s\to K$ form factors, defined in Eqs.~(\ref{eq-ezf0}) and (\ref{eq-ezf+}) and ({\it bottom}) the associated covariance matrix.}
        	\begin{tabular}{lcccccc}
        	\hline\hline
        	\T Coefficient		&		&		&		&		&		& Value		\\ [1ex]
        	\hline
        	\T $b^{(0)}_1$ 		&		&		&		&		&		& 0.315(129)	\\ [-0.2ex]
	\T $b^{(0)}_2$		&		&		&		&		&		& 0.945(1.305)	\\  [-0.2ex]
	\T $b^{(0)}_3$		&		&		&		&		&		& 2.391(4.671)	\\  [-0.2ex]
	\T $b^{(+)}_0$		&		&		&		&		&		& 0.3680(214)	\\  [-0.2ex]
	\T $b^{(+)}_1$		&		&		&		&		&		& -0.750(193)	\\  [-0.2ex]
	\T $b^{(+)}_2$		&		&		&		&		&		& 2.720(1.458)	\\ [0.5ex]
	\hline\hline
	\\[-2.0ex]
          \T  				& $b^{(0)}_1$	& $b^{(0)}_2$	& $b^{(0)}_3$	& $b^{(+)}_0$	& $b^{(+)}_1$	& $b^{(+)}_2$	 \\ [0.5ex]
	\hline
	& & & & & & \\ [-3ex]
	\T $b^{(0)}_1$		& $1.676\times 10^{-2}$  	& $1.462\times 10^{-1}$   	&  $4.453\times 10^{-1}$ 	& $1.165\times 10^{-3}$ 	& $2.140\times 10^{-2}$ 	&  $1.434\times 10^{-1}$ 	\\  [-0.2ex]
	\T $b^{(0)}_2$ 		& 		   			& $1.702$				&  $5.852$ 			& $9.481\times 10^{-3}$ 	&  $2.255\times 10^{-1}$ 	&  $1.539$ 			\\  [-0.2ex]
	\T $b^{(0)}_3$ 		& 		   			&  		  			& $2.181\times 10^{1}$ 	&  $2.963\times 10^{-2}$ 	&  $7.472\times 10^{-1}$ 	&  $5.325$ 			\\  [-0.2ex]
	\T $b^{(+)}_0$ 		&  		 			& 		  			& 			 		& $4.577\times 10^{-4}$ 	& $1.157\times 10^{-3}$ 	& $-1.309\times 10^{-3}$ \\  [-0.2ex]
	\T $b^{(+)}_1$ 		&  		  			& 					& 			 		& 		 			&  $3.721\times 10^{-2}$ 	&  $1.858\times 10^{-1}$ \\  [-0.2ex]
	\T $b^{(+)}_2$ 		&  		  			& 					& 		 	 		&  		 			& 					&  $2.124$			\\   [0.5ex]
	\hline\hline
        \end{tabular}
            \label{tab-coeffs}
\end{table*}
In the physical limit our form factor results are parametrized in a BCL~\cite{Bourrely:2010} form with coefficients $b_k^{(0,+)}$ [see Eq.~(\ref{eq-limit})].  Including the kinematic constraint and terms through order $z^3$, we have
\begin{align}
P_0(q^2) f_0(q^2) &= \sum_{k=1}^{3}b_k^{(0)} (z^k - z(0)^k) \nonumber \\
 + & \sum_{k=0}^{2} b_k^{(+)} \left[z(0)^k - (-1)^{k-3} \frac{k}{3} z(0)^3\right] , \label{eq-ezf0} \\
P_+(q^2) f_+(q^2) &= \sum_{k=0}^{2} b_k^{(+)} \left[z^k - (-1)^{k-3} \frac{k}{3} z^3\right] , \label{eq-ezf+}
\end{align}
where 
\begin{eqnarray}
z(q^2) &=& \frac{\sqrt{t_+ - q^2} - \sqrt{t_+-t_0} }{\sqrt{t_+ - q^2} + \sqrt{t_+-t_0} } ,\\
t_+ &=& (M_{B_s}+M_K)^2 ,\\
t_0 &=& (M_{B_s}+M_K) (\sqrt{M_{B_s}} - \sqrt{M_K})^2, \\
P_{0,+}(q^2) &=& 1-q^2/M_{0,+}^2,
\end{eqnarray}
and the resonance masses are $M_0=5.6794(10)\,{\rm GeV}$ and $M_+=5.32520(48)\,{\rm GeV}$.  
The values of the coefficients $b_k^{(0,+)}$, derived from the extrapolation fit results of Sec.~\ref{sec-Extrap}, and the associated covariance matrix, are given in Table~\ref{tab-coeffs}.  Note that it is necessary to take into account the correlations among the coefficients to correctly reproduce the form factor errors.

\section{Phenomenology}
\label{sec-pheno}
%%%%%%%%%%%%%%%%%%%%%%%%%%%%%%%%%%%%%%%%%%%%%%%%%%%%%%%
%%%%%%%%%%%%%%%%%%%%%%%%%%%%%%%%%%%%%%%%%%%%%%%%%%%%%%%
%%%%%%%%%%%%%%%%%%%%%%%%%%%%%%%%%%%%%%%%%%%%%%%%%%%%%%% 
With the benefit of {\it ab initio} form factors from lattice QCD, we explore the standard model implications of our results.  In this section we make standard model predictions for several observables related to the $B_s\to K\ell\nu$ decay for $\ell=\mu$ and $\tau$.

\begin{figure}[!t]
%{\scalebox{1.0}{\includegraphics[angle=270,width=0.5\textwidth]{/Users/cmb/analysis/Observables/BstoK_Etas/muon/BstoKmunu_v9a2_dZetadq2_v_q2-eps-converted-to.pdf}}}
{\scalebox{1.0}{\includegraphics[angle=270,width=0.5\textwidth]{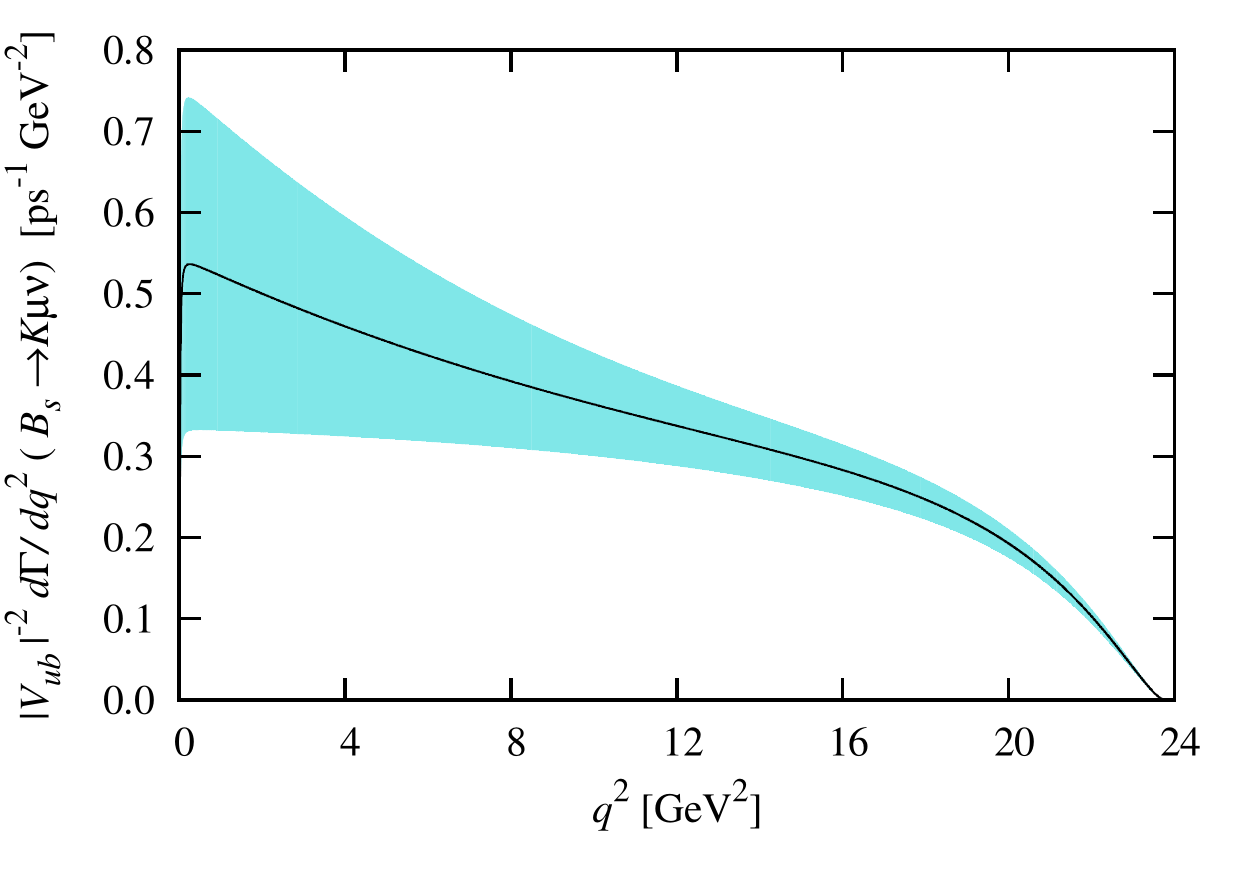}}}
\\
%{\scalebox{1.0}{\includegraphics[angle=270,width=0.5\textwidth]{/Users/cmb/analysis/Observables/BstoK_Etas/tau/BstoKtaunu_v9a2_dZetadq2_v_q2-eps-converted-to.pdf}}}
{\scalebox{1.0}{\includegraphics[angle=270,width=0.5\textwidth]{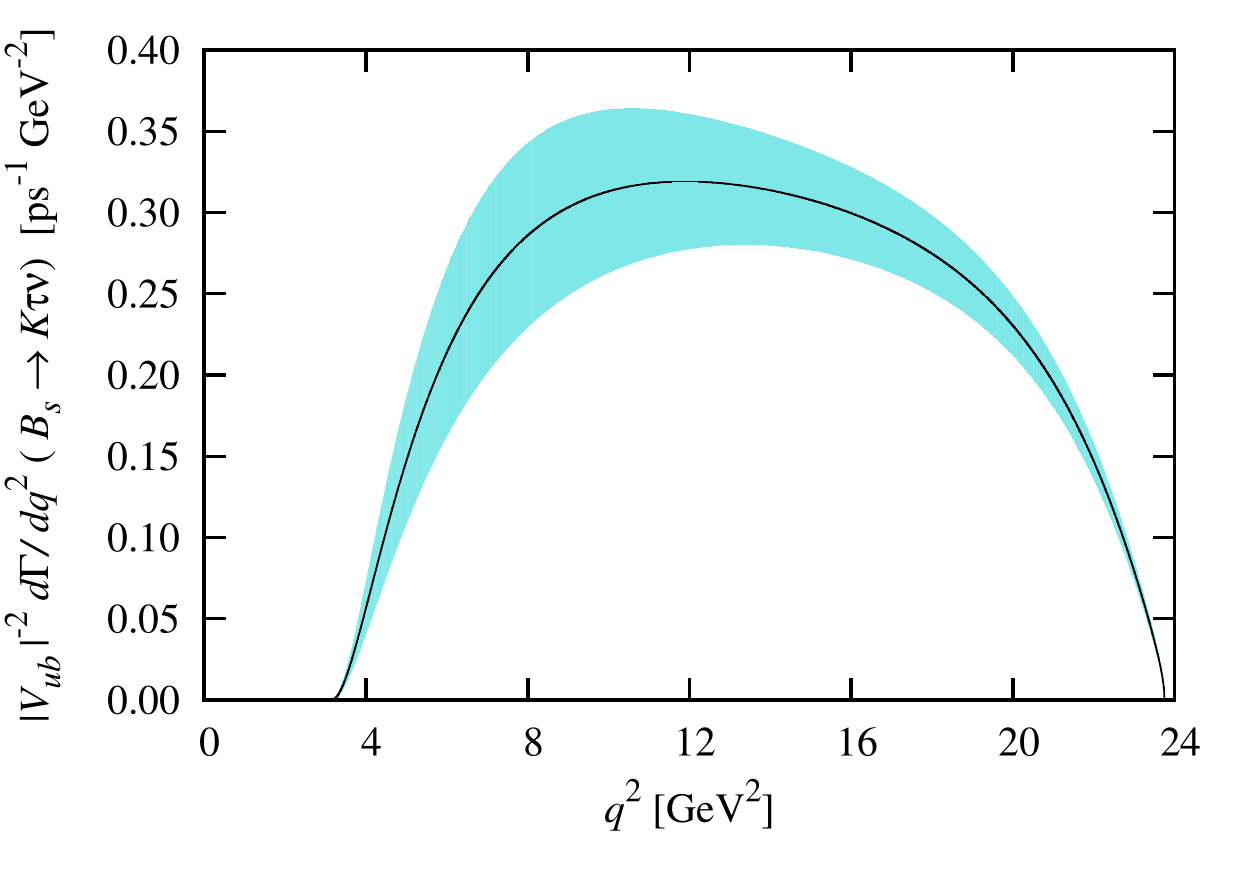}}}
\caption{(color online). Predicted differential decay rates, divided by $|V_{ub}|^2$, for ({\it top}) $B_s\to K \mu \nu$ and ({\it bottom}) $B_s\to K\tau\nu$.}
\label{fig-dZetadq2_mu}
\end{figure}

\begin{figure}[!t]
%{\scalebox{1.0}{\includegraphics[angle=270,width=0.5\textwidth]{/Users/cmb/analysis/Observables/BstoK_Etas/muon/BstoKmunu_v9a2_dBdq2_v_q2_incl_v_Belleexcl_Vubs-eps-converted-to.pdf}}}
{\scalebox{1.0}{\includegraphics[angle=270,width=0.5\textwidth]{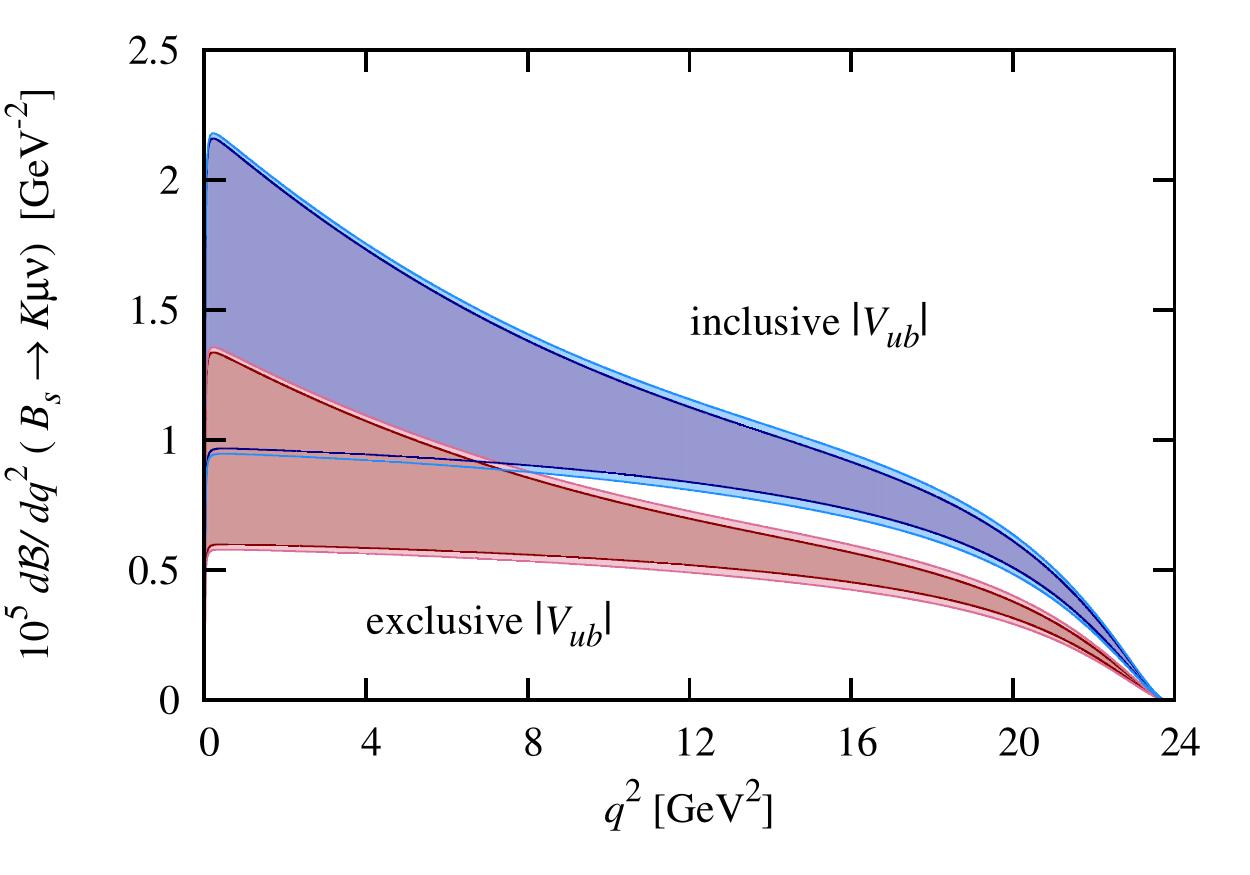}}}
\\
%{\scalebox{1.0}{\includegraphics[angle=270,width=0.5\textwidth]{/Users/cmb/analysis/Observables/BstoK_Etas/tau/BstoKtaunu_v9a2_dBdq2_v_q2_incl_v_Belleexcl_Vubs-eps-converted-to.pdf}}}
{\scalebox{1.0}{\includegraphics[angle=270,width=0.5\textwidth]{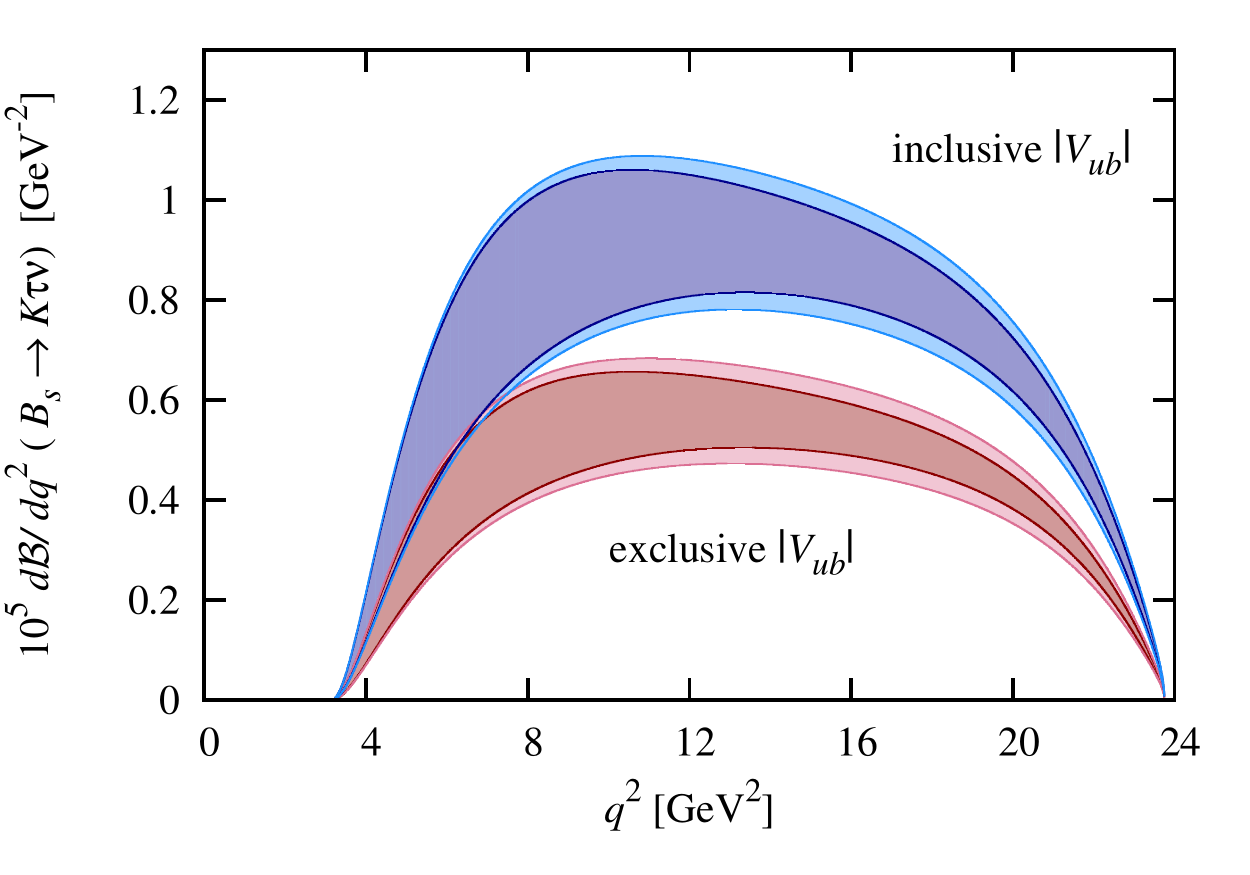}}}
\caption{(color online). Predicted differential branching fractions for the ({\it top}) $B_s\to K\mu\nu$ and ({\it bottom}) $B_s\to K\tau\nu$ decays using inclusive and exclusive semileptonic determinations of $|V_{ub}|$.  In each band, the light outer band includes all sources of error and the dark interior band neglects the uncertainty in $|V_{ub}|$.}
\label{fig-dBdq2_diffVub}
\end{figure}

The standard model $B_s\to K \ell \nu$ differential decay rate is related to the form factors by
\begin{multline}
\frac{d\Gamma}{dq^2} = \frac{G_F^2 |V_{ub}|^2}{24\pi^3 M^2_{B_s}} \Big(1-\frac{m_\ell^2}{q^2}\Big)^2\ |{\bf p}_K| \bigg[  \Big( 1+\frac{m_\ell^2}{2q^2} \Big) M^2_{B_s} {\bf p}_K^2 |f_+|^2 \\
+\ \frac{3m_\ell^2}{8q^2} (M^2_{B_s} - M^2_K)^2 |f_0|^2\bigg].
\end{multline}
In Fig.~\ref{fig-dZetadq2_mu} we plot predicted differential decay rates for $B_s\to K \mu \nu$ and $B_s\to K \tau \nu$, divided by $|V_{ub}|^2$, over the full kinematic range of $q^2$.
The ratio $\Gamma/|V_{ub}|^2$ can be combined with experimental results for the decay rates, typically differential decay rates integrated over $q^2$ bins, to allow the determination of $|V_{ub}|$.  In Eqs.~(\ref{eq-Zetamu}) and (\ref{eq-Zetatau}) we give numerical results for $d\Gamma/dq^2$, integrated over the kinematically accessible regions of $q^2$,
\begin{eqnarray}
\Gamma(B_s\to K\mu\nu)/|V_{ub}|^2 &=& 7.75(1.52)\ {\rm ps}^{-1}, \label{eq-Zetamu} \\
\Gamma(B_s\to K\tau\nu)/|V_{ub}|^2 &=& 4.92(0.60)\ {\rm ps}^{-1}. \label{eq-Zetatau}
\end{eqnarray}
Combining our form factor results with the current\footnote{For inclusive $|V_{ub}|$ we take the value from the Particle Data Group~\cite{PDG:2012}.  For the exclusive determination we use the ``global lattice~+~Belle" results reported by the FLAG-2 collaboration~\cite{FLAG2}.} inclusive and exclusive semileptonic determinations of $|V_{ub}|$,
\begin{eqnarray}
\text{exclusive } |V_{ub}| &=& 3.47(22) \times 10^{-3}, \label{eq-excl}  \\
\text{inclusive } |V_{ub}|  &=& 4.41(22) \times 10^{-3}, \label{eq-incl}
\end{eqnarray}
we demonstrate in Fig.~\ref{fig-dBdq2_diffVub} the potential of this decay to shed light on this $\sim\!3\sigma$ discrepancy.  In this and subsequent figures, dark interior bands represent the error in the differential branching fractions omitting the error associated with $|V_{ub}|$.  Experimental errors commensurate with these predictions, especially for the $B_s\to K\tau\nu$ decay or at large $q^2$ for the $B_s\to K\mu\nu$ decay, would allow differentiation between the current inclusive and exclusive values of $|V_{ub}|$.

\begin{figure}[!t]
%{\scalebox{1.0}{\includegraphics[angle=270,width=0.52\textwidth]{/Users/cmb/analysis/Observables/BstoK_Etas/ratio_tau_mu/BstoK_Rtaumu_v9a2_v_q2-eps-converted-to.pdf}}}
{\scalebox{1.0}{\includegraphics[angle=270,width=0.52\textwidth]{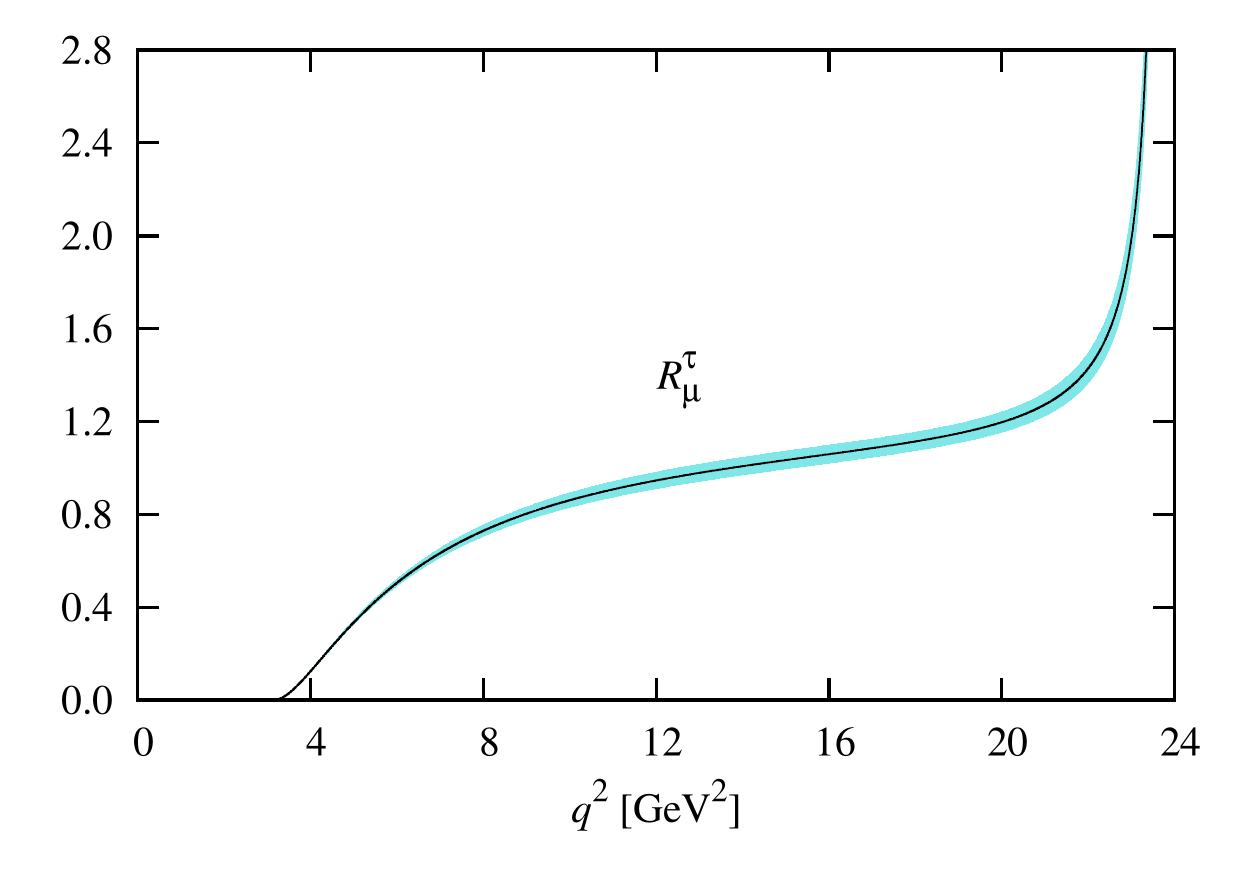}}}
\caption{(color online). Predicted differential branching fraction ratio.}
\label{fig-Rtaumu}
\end{figure}

Decays that couple to the $\tau$ have increased dependence on the scalar form factor and to new physics models with scalar states (see, e.g., Refs.~\cite{Tsai:1997, Chen:2006} for a discussion of new physics in the closely related decay $B\to \pi \tau \nu$).
The ratio of the $B_s\to K\tau\nu$ differential branching fraction to that for $B_s\to K\mu\nu$,
\begin{equation}
R^\tau_\mu(q_{\rm low}^2, q_{\rm high}^2) = \frac{  \int_{q_{\rm low}^2}^{q_{\rm high}^2}dq^2\ d\mathcal{B}/dq^2(B_s\to K \tau\nu) }{\int_{q_{\rm low}^2}^{q_{\rm high}^2}dq^2\ d\mathcal{B}/dq^2(B_s\to K \mu\nu) },
\label{eq-Rtaumu}
\end{equation}
is therefore a potentially sensitive probe of new physics.  Integrating over the full kinematic range, we find
\begin{equation}
R_\mu^\tau(m_\mu^2, q^2_{\rm max}) = 0.695(50),
\label{eq-R}
\end{equation}
where $q^2_{\rm max}=(M_{B_s} - M_K)^2$.  
%We plot the standard model prediction for this ratio over the full kinematic range of $q^2$ in Fig.~\ref{fig-Rtaumu}.
We plot the standard model prediction for this ratio, as a function of $q^2=(q^2_{\rm low}+q^2_{\rm high})/2$, over the full kinematic range in Fig.~\ref{fig-Rtaumu}.

\begin{figure}[!t]
%{\scalebox{1.0}{\includegraphics[angle=270,width=0.5\textwidth]{/Users/cmb/analysis/Observables/BstoK_Etas/muon/BstoKmunu_v9a2_AFB_v_q2-eps-converted-to.pdf}}}
{\scalebox{1.0}{\includegraphics[angle=270,width=0.5\textwidth]{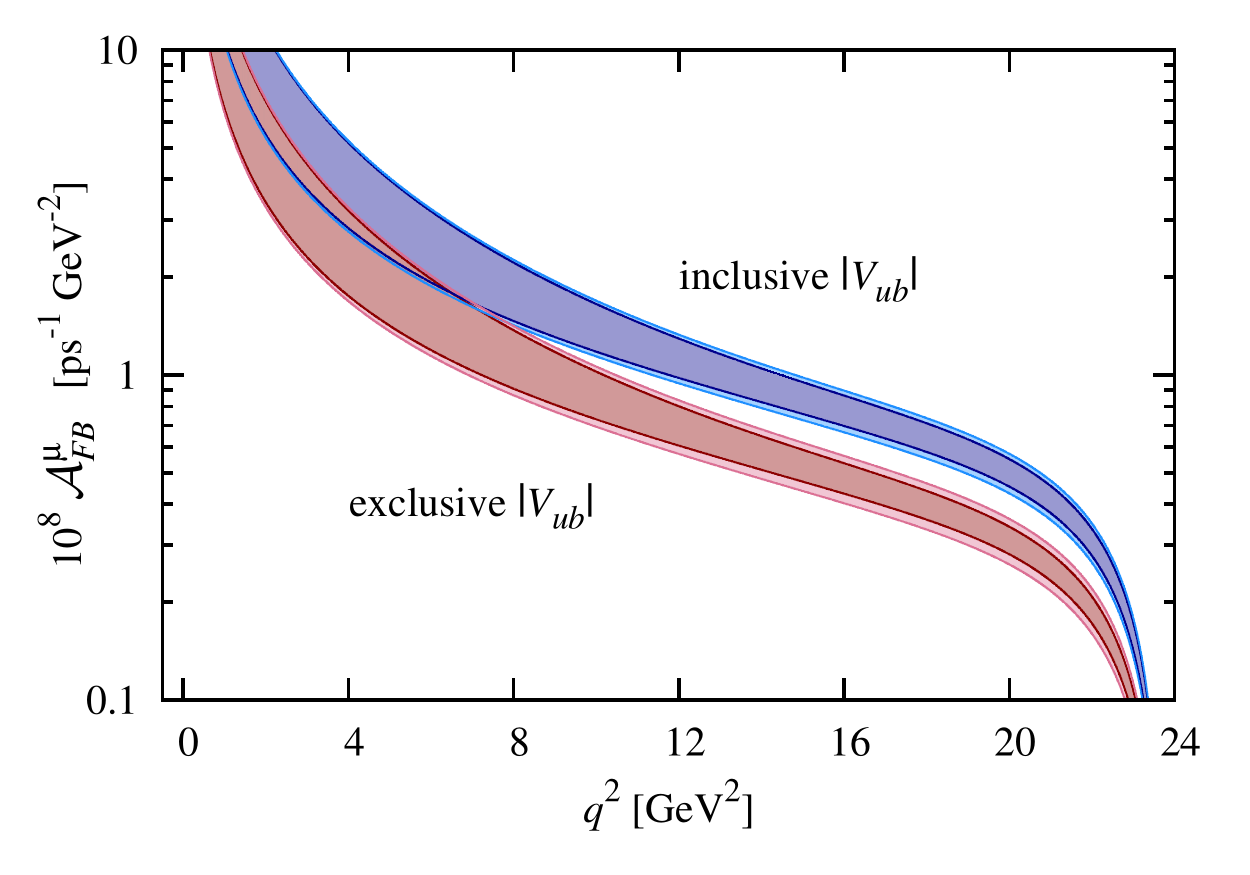}}}
\\
%{\scalebox{1.0}{\includegraphics[angle=270,width=0.5\textwidth]{/Users/cmb/analysis/Observables/BstoK_Etas/tau/BstoKtaunu_v9a2_AFB_v_q2-eps-converted-to.pdf}}}
{\scalebox{1.0}{\includegraphics[angle=270,width=0.5\textwidth]{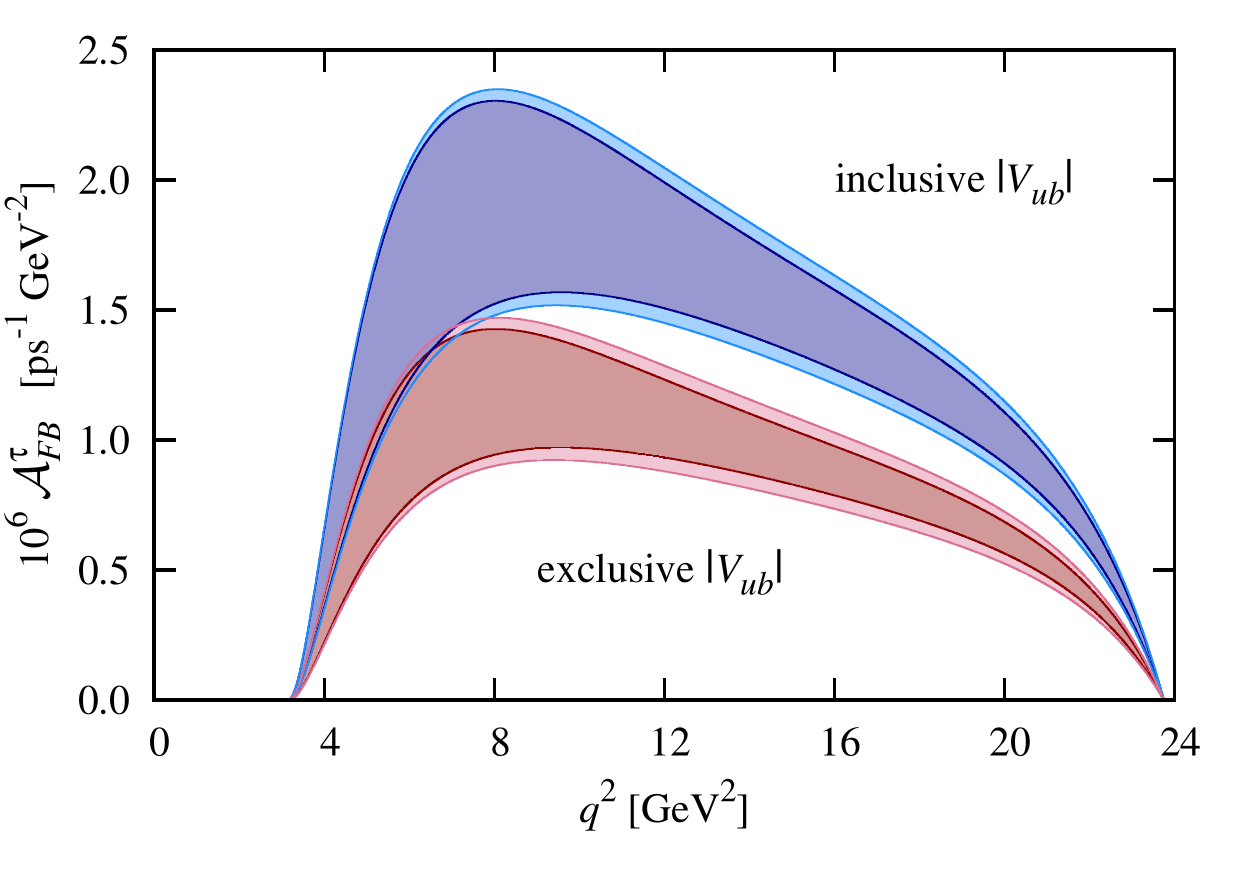}}}
\caption{(color online). Differential decay rate forward-backward asymmetries for the ({\it top}) $B_s\to K\mu\nu$ and ({\it bottom}) $B_s\to K\tau\nu$ decays using inclusive and exclusive semileptonic determinations of $|V_{ub}|$.  Light outer bands includes all sources of error and the dark interior bands neglect uncertainty in $|V_{ub}|$.}
\label{fig-AFB_diffVub}
\end{figure}

The angular dependence of the differential decay rate, neglecting final state electromagnetic interactions, is given by
%\begin{multline}
%\frac{d^2\Gamma}{dq^2\, d\cos\theta_\ell} = \frac{G_F^2 |V_{ub}|^2}{128\pi^3 M_{B_s}^2}\, \Big(1-\frac{m_\ell^2}{q^2}\Big)^2 |{\bf p}_K|  \\
%\times \Big[ 4M_{B_s}^2 | {\bf p}_K |^2 \Big(\sin^2\theta_\ell + \frac{m_\ell^2}{q^2} \cos^2\theta_\ell \Big) f_+^2 \\
%+ \frac{4m_\ell^2}{q^2} (M^2_{B_s} - M^2_K) M_{B_s} | {\bf p}_K | \cos\theta_\ell f_0 f_+  \\
%+  \frac{m_\ell^2}{q^2} (M^2_{B_s} - M^2_K)^2 f_0^2 \Big],
%\end{multline}
\begin{eqnarray}
\frac{d^2\Gamma}{dq^2\, d\cos\theta_\ell} &=& \frac{G_F^2 |V_{ub}|^2}{128\pi^3 M_{B_s}^2}\, \Big(1-\frac{m_\ell^2}{q^2}\Big)^2 |{\bf p}_K| \nonumber \\
&&\times \Big[ 4M_{B_s}^2 | {\bf p}_K |^2 \Big(\sin^2\theta_\ell + \frac{m_\ell^2}{q^2} \cos^2\theta_\ell \Big) f_+^2 \nonumber \\
&&+ \frac{4m_\ell^2}{q^2} (M^2_{B_s} - M^2_K) M_{B_s} | {\bf p}_K | \cos\theta_\ell f_0 f_+ \nonumber \\
&&+  \frac{m_\ell^2}{q^2} (M^2_{B_s} - M^2_K)^2 f_0^2 \Big],
\end{eqnarray}
where $\theta_\ell$ is defined, in the $q^2$ rest frame (i.e. where ${\bf p}_\ell + {\bf p}_\nu$ is zero), as the angle between the final state lepton and the $B_s$ meson.  From this angular dependence we can extract a forward-backward asymmetry~\cite{Meissner:2013},
\begin{eqnarray}
\mathcal{A}^\ell_{\rm FB}(q^2) &=& \Bigg[ \int_0^1 - \int_{-1}^0 \Bigg] d\cos\theta_\ell \frac{d^2\Gamma}{dq^2\, d\cos\theta_\ell} \\
&=& \frac{G_F^2 |V_{ub}|^2}{32\pi^3 M_{B_s}}\, \Big(1-\frac{m_\ell^2}{q^2}\Big)^2 |{\bf p}_K|^2 \nonumber \\
& & \times \frac{m_\ell^2}{q^2} (M^2_{B_s} - M^2_K) f_0 f_+,
\end{eqnarray}
which is suppressed in the standard model by a factor of $m_\ell^2/q^2$.
In Fig.~\ref{fig-AFB_diffVub} we show standard model predictions for the forward-backward asymmetry using the inclusive and exclusive values for $|V_{ub}|$.
Integrating over the full kinematic range of $q^2$ gives
\begin{eqnarray}
\int_{m_\mu^2}^{q^2_{\rm max}} dq^2\ \mathcal{A}^\mu_{\rm FB}(q^2)/|V_{ub}|^2 &=& 0.052(17)\ {\rm ps}^{-1}, \label{eq-AFBmu} \\
\int_{m_\tau^2}^{q^2_{\rm max}} dq^2\ \mathcal{A}^\tau_{\rm FB}(q^2)/|V_{ub}|^2 &=& 1.40(20)\ {\rm ps}^{-1}. \label{eq-AFBtau}
\end{eqnarray}

\begin{figure}[!t]
%{\scalebox{1.0}{\includegraphics[angle=270,width=0.5\textwidth]{/Users/cmb/analysis/Observables/BstoK_Etas/muon/BstoKmunu_v9a2_AFBn_v_q2-eps-converted-to.pdf}}}
{\scalebox{1.0}{\includegraphics[angle=270,width=0.5\textwidth]{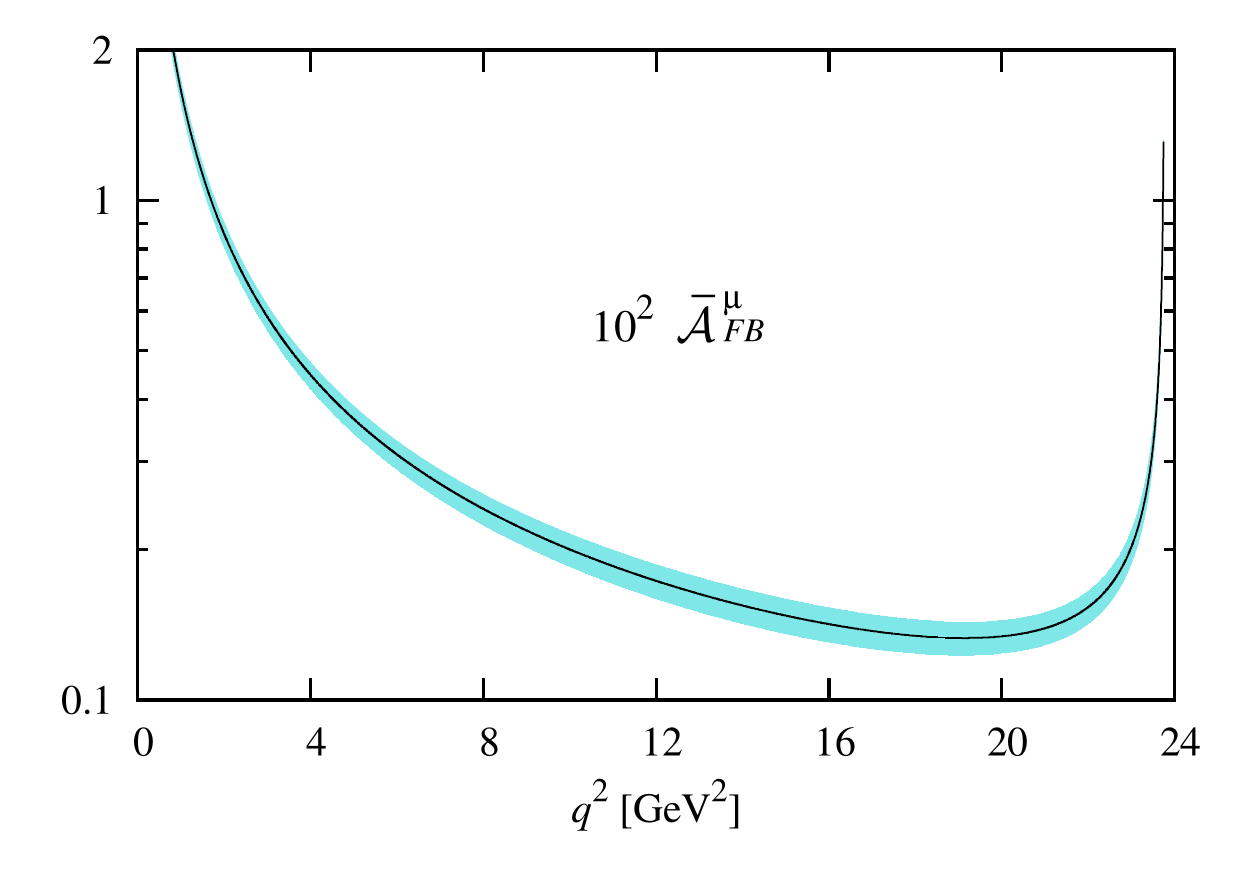}}}
\\
%{\scalebox{1.0}{\includegraphics[angle=270,width=0.5\textwidth]{/Users/cmb/analysis/Observables/BstoK_Etas/tau/BstoKtaunu_v9a2_AFBn_v_q2-eps-converted-to.pdf}}}
{\scalebox{1.0}{\includegraphics[angle=270,width=0.5\textwidth]{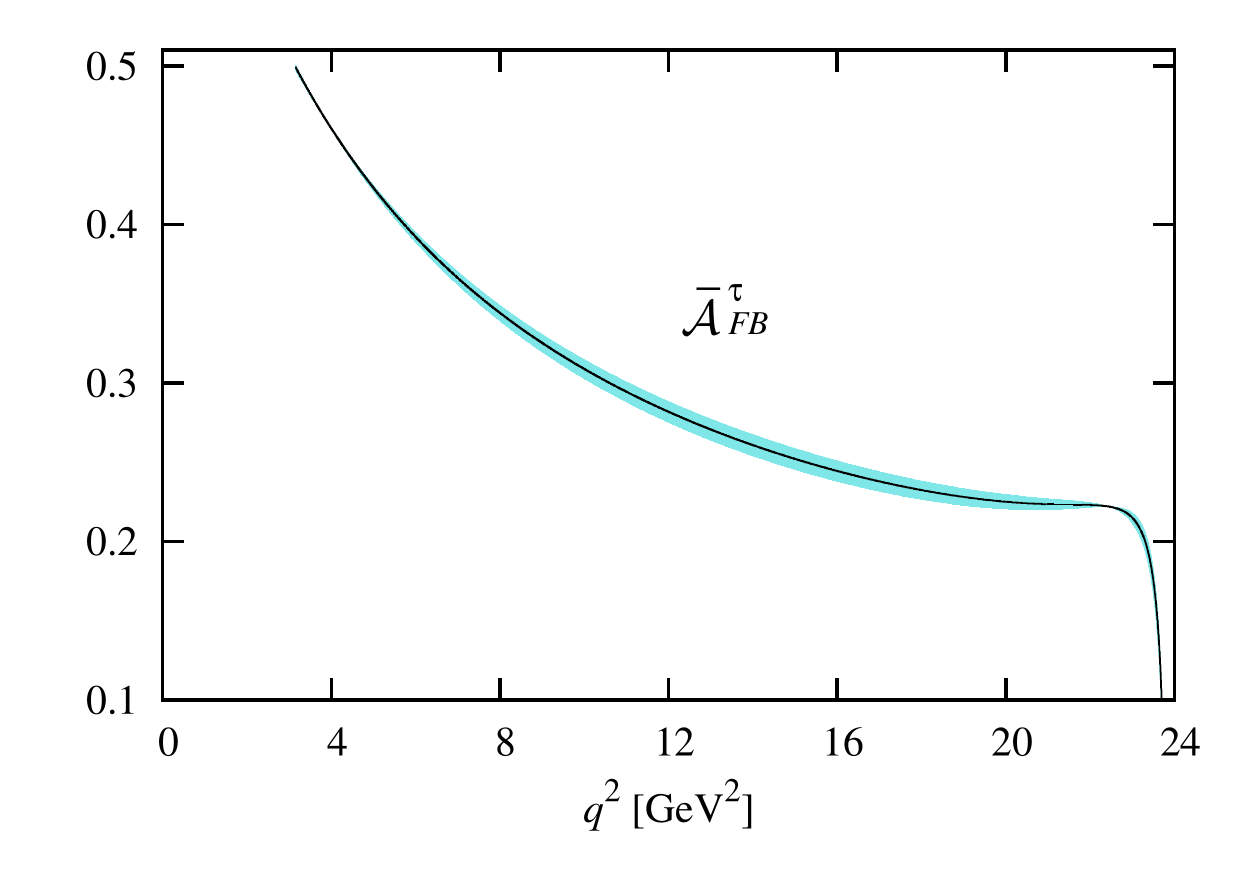}}}
\caption{(color online). Normalized differential decay rate forward-backward asymmetries for the ({\it top}) $B_s\to K\mu\nu$ and ({\it bottom}) $B_s\to K\tau\nu$ decays.}
\label{fig-AFBn}
\end{figure}

Normalizing the forward-backward asymmetry by the differential decay rate removes $|V_{ub}|$ ambiguity and most hadronic uncertainties,
\begin{equation}
\bar{\mathcal{A}}^\ell_{\rm FB}(q^2_{\rm low}, q^2_{\rm high}) = \frac{\int_{q_{\rm low}^2}^{q_{\rm high}^2}dq^2\  \mathcal{A}^\ell_{\rm FB}(q^2)}{\int_{q_{\rm low}^2}^{q_{\rm high}^2} dq^2\ d\Gamma/dq^2},
\label{eq-AFBn}
\end{equation}
and represents the probability the lepton will have a momentum component, in this frame, in the direction of motion of the parent $B_s$ meson.
Integrating over $q^2$ yields
\begin{eqnarray}
\bar{\mathcal{A}}^\mu_{\rm FB}(m^2_\mu, q^2_{\rm max}) &=& 0.0066(10), \\
\bar{\mathcal{A}}^\tau_{\rm FB}(m^2_\tau, q^2_{\rm max}) &=& 0.284(17) ,
\end{eqnarray}
with central values equal to those obtained by taking the ratio of results from Eqs.~(\ref{eq-AFBmu}) and (\ref{eq-AFBtau}) with those from Eqs.~(\ref{eq-Zetamu}) and (\ref{eq-Zetatau}).  The errors, however, are $\sim\!3\times$ smaller when correlations are accounted for.
The normalized standard model asymmetries are plotted in Fig.~\ref{fig-AFBn} as a function of $q^2$.

\begin{figure}[!t]
%{\scalebox{1.0}{\includegraphics[angle=270,width=0.5\textwidth]{/Users/cmb/analysis/Observables/BstoK_Etas/tau/BstoKtaunu_v9a2_Apol_v_q2-eps-converted-to.pdf}}}
{\scalebox{1.0}{\includegraphics[angle=270,width=0.5\textwidth]{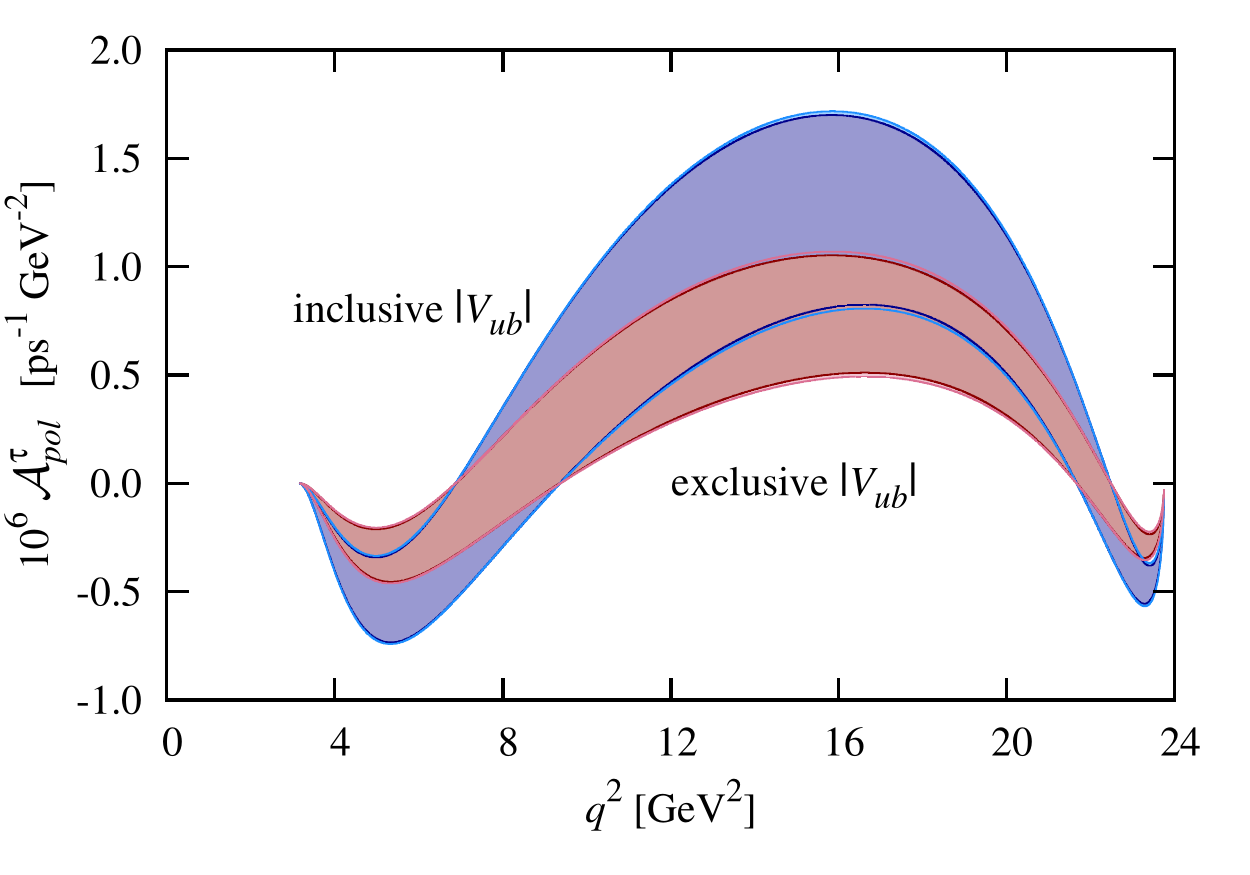}}}
\caption{(color online). Standard model $\tau$-polarization distribution for the differential decay rate of $B_s\to K\tau\nu$.}
\label{fig-Apol_tau}
\end{figure}

\begin{figure}[!t]
%{\scalebox{1.0}{\includegraphics[angle=270,width=0.5\textwidth]{/Users/cmb/analysis/Observables/BstoK_Etas/muon/BstoKmunu_v9a2_Apoln_v_q2-eps-converted-to.pdf}}}
{\scalebox{1.0}{\includegraphics[angle=270,width=0.5\textwidth]{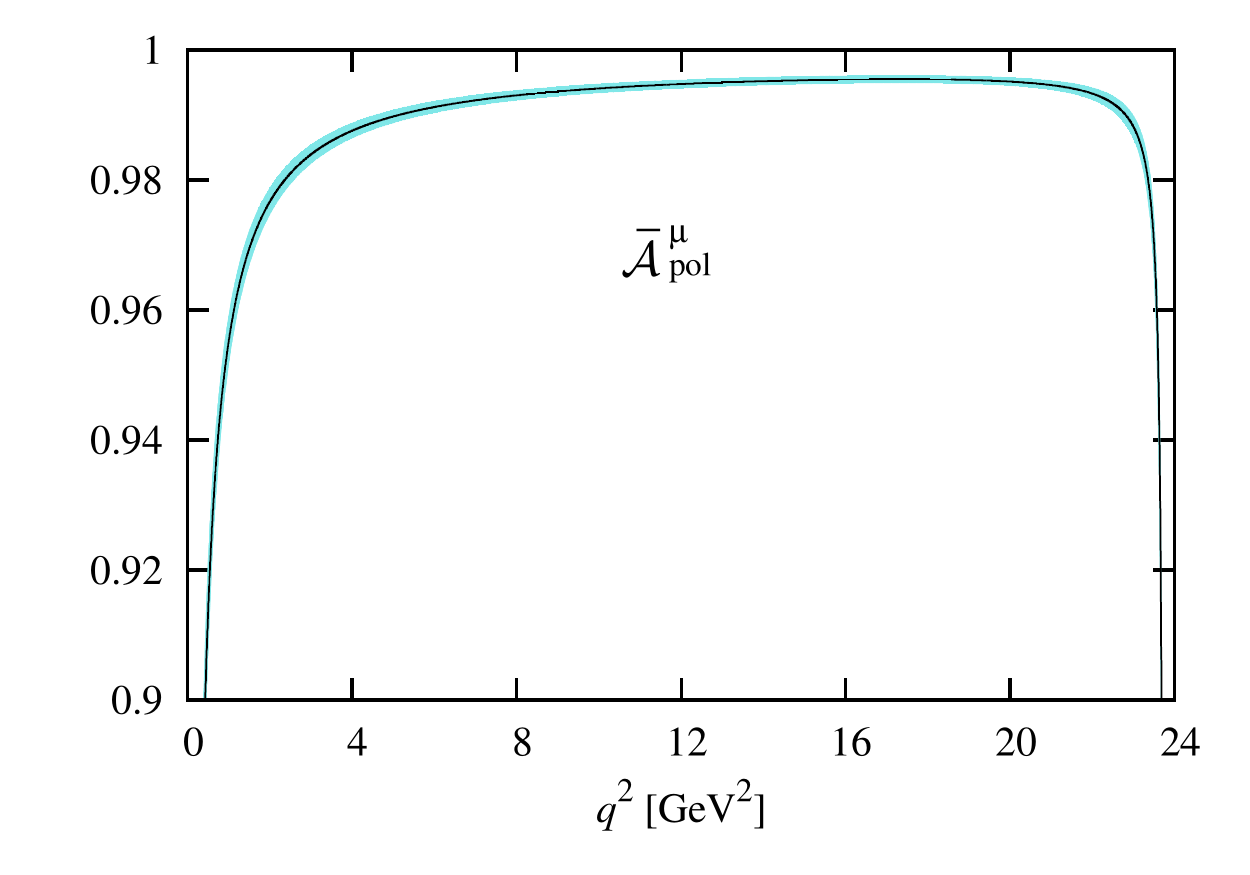}}}
\\
%{\scalebox{1.0}{\includegraphics[angle=270,width=0.5\textwidth]{/Users/cmb/analysis/Observables/BstoK_Etas/tau/BstoKtaunu_v9a2_Apoln_v_q2-eps-converted-to.pdf}}}
{\scalebox{1.0}{\includegraphics[angle=270,width=0.5\textwidth]{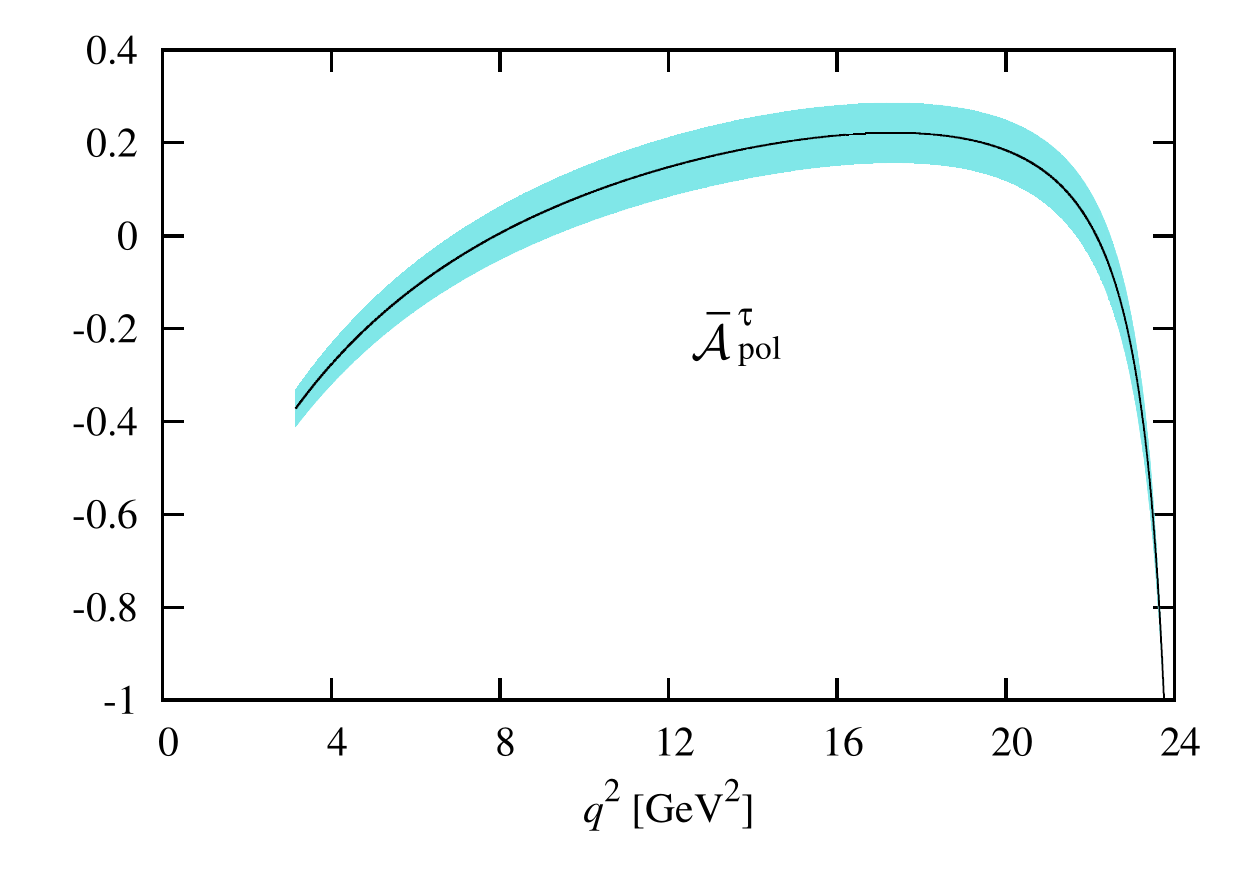}}}
\caption{(color online). Standard model $\ell$-polarization fraction for the differential decay rate of $B_s\to K\ell\nu$, for $\ell=\mu, \tau$.}
\label{fig-Apoln}
\end{figure}

The production of right-handed final state leptons is helicity-suppressed in the standard model, providing a probe of new physics via helicity-violating interactions.
The standard model differential decay rates for left-handed (LH) and right handed (RH) polarized final state leptons in $B_s\to K\ell\nu$ decays is~\cite{Meissner:2013}
%\begin{multline}
%\frac{d\Gamma({\rm LH})}{dq^2} = \frac{G_F^2 |V_{ub}|^2 |{\bf p}_K|^3}{24\pi^3}  \Big(1-\frac{m_\ell^2}{q^2}\Big)^2 f_+^2\ , \\
%\frac{d\Gamma({\rm RH})}{dq^2} = \frac{G_F^2 |V_{ub}|^2 |{\bf p}_K|}{24\pi^3} \frac{m_\ell^2}{q^2}  \Big(1-\frac{m_\ell^2}{q^2}\Big)^2 \hspace{0.575in} \\
%\times \Bigg[\frac{3}{8} \frac{(M^2_{B_s}-M^2_K)^2}{M^2_{B_s}} f^2_0 + \frac{1}{2} |{\bf p}_K|^2 f^2_+ \Bigg],
%\end{multline}
\begin{eqnarray}
\frac{d\Gamma({\rm LH})}{dq^2} &=& \frac{G_F^2 |V_{ub}|^2 |{\bf p}_K|^3}{24\pi^3}  \Big(1-\frac{m_\ell^2}{q^2}\Big)^2 f_+^2\ , \nonumber \\
\frac{d\Gamma({\rm RH})}{dq^2} &=& \frac{G_F^2 |V_{ub}|^2 |{\bf p}_K|}{24\pi^3} \frac{m_\ell^2}{q^2}  \Big(1-\frac{m_\ell^2}{q^2}\Big)^2 \\
&& \times \Bigg[\frac{3}{8} \frac{(M^2_{B_s}-M^2_K)^2}{M^2_{B_s}} f^2_0 + \frac{1}{2} |{\bf p}_K|^2 f^2_+ \Bigg], \nonumber
\end{eqnarray}
and the $\ell$-polarization distribution is given by the difference
\begin{equation}
\mathcal{A}_{\rm pol}^\ell(q^2) = \frac{d\Gamma({\rm LH})}{dq^2} - \frac{d\Gamma({\rm RH})}{dq^2}.
\end{equation}
We plot the $\tau$-polarization distribution, again using the inclusive and exclusive values of $|V_{ub}|$ from Eqs.~(\ref{eq-excl}) and (\ref{eq-incl}), in Fig.~\ref{fig-Apol_tau}.
Because of their relatively small mass, muons produced in the decay are predominantly left-handed and the plot of $\mathcal{A}_{\rm pol}^\mu$ is equivalent to the total differential decay rate.  Integrating the $\ell$-polarization distributions over $q^2$ gives
\begin{eqnarray}
\int_{m_\mu^2}^{q^2_{\rm max}} dq^2\ \mathcal{A}^\mu_{\rm pol}(q^2)/|V_{ub}|^2 &=& 7.61(1.60)\ {\rm ps}^{-1}, \label{eq-Apolmu} \\
\int_{m_\tau^2}^{q^2_{\rm max}} dq^2\ \mathcal{A}^\tau_{\rm pol}(q^2)/|V_{ub}|^2 &=& 0.52(32)\ {\rm ps}^{-1}. \label{eq-Apoltau}
\end{eqnarray}
As with the forward-backward asymmetry, we normalize the $\ell$-polarization distribution by the differential decay rate to remove ambiguity associated with $|V_{ub}|$ and hadronic uncertainties.  The resulting polarization fraction~\cite{Meissner:2013} is defined by
\begin{equation}
\bar{\mathcal{A}}_{\rm pol}^\ell(q^2_{\rm low}, q^2_{\rm high}) = \frac{\int_{q^2_{\rm low}}^{q^2_{\rm high}} dq^2\ \mathcal{A}_{\rm pol}^\ell(q^2)}{\int_{q^2_{\rm low}}^{q^2_{\rm high}} dq^2\  d\Gamma / dq^2}.
\label{eq-Apoln}
\end{equation}
Integrating over $q^2$ we find the standard model prediction for the fraction of polarized leptons to be
\begin{eqnarray}
\bar{\mathcal{A}}_{\rm pol}^\mu(m^2_\mu, q^2_{\rm max}) &=& 0.982({}_{-79}^{+18}), \\ 
\bar{\mathcal{A}}_{\rm pol}^\tau(m^2_\tau, q^2_{\rm max}) &=& 0.105(63),
\end{eqnarray}
where the error associated with the numerical integration of $\bar{\mathcal{A}}_{\rm pol}^\mu$ ($\pm0.079$) has been truncated to satisfy the constraint that $\bar{\mathcal{A}}_{\rm pol}^\ell<1$.
The $q^2$ dependence of the $\ell$-polarization fraction is plotted in Fig.~\ref{fig-Apoln}.

\section{Summary and Outlook}
\label{sec-summary}
%%%%%%%%%%%%%%%%%%%%%%%%%%%%%%%%%%%%%%%%%%%%%%%%%%%%%%%
%%%%%%%%%%%%%%%%%%%%%%%%%%%%%%%%%%%%%%%%%%%%%%%%%%%%%%%
%%%%%%%%%%%%%%%%%%%%%%%%%%%%%%%%%%%%%%%%%%%%%%%%%%%%%%%
Using NRQCD $b$ and HISQ light and strange valence quarks with the MILC $2+1$ dynamical asqtad configurations, we report on the first lattice QCD calculation of the form factors for the semileptonic decay $B_s \to K \ell\nu$.  

With the help of a new technique, called chaining, we fit the $B_s\to K$ correlator data simultaneously with data for the fictitious decay $B_s\to\eta_s$.  Fitting these data simultaneously accounts for correlations\,---\,useful for constructing ratios of form factors.  We extrapolate our lattice form factor results to the continuum, to physical quark mass, and over the full kinematic range of $q^2$ using a combination of the modified $z$~expansion and HPChPT that we refer to as the HPChPT $z$~expansion.

We then make standard model predictions for:
\begin{enumerate}[ {(}i{)} ] \itemsep-0.3em
\item differential decay rates divided by $|V_{ub}|^2$, an observable that, when combined with experiment, will allow an alternative semileptonic exclusive determination of $|V_{ub}|$,
\item differential branching fractions using both the inclusive and exclusive semileptonic $B\to\pi\ell\nu$ determinations of $|V_{ub}|$,
\item the ratio of differential branching fractions $R^\tau_\mu(q^2)$,
\item the forward-backward asymmetry, using inclusive and exclusive values of $|V_{ub}|$,
\item the normalized forward-backward asymmetry, 
\item the $\tau$-polarization distribution in the differential decay rate for $B_s\to K\tau\nu$, and
\item the $\ell$-polarization fraction in the differential decay rate for $B_s\to K\ell\nu$, for $\ell=\mu, \tau$.
\end{enumerate}

In Appendix~\ref{sec-Ratio} we construct ratios of form factors for $B_s\to K$ with those for $B_s\to\eta_s$.  In combination with a future calculation of $B_s\to\eta_s$ using HISQ $b$, these ratios can provide a nonperturbative determination of the $b\to u$ current matching factor.  This would be relevant for both $B_s\to K \ell \nu$ and $B\to\pi \ell \nu$ simulations using NRQCD $b$ quarks.

Our results, built on first principles lattice QCD form factors, greatly clarify standard model expectations~\cite{Meissner:2013} based on model estimates of form factors~\cite{Wang:2012, Faustov:2013, Verma:2012}, most notably at large $q^2$.
Combining our form factors, which are most precise at large $q^2$, with model calculations, typically more reliable at low $q^2$, would result in a more precise determination of $f_0$ and $f_+$.  We are studying the possibility of further refining $B_s\to K\ell\nu$ standard model predictions using such form factors.

\section*{ Acknowledgements}
%%%%%%%%%%%%%%%%%%%%%%%%%%%%%%%%%%%%%%%%%%%%%%%%%%%%%%%%%%%%%%%%%%%
%%%%%%%%%%%%%%%%%%%%%%%%%%%%%%%%%%%%%%%%%%%%%%%%%%%%%%%%%%%%%%%%%%%
%%%%%%%%%%%%%%%%%%%%%%%%%%%%%%%%%%%%%%%%%%%%%%%%%%%%%%%%%%%%%%%%%%%
This research was supported by the DOE and NSF.
We thank the MILC collaboration for making their asqtad $N_f=2+1$ gauge field configurations available.  
Computations were carried out at the Ohio Supercomputer Center and on facilities of the USQCD collaboration funded by the Office of Science of the U.S. DOE.

%================================= APPENDICES =================================
\appendix

%\clearpage
\section{Fitting Basics}
\label{app-basics}
Here we describe in more detail two aspects of our statistical analysis: 1) the 
definition of our error budgets for fit results; and 2) the technique
for chained fits of multiple data sets. We also discuss a general procedure
for testing fit procedures. 
These are general techniques applicable to many types of fitting problems~\cite{software}.
%The fitting software used in this paper is available online: \href{https://github.com/gplepage/lsqfit}{\tt https://github.com/gplepage/lsqfit}~\cite{lsqfit} is a general package for nonlinear least squares fitting, and \href{https://github.com/gplepage/corrfitter}{\tt https://github.com/gplepage/corrfitter}~\cite{corrfitter} is a general package for fitting 2-point and 3-point correlators in lattice QCD. This software implements the strategies discussed in Appendix A of the present paper.
Finally we illustrate these ideas with an example drawn from this paper.

\subsection{Fits and Error Budgets}
The formal structure of a least-squares problem involves fitting input 
data $y_i$ with functions $f_i(p)$ by adjusting fit parameters $p_\alpha$
to minmize
\begin{equation}
    \chi^2(p) = \sum_{ij} \Delta y(p)_i \left(\mathrm{cov}_y^{-1}\right)_{ij}
    \Delta y(p)_j,
\end{equation}
where $\mathrm{cov}_{ij}$ is the covariance matrix for the input data and
\begin{equation}
    \Delta y(p)_i \equiv f_i(p) - y_i.
\end{equation}
There are generally two types of input data\,---\,actual data, and prior
data for each fit parameter\,---\,but we lump these together here since 
they enter $\chi^2(p)$ in the same way. So the sums here over~$i$ and~$j$ 
are over all data and priors. Note that priors and data may be correlated 
in some problems.

The best-fit parameters $\overline p_\alpha$ are those that minimize~$\chi^2$:
\begin{equation}
    \partial_\alpha \chi^2(\overline p) = 2
    \sum_{ij} \partial_\alpha f_i(\overline p) \left(\mathrm{cov}_y^{-1}\right)_{ij}
    \Delta y(\overline p)_j = 0
\label{eq:pbar}
\end{equation}
where the derivative $\partial_\alpha \equiv \partial/\partial \overline p_\alpha$.
The inverse covariance matrix, $\partial_\alpha\partial_\beta \chi^2(\overline p) / 2$,
for the $\overline p_\alpha$ is then given by
\begin{equation}
    \left(\mathrm{cov}_p^{-1}\right)_{\alpha\beta} 
    = \sum_{ij} \partial_\alpha f_i(\overline p) \left(\mathrm{cov}_y^{-1}\right)_{ij}
    \partial_\beta f_j(\overline p) + \mathcal{O}(\Delta y),
    \label{eq:covp}
\end{equation}
where we neglect terms proportional to $\Delta y$ (which makes sense for reasonable fits to accurate
data). This is the conventional result.

The uncertainties in the $\overline p_\alpha$ are due to the uncertainties in
the input data~$y_i$, and, for very accurate data, depend linearly upon 
$\mathrm{cov}_y$. The relationship can be demonstrated by differentiating
Eq.~(\ref{eq:pbar}) with respect to $y_j$ to obtain
\begin{equation}
    \sum_\beta \left(\mathrm{cov}_p^{-1}\right)_{\alpha\beta}
    \frac{\partial \overline p_\beta}{\partial y_j}
    = \sum_i\partial_\alpha f_i(\overline p) \left(\mathrm{cov}_y^{-1}\right)_{ij} + \mathcal{O}(\Delta y),
\end{equation}
where again we neglect terms proportional to $\Delta y$. Solving for 
$\partial \overline p_\beta/\partial y_j$ gives:
\begin{equation}
    \frac{\partial \overline p_\beta}{\partial y_j} = \sum_{\alpha i}
    \left(\mathrm{cov}_p\right)_{\beta\alpha} \partial_\alpha f_i(\overline p) \left(\mathrm{cov}_y^{-1}\right)_{ij}
    \label{eq:dpdy}
\end{equation}
In the high-statistics, small-error limit the covariances in the $\overline p_\alpha$ are related to those in the~$y_i$ by the standard formula
\begin{equation}
    \left(\mathrm{cov}_p\right)_{\alpha\beta} 
    = \sum_{ij} \frac{\partial \overline p_\alpha}{\partial y_i}
    \left(\mathrm{cov}_y\right)_{ij} 
    \frac{\partial \overline p_\beta}{\partial y_j},
    \label{eq:covpy}
\end{equation}
and, indeed, substituting Eq.~(\ref{eq:dpdy}) into this equation 
reproduces Eq.~(\ref{eq:covp}) for~$\mathrm{cov}_p$. 

Eqs.~(\ref{eq:dpdy}) and~(\ref{eq:covpy}) allow us to express the error 
$\sigma_g$ for a function $g(\overline p)$ of the best-fit parameter values 
in terms of the input errors:
\begin{equation}
    \sigma_g^2 \equiv \sum_{\alpha\beta} \partial_\alpha g(\overline p)
    \left(\mathrm{cov}_p\right)_{\alpha\beta}    
    \partial_\beta g(\overline p)
    = \sum_{ij} c_{ij} \left(\mathrm{cov}_y\right)_{ij} 
    \label{eq:covg}
\end{equation}
where
\begin{equation}
    c_{ij} \equiv \sum_{\alpha\beta} 
    \partial_\alpha g(\overline p)
    \frac{\partial \overline p_\alpha}{\partial y_i}
    \frac{\partial \overline p_\beta}{\partial y_j}
    \partial_\beta g(\overline p).
\end{equation}
and Eq.~(\ref{eq:dpdy}) is used to evaluate $\partial \bar p_\alpha / \partial y_i$.
We can then decompose $\sigma_g^2$ into separate contributions coming from 
the different block-diagonal submatrices of $\mathrm{cov}_y$. These
contributions to $\sigma_g$ constitute the error budget for~$g(\overline p)$.

The $c_{ij}$s in Eq.~(\ref{eq:covg}) depend upon both the $y_i$ and their 
covariance matrix, but that dependence can be neglected to leading order 
in $\mathrm{cov}_y$. Consequently Eq.~(\ref{eq:covg}) can be used to estimate
the impact on $\sigma_g$ 
of possible modifications to any element of~$\mathrm{cov}_y$.

Note that the data's covariance matrix $\mathrm{cov}_y$ can be quite singular 
if there are strong correlations in the data. This can make it 
numerically difficult to invert the matrix for use in~$\chi^2(p)$. This 
problem is typically dealt with by using a \emph{singular value decomposition}
(SVD) to regulate the most singular components of the covariance matrix. In our
fits we rescale the covariance matrix by its diagonal elements to obtain
the correlation matrix, which we then diagonalize. We introduce a minimum
eigenvalue by setting any smaller eigenvalue equal to the
minimum. We then reconstitute the correlation matrix, and rescale it back 
into a (less singular) covariance matrix which we use in the fit. 
This procedure, in effect, increases the error 
in the data and so increases the uncertainties in the final fit results; it
is a conservative move.

It is common when using SVD to discard eigenmodes
corresponding to the small eigenvalues. This is equivalent to setting the 
variance associated with these modes to infinity in the fit. 
In our implementation, all eigenmodes are retained, but the small eigenvalues
are replaced by a (larger) minimum eigenvalue. This is a more realistic
estimate for the variances of these modes\,---\,that is, more realistic 
than setting them to infinity\,---\,and gives more accurate fit results.

\subsection{Chained Fits}
Chained fits simplify fits of multiple data sets whose fit functions
share fit parameters by allowing us to fit each data set separately. 
To illustrate, consider two sets of data,
$y_{i}(A)$ and $y_{j}(B)$, that we fit with functions  $f_{i}(A, p)$ and
$f_{j}(B, p)$, respectively\,---\,both functions of the same fit parameters
$p_\alpha$ (unlike the previous section, here we do not lump the priors in with the $y$s). 
The fit procedure is straightforward in  a Bayesian framework if
$y(A)$ and $y(B)$ are statistically uncorrelated. We first fit, say, 
data set $y(A)$ to obtain best-fit estimates $\overline
p(A)$  for the parameters and an estimate
$\mathrm{cov}_{p(A)}$ for the  parameters' covariance matrix. We then fit
data set $y(B)$, but using $\overline p(A)$ 
and $\mathrm{cov}_{p(A)}$ to form the prior for the fit parameters. 

This two-step fit merges the information contained in~$y(A)$ with that from~$y(B)$ by feeding
the information from the first fit into the second fit as prior 
information.
The order in which the data sets are fit doesn't matter in the high-statistics
(Gaussian) limit; with larger errors, it is better to fit the more accurate
data set first. The $\chi^2$ for the two-step fit is the sum of the 
$\chi^2$s for each step.

The situation is slightly more complicated if $y(A)$ and $y(B)$ 
are correlated. Then the best-fit parameters $\overline p(A)$ from
the first fit above are correlated with the second data set~$y(B)$. 
The  $p(A)$-$y(B)$ covariance can be computed from
\begin{equation}
    \mathrm{cov}_{p(A) y(B)} \equiv \sum_{y(A)}
    \frac{\partial \overline p(A)}{\partial y(A)} \mathrm{cov}_{y(A) y(B)}
\end{equation}
using Eq.~(\ref{eq:dpdy}) in the previous section. This correlation must 
be included in the second fit, to data set~$y(B)$. So the second fit uses the
best-fit parameters $\overline p(A)$ from the first fit to construct the prior, 
together with $\mathrm{cov}_{p(A)}$ for parameter-parameter covariances and
$\mathrm{cov}_{p(A)y(B)}$ for parameter-data covariances.

We refer to a sequential fit of multiple data sets, where the best-fit parameters
and covariance matrix from one fit are used as the prior for the next fit, 
as a \emph{chained fit}. It is essential in such fits to account for possible
correlations between the priors (from previous fits) and the data being fit at each stage. 
The results of a chained fit should agree with those of a simultaneous fit 
in the limit of large (i.e., Gaussian) statistics.

\subsection{Testing Fits}
It is generally useful to have ways of testing particular fit strategies. 
One simple approach to testing
is to create multiple fake data sets that are very similar to the actual data 
being fit, but where the exact values for the fit parameters are known ahead 
of time. Running several such data sets through an analysis code tells you
very quickly whether, for example, your analysis code gives results that are
correct to within one sigma 68\% of the time, as is desired.

It is easy to create fake data sets of this sort. One simple recipe is the
following:
\begin{enumerate}
    \item Fit the actual data to obtain a set of parameter values 
    $p_{\alpha}^*$ such that the fit function $f_i(p^*)$ closely matches
    the mean values $y_i$ of the actual data. Calculate the difference
    between the actual means of the data and the fit values for $p=p^*$:
    \begin{equation}
        \delta y_i \equiv f_i(p^*) - y_i.
        \label{eq:dy}
    \end{equation}

    \item Create a bootstrap copy $y_i^\mathrm{bs}$ of the original data and replace
    its mean values by:
    \begin{equation}
        y^*_i = y^\mathrm{bs}_i + \delta y_i
    \end{equation}
    The fake data set then consists of the mean values $y^*_i$ and 
    the covariance matrix $\mathrm{cov}_y$ of the original data. 
    The role of the bootstrap here is to generate fluctuations in the means with the same distribution as the original data.
    These data sets will fluctuate around central 
    values $f_i(p^*)$ rather than the original means of the data.

    \item Repeat the second step to create any number of additional fake data
    sets. 
\end{enumerate}
Each fake data set is fit using the same procedure that was used to analyze the
original data. The results for the fit parameters are compared with the parameter
values $p^*$ used to define the correction $\delta y_i$ [Eq.~(\ref{eq:dy})],
since, by construction, these are the \emph{correct} values for the parameters 
in the fake data. 

Typically only a handful of parameters from a fit are of interest. Their
best-fit values from different fake data sets will differ, but they 
should all agree with the $p^*$ values to within the errors generated 
by the fake fit (that is, to within one sigma 68\% of the time, 
two sigma 95\% of the time, and so on). Such tests can reveal, for example,
potential problems coming from poor priors or inadequate SVD cuts, or
biases in particular combinations of fit parameters.

\begin{figure*}[t]
%{\scalebox{0.9}{\includegraphics[angle=0,width=1.0\textwidth]{/Users/cmb/fitting/09July2013/fitscripts/plots/BstoK_Etas/C3_BsKEtas_23pt_state0_compare_BsEtas-eps-converted-to.pdf}}}
{\scalebox{0.9}{\includegraphics[angle=0,width=1.0\textwidth]{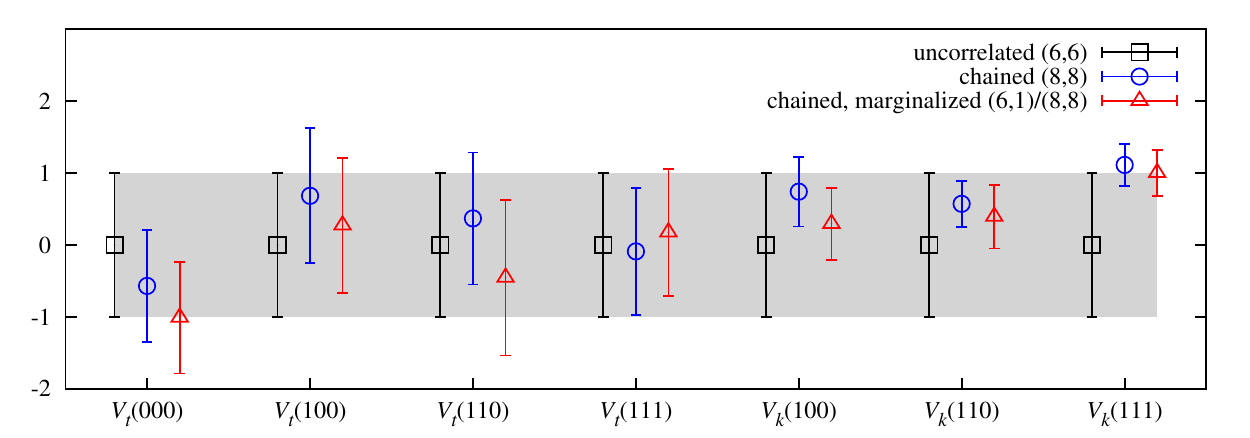}}}
\caption{(color online). $B_s\to\eta_s$ three point ground state amplitudes, for varying currents and momenta, as obtained from different fitting strategies described in the text.  Plotted central values indicate the number of standard deviations by which a fit result differs from an ``uncorrelated" fit.  The size of the error bars is the ratio of the plotted fit error to that from an uncorrelated fit.}
\label{fig-chain_compare}
\end{figure*}
\subsection{Example}
We compare chained and unchained fit results in Fig.~\ref{fig-chain_compare}.   
Because unchained fits to very large data sets are unreliable, for purposes of comparison we divide the data into the smallest subsets that allow the extraction of individual matrix elements.  Such fits are uncorrelated in that they neglect correlations among data at different momenta, for different currents, and among the two decays.
The uncorrelated fits include only one decay mode ($B_s\to K$ or $B_s\to\eta_s$), data for only one simulation momentum (000, 100, 110, or 111), and only one current ($V_t$ or $V_k$).   
These fits are still complicated, however, as they require the minimum amount of data needed to extract a single matrix element. 
This minimum number of correlators consists of parent and daughter two point and three point data, i.e. $B_s\to B_s$, $\eta_s\to\eta_s(000)$, and $B_s\to V_t \to \eta_s(000)$.  
Including correlations results in marked improvement in the accuracy of matrix elements obtained from the noisiest data\,---\,that for $V_k$ at large momenta.
This improvement can be traced to correlations of these data with the more precise data for $V_t$ (for the same decay and at a common momentum), as demonstrated in Fig.~\ref{fig-correlations}.

In addition to properly accounting for correlations in the data, chaining reduces the time required to perform the fits.  
While the uncorrelated fits required a total of 1 hour 14 minutes, the chained (8,8) fit required only 24 minutes.
The use of marginalization significantly reduces the time required.  The chained and marginalized (6,1)/(8,8) fit required only 57 seconds.

%\clearpage
\section{Correlator Fit Results}
\label{sec-corrfit_results}
%%%%%%%%%%%%%%%%%%%%%%%%%%%%%%%%%%%%%%%%%%%%%%%%%%%%%%%
%%%%%%%%%%%%%%%%%%%%%%%%%%%%%%%%%%%%%%%%%%%%%%%%%%%%%%%
%%%%%%%%%%%%%%%%%%%%%%%%%%%%%%%%%%%%%%%%%%%%%%%%%%%%%%%
The method for selecting priors for correlator fits was described in detail in Appendix B of Ref.~\cite{Bouchard:2013a}.  We use the same method in this analysis.
Tables~\ref{tab-MBcorrfit},~\ref{tab-MKcorrfit},~and~\ref{tab-MEtascorrfit} tabulate priors and fit results for ground state energies.  
They compares results obtained from fits to two point correlation function data to those from simultaneous fits to two and three point correlation function data, as described in Sec.~\ref{sec-CorrFits}.  
The combined fits show improved precision for the $B_s$ meson mass and the larger momenta daughter meson energies, suggesting that the three point correlation function data provide additional information to the fit.  
Within errors, the two point and simultaneous two and three point fit results are consistent.
\setlength{\tabcolsep}{0.12in}
\begin{table}[t]
\caption{$B_s$ priors and fit results for $aE^{{\rm sim}(0)}_{B_s}$.}
\begin{tabular}{llll}
	\hline\hline
	\T Ensemble	& Prior		& 2pt			& 2+3pt		\\ [0.5ex]
	\hline
	\T C1	& 0.537(53)	& 0.53780(72)	& 0.53801(31)	\\
	\T C2	& 0.54(6)		& 0.54360(84)	& 0.54234(35)	\\
	\T C3	& 0.54(8)		& 0.5362(15)	& 0.53575(36)	\\
	\T F1		& 0.405(55)	& 0.4081(13)	& 0.40869(21)	\\
	\T F2		& 0.407(60)	& 0.40770(64)	& 0.40710(23)	\\ [0.5ex]
	\hline\hline
\end{tabular}
\label{tab-MBcorrfit}
\end{table}
\setlength{\tabcolsep}{0.4in}
\begin{table*}[t]
\caption{$K$ priors and fit results.  For each ensemble, the first row lists priors, the second row gives two point correlator fit results, and the third row shows simultaneous two and three point correlator fit results.}
\begin{tabular}{lllll}
	\hline\hline
	\T Ensemble	& $aM^{(0)}_K$			& $aE^{(0)}_{K(100)}$	& $aE^{(0)}_{K(110)}$	& $ aE^{(0)}_{K(111)}$ \\ [0.5ex]
	\hline
	\T C1	& 0.312(17)				& 0.41(11)			& 0.48(23)			& 0.55(28)	\\ [-0.4ex]
	\T 		& 0.31211(15)				& 0.40657(58)			& 0.48461(76)			& 0.5511(16)	\\  [-0.4ex]
	\T		& 0.31195(14)				& 0.40661(49)			& 0.48408(63)			& 0.5513(13)	\\ [-0.4ex]
	\T C2	& 0.329(24)				& 0.45(15)			& 0.55(15)			& 0.61(31)	\\  [-0.4ex]
	\T 		& 0.32863(18)				& 0.45406(85)			& 0.5511(16)			& 0.6261(75)	\\  [-0.4ex]
	\T		& 0.32870(16)				& 0.45434(73)			& 0.5506(11)			& 0.6273(35)	\\ [-0.4ex]
	\T C3	& 0.356(25)				& 0.475(75)			& 0.58(20)			& 0.65(30)	\\  [-0.4ex]
	\T 		& 0.35717(22)				& 0.47521(85)			& 0.5723(11)			& 0.6524(30)	\\  [-0.4ex]
	\T		& 0.35744(21)				& 0.47507(71)			& 0.57218(80)			& 0.6539(18)	\\ [-0.4ex]
	\T F1		& 0.229(60)				& 0.32(24)			& 0.39(34)			& 0.43(40)	\\  [-0.4ex]
	\T 		& 0.22865(11)				& 0.32024(66)			& 0.39229(86)			& 0.4515(25)	\\  [-0.4ex]
	\T		& 0.22861(12)				& 0.32020(61)			& 0.39192(82)			& 0.4528(16)	\\ [-0.4ex]
	\T F2		& 0.246(36)				& 0.33(23)			& 0.40(30)			& 0.47(37)	\\  [-0.4ex]
	\T 		& 0.24577(13)				& 0.33322(52)			& 0.40214(73)			& 0.4623(14)	\\  [-0.4ex]
	\T		& 0.24566(13)				& 0.33310(50)			& 0.40184(72)			& 0.4624(11)	\\ [0.5ex]
	\hline\hline
\end{tabular}
\label{tab-MKcorrfit}
\end{table*}
\setlength{\tabcolsep}{0.39in}
\begin{table*}[t]
\caption{Like Table~\ref{tab-MKcorrfit} but for the $\eta_s$.}
\begin{tabular}{lllll}
	\hline\hline
	\T Ensemble	& $aM^{(0)}_{\eta_s}$		& $aE^{(0)}_{\eta_s(100)}$	& $aE^{(0)}_{\eta_s(110)}$	&$aE^{(0)}_{\eta_s(111)}$\\[0.5ex]
	\hline
	\T C1	& 0.411(9)				& 0.487(12)				& 0.553(50)				& 0.61(11)		 \\  [-0.4ex]
	\T 		& 0.41111(12)				& 0.48736(23)				& 0.55311(29)				& 0.61148(60)		 \\  [-0.4ex]
	\T		& 0.41107(11)				& 0.48726(23)				& 0.55294(29)				& 0.61135(52)		 \\  [-0.4ex]
	\T C2	& 0.415(12)				& 0.52(5)					& 0.61(11)				& 0.68(23)		 \\  [-0.4ex]
	\T 		& 0.41445(17)				& 0.51949(46)				& 0.6063(12)				& 0.6797(31)		 \\  [-0.4ex]
	\T		& 0.41446(15)				& 0.51934(44)				& 0.60647(67)				& 0.6794(18)		 \\  [-0.4ex]
	\T C3	& 0.412(20)				& 0.518(40)				& 0.61(12)				& 0.69(35)		 \\  [-0.4ex]
	\T 		& 0.41180(23)				& 0.51757(63)				& 0.60723(78)				& 0.6831(23)		 \\  [-0.4ex]
	\T		& 0.41175(20)				& 0.51742(57)				& 0.60720(67)				& 0.6843(14)		 \\  [-0.4ex]
	\T F1		& 0.294(24)				& 0.37(10)				& 0.43(23)				& 0.48(34)		 \\  [-0.4ex]
	\T 		& 0.294109(93)			& 0.36965(31)				& 0.43278(45)				& 0.4867(13)		 \\  [-0.4ex]
	\T		& 0.294066(88)			& 0.36988(26)				& 0.43301(38)				& 0.48729(88)		 \\  [-0.4ex]
	\T F2		& 0.293(30)				& 0.369(89)				& 0.43(18)				& 0.49(30)		 \\  [-0.4ex]
	\T 		& 0.29315(12)				& 0.36939(35)				& 0.43259(45)				& 0.48810(87)		 \\  [-0.4ex]
	\T		& 0.29310(12)				& 0.36927(35)				& 0.43197(48)				& 0.48729(97)		 \\ [0.5ex]
\hline\hline
\end{tabular}
\label{tab-MEtascorrfit}
\end{table*}
\setlength{\tabcolsep}{0.18in}
\begin{table}[t]
\caption{Group I priors for the HPChPT $z$~expansion for $f^{B_s K}_{0,+}$ and $f^{B_s\eta_s}_{0,+}$.  Quantities listed in five consecutive rows have ensemble-dependent values corresponding to C1, C2, C3, F1, and F2.}
\begin{tabular}{lcccccc}
\hline\hline	
	\T Group I  					& Prior			& Fit    		\\ [0.5ex]
	\hline
	\T $r_1$ [fm]					& 0.3133(23)		& 0.3133(23)  \\  [-0.2ex]
	\T $g_{B^*B\pi}$				& 0.51(20)		& 0.53(20)  \\ [-0.2ex]
	\T $M_{\eta_s^{\rm phys}}$ [GeV]	& 0.6858(40)		& 0.6858(40) \\ [-0.2ex]
	\T $\Delta^{B_sK}_0$ [GeV]		& 0.3127(10)		& 0.3126(10)  \\ [-0.2ex]
	\T $\Delta^{B_sK}_+$ [GeV]		& -0.04157(42)		& -0.04157(42)	  \\ [-0.2ex]
	\T $\Delta^{B_s\eta_s}_0$ [GeV]	& 0.4000(10)		& 0.4000(10)  \\ [-0.2ex]
	\T $\Delta^{B_s\eta_s}_+$ [GeV]	& 0.0487(22)		& 0.0487(22)  \\ [-0.2ex]
	\T $m_\parallel$				& 0.00(4)			& 0.000(40)	\\ [-0.2ex]
	\T $m_\perp$					& 0.00(4)			& 0.001(40)	\\	 [-0.2ex]
	\T $r_1/a$						& 2.647(3)		& 2.6465(30)  \\ [-0.2ex]
	\T $r_1/a$						& 2.618(3)		& 2.6186(30)  \\ [-0.2ex]
	\T $r_1/a$						& 2.644(3)		& 2.6438(30)  \\ [-0.2ex]
	\T $r_1/a$						& 3.699(3)		& 3.6992(30)  \\ [-0.2ex]
	\T $r_1/a$						& 3.712(4)		& 3.7117(40)  \\  [-0.2ex]
	\T $aM_{B_s}$					& 3.2303(12)		& 3.2300(12)  \\ [-0.2ex]
	\T $aM_{B_s}$					& 3.2663(13)		& 3.2668(12)  \\ [-0.2ex]
	\T $aM_{B_s}$					& 3.2336(13)		& 3.2333(12)  \\  [-0.2ex]
	\T $aM_{B_s}$					& 2.30849(89)		& 2.30841(87)  \\ [-0.2ex]
	\T $aM_{B_s}$					& 2.30035(90)		& 2.30048(88)  \\ [-0.2ex]
	\T $aM_K^{\rm HISQ}$			& 0.31195(14)		& 0.31196(14)  \\ [-0.2ex]
	\T $aM_K^{\rm HISQ}$			& 0.32870(17)		& 0.32868(17)  \\ [-0.2ex]
	\T $aM_K^{\rm HISQ}$			& 0.35744(21)		& 0.35746(21)	  \\ [-0.2ex]
	\T $aM_K^{\rm HISQ}$			& 0.22861(12)		& 0.22861(12)  \\ [-0.2ex]
	\T $aM_K^{\rm HISQ}$			& 0.24566(13)		& 0.24565(13)  \\ [-0.2ex]
	\T $aM_K^{\rm asqtad}$			& 0.36530(29)		& 0.36532(29)\\ [-0.2ex]
	\T $aM_K^{\rm asqtad}$			& 0.38331(24)		& 0.38331(24)	\\ [-0.2ex]
	\T $aM_K^{\rm asqtad}$			& 0.40984(21)		& 0.40983(21)	\\ [-0.2ex]
	\T $aM_K^{\rm asqtad}$			& 0.25318(19)		& 0.25316(19)	\\ [-0.2ex]
	\T $aM_K^{\rm asqtad}$			& 0.27217(21)		& 0.27219(21)	\\ [-0.2ex]
	\T $aM_\pi^{\rm HISQ}$			& 0.15988(12)		& 0.15988(12)	\\ [-0.2ex]
	\T $aM_\pi^{\rm HISQ}$			& 0.21097(16)		& 0.21097(16)	\\ [-0.2ex]
	\T $aM_\pi^{\rm HISQ}$			& 0.29309(22)		& 0.29309(22)	\\ [-0.2ex]
	\T $aM_\pi^{\rm HISQ}$			& 0.13453(11)		& 0.13453(11)	\\ [-0.2ex]
	\T $aM_\pi^{\rm HISQ}$			& 0.18737(13)		& 0.18736(13)	\\ [-0.2ex]
	\T $aM_\pi^{\rm asqtad}$			& 0.15971(20)		& 0.15971(20)	\\ [-0.2ex]
	\T $aM_\pi^{\rm asqtad}$			& 0.22447(17)		& 0.22447(17)	\\ [-0.2ex]
	\T $aM_\pi^{\rm asqtad}$			& 0.31125(16)		& 0.31125(16)	\\ [-0.2ex]
	\T $aM_\pi^{\rm asqtad}$			& 0.14789(18)		& 0.14789(18)	\\ [-0.2ex]
	\T $aM_\pi^{\rm asqtad}$			& 0.20635(18)		& 0.20365(18)	\\  [-0.2ex]
	\T $aM_{\eta_s}^{\rm HISQ}$		& 0.41107(11)		& 0.41109(11)	\\ [-0.2ex]
	\T $aM_{\eta_s}^{\rm HISQ}$		& 0.41447(15)		& 0.41442(15)	\\ [-0.2ex]
	\T $aM_{\eta_s}^{\rm HISQ}$		& 0.41176(20)		& 0.41177(20)	\\ [-0.2ex]
	\T $aM_{\eta_s}^{\rm HISQ}$		& 0.294066(89)	& 0.294053
	(89)	\\
	\T $aM_{\eta_s}^{\rm HISQ}$		& 0.29310(12)		& 0.29312(12)	\\ [0.9ex]
\hline\hline
\end{tabular}
\label{tab-modzpriorsI}
\end{table}
%
%\clearpage

%\clearpage
\section{HPChPT $z$ Expansion Fit Results}
\label{sec-HPChPTz_results}
%%%%%%%%%%%%%%%%%%%%%%%%%%%%%%%%%%%%%%%%%%%%%%%%%%%%%%%
%%%%%%%%%%%%%%%%%%%%%%%%%%%%%%%%%%%%%%%%%%%%%%%%%%%%%%%
%%%%%%%%%%%%%%%%%%%%%%%%%%%%%%%%%%%%%%%%%%%%%%%%%%%%%%%
Group I parameters listed in Table~\ref{tab-modzpriorsI} insert error in the fit based on uncertainty associated with input parameters -- quantities not determined by the data.
Priors for $r_1$ and $M_{\eta^{\rm phys}_s}$ are taken from Ref.~\cite{Davies:2009}.
We base our prior choice for the $BB^*\pi$ coupling $g_{BB^*\pi}$ on the combined works in Ref.~\cite{gBBstarPi}.
Resonance masses for the Blaschke factors $P_{0,+}$ introduced in Eq.~(\ref{eq-basicz}) are calculated relative to the $B_s$ meson mass in our simulations,
\begin{eqnarray}
M^{B_sK}_0 &=& M_{B_s} - (M_{B_s}-M_B) + 400(1)\, {\rm MeV}, \\
M^{B_sK}_+ &=& M_{B_s} - (M_{B_s}-M_B) + \Delta^{\rm hyperfine}_B, \\
M^{B_s\eta_s}_0 &=& M_{B_s} + 400(1)\, {\rm MeV}, \\
M^{B_s\eta_s}_+ &=& M_{B_s} + \Delta^{\rm hyperfine}_{B_s},
\end{eqnarray}
and we refer to the shift relative to $M_{B_s}$ as $\Delta^{B_sK, B_s\eta_s}_{0,+}$.
The $M_{B_s}-M_B$ and hyperfine splittings are taken from the PDG~\cite{PDG:2012}.
We tested increasing the uncertainty in the location of the scalar pole, which we have taken to be $400(1)\, {\rm MeV}$ above the $J^P=0^-$ state.
A splitting of $400(50)\, {\rm MeV}$ gives identical results for the form factors, in both the central value and error, but accommodates for part of the error in $f_0$ via allowed uncertainty in $M_0$.
To reconstruct the form factors in this case, correlations between $P_0$ and the coefficients of the $z$~expansion must be accounted for.  
By effectively fixing $M_0$ we arrive at the same fit results and can neglect uncertainty in $P_0$ and correlations with the coefficients.
The 4\% uncertainty associated with the perturbative matching is accounted for by $m_{\parallel}$ and $m_\perp$, where we use prior central values of zero and width 0.04, as explained by Eq.~(\ref{eq-matcherr}) and surrounding text.  Matrix elements for $B_s\to K$ and $B_s\to\eta_s$ use the same matching factors so we use common $m_{\parallel,\perp}$ for both data sets.
We use values for $r_1/a$ from Ref.~\cite{Bazavov:2010} and $M_{\pi,K}^{\rm asqtad}$ from Ref.~\cite{Aubin:2004}.
We use values for $M_{\pi, K,\eta_s}^{\rm HISQ}$ and $M_{B_s}$ from best fit results in this and an ongoing $B\to\pi$ analysis using HISQ valence quarks. 

The Group II parameters of Table~\ref{tab-modzpriorsII} are quantities determined by the fit.  
We choose priors for $a_k$ to be $0\pm5$, based roughly on the unitarity constraint, and verified that fit results are insensitive to variations in the prior width from 1 to 10.
Chiral analytic terms are written in terms of dimensionless parameters that are naturally $\mathcal{O}(1)$.  For this reason we use priors of zero with width one for $c_1$ and $c_3$.  Based on previous analyses using the same ensembles we know that sea-quark effects are smaller than those of the valence quarks, so we choose priors for $c_2$ to be $0 \pm 0.3$.
The leading order HISQ discretization effects are $\mathcal{O}(\alpha_s a^2)$, so for the coefficients $d_1$ and $e_1$ which characterize the $\mathcal{O}(a^2)$ discretization effects, we choose priors of $0\pm 0.3$.  Coefficients $d_2$ and $e_2$ characterize $\mathcal{O}(a^4)$ effects and we use $0\pm 1$.  The coefficients $h$ and $l$ characterize light- and heavy-quark mass-dependent discretization effects.  These terms are written in terms of $\mathcal{O}(1)$ quantities and we take the coefficients to have priors of $0\pm1$.
\setlength{\tabcolsep}{0.05in}
\begin{table*}[t]
\caption{Group II priors and fit results for the simultaneous HPChPT $z$~expansion for $f_{0,+}^{B_sK}$ and $f_{0,+}^{B_s\eta_s}$.}
\begin{tabular}{lcccccccccccc}
\hline\hline
	\T				& 		& \multicolumn{4}{c}{Fit result} 											& 			&			&		& \multicolumn{4}{c}{Fit result} 									\\
	\T Group II   		& Prior	& $f_0^{B_sK}$	& $f_+^{B_sK}$	& $f_0^{B_s\eta_s}$	& $f_+^{B_s\eta_s}$	& \hspace{0.1in} & Group II 	& Prior	& $f_0^{B_sK}$	& $f_+^{B_sK}$ 	& $f_0^{B_s\eta_s}$	& $f_+^{B_s\eta_s}$\\ [0.5ex]
	\hline
	\\ [-3ex]
	\T $a_0$			& 0(5)	& 0.24(10)		& 0.284(32)		& 0.04(12)		& 0.293(30)		& 			& $h_1^{(0)}$	& 0(1)	& 0.31(92)		& 0.37(92)		& 0.0(1.0)			& 0.22(95)	\\ [-0.0ex]
	\T $a_1$			& 0(5)	& 0.7(1.0)			& -0.58(16)		& 0.0(1.2)			& -0.99(18)		& 			& $h_1^{(1)}$	& 0(1)	& 0.0(1.0)			& 0.0(1.0)			& 0.0(1.0)			& 0.0(1.0)		\\ [-0.0ex]
	\T $a_2$			& 0(5)	& 1.9(3.6)			& 2.1(1.1)			& 2.1(4.3)			& 3.2(1.7)			& 			& $h_1^{(2)}$	& 0(1)	& 0.0(1.0)			& 0.0(1.0)			& 0.0(1.0)			& 0.0(1.0)		\\ [-0.0ex]
	\T $c_1^{(0)}$		& 0(1)	& 0.01(60)		& 0.07(11)		& -0.23(99)		& -0.16(15)		& 			& $h_2^{(0)}$	& 0(1)	& 0.20(0.99)		& 0.02(99)		& 0.0(1.0)			& -0.15(99)	\\ [-0.0ex]
	\T $c_1^{(1)}$		& 0(1)	& 0.11(90)		& -0.16(38)		& 0.0(1.0)			& -0.39(25)		& 			& $h_2^{(1)}$	& 0(1)	& 0.0(1.0)			& 0.0(1.0)			& 0.0(1.0)			& 0.0(1.0)		\\ [-0.0ex]
	\T $c_1^{(2)}$		& 0(1)	& -0.04(99)		& -0.62(85)  		& 0.11(98)		& -1.25(86)		& 			& $h_2^{(2)}$	& 0(1)	& 0.0(1.0)			& 0.0(1.0)			& 0.0(1.0)			& 0.0(1.0)		\\ [-0.0ex]
	\T $c_2^{(0)}$		& 0(0.3)	& -0.24(27)		& 0.05(29)		& -0.03(27)		& 0.15(29)		& 			& $h_3^{(0)}$	& 0(1)	& 0.0(1.0)			& 0.1(1.0)			& 0.0(1.0)			& 0.0(1.0)		\\ [-0.0ex]
	\T $c_2^{(1)}$		& 0(0.3)	& 0.00(30)		& -0.02(30)		& 0.00(30)		& -0.01(30)		&			& $h_3^{(1)}$	& 0(1)	& 0.0(1.0)			& 0.0(1.0)			& 0.0(1.0)			& 0.0(1.0)		\\ [-0.0ex]
	\T $c_2^{(2)}$		& 0(0.3)	& 0.00(30)		& -0.01(30)		& 0.00(30)		& -0.01(30)		&			& $h_3^{(2)}$	& 0(1)	& 0.0(1.0)			& 0.0(1.0)			& 0.0(1.0)			& 0.0(1.0)		\\ [-0.0ex]
	\T $c_3^{(0)}$		& 0(1)	& 0.37(99)		& -1.22(74)		& 0.0(1.0)			& -0.19(70)		&			& $h_4^{(0)}$	& 0(1)	& 0.0(1.0)			& 0.0(1.0)			& 0.0(1.0)			& 0.0(1.0)		\\ [-0.0ex]
	\T $c_3^{(1)}$		& 0(1)	& 0.0(1.0)			& 0.34(97)		& 0.0(1.0)			& -0.24(94)		&			& $h_4^{(1)}$	& 0(1)	& 0.0(1.0)			& 0.0(1.0)			& 0.0(1.0)			& 0.0(1.0)		\\ [-0.0ex]
	\T $c_3^{(2)}$		& 0(1)	& 0.1(1.0)			& -0.20(99)		& 0.0(1.0)			& 0.00(99)		&			& $h_4^{(2)}$	& 0(1)	& 0.0(1.0)			& 0.0(1.0)			& 0.0(1.0)			& 0.0(1.0)		\\ [-0.0ex]
	\T $d_1^{(0)}$		& 0(0.3)	& 0.16(18)		& -0.20(22)		& -0.02(21)		& -0.15(24)		& 			& $l_1^{(0)}$	& 0(1)	& 0.64(0.97)		& 0.18(0.98)		& 0.0(1.0)			& 0.24(98)	\\ [-0.0ex]
	\T $d_1^{(1)}$		& 0(0.3)	& -0.01(30)		& -0.06(30)		& 0.00(30)		& -0.05(29)		& 			& $l_1^{(1)}$	& 0(1)	& 0.0(1.0)			& 0.0(1.0)			& 0.0(1.0)			& 0.0(1.0)		\\ [-0.0ex]
	\T $d_1^{(2)}$		& 0(0.3)	& 0.01(30)		& -0.03(30)		& 0.00(30)		& -0.01(30)		& 			& $l_1^{(2)}$	& 0(1)	& 0.0(1.0)			& 0.0(1.0)			& 0.0(1.0)			& 0.0(1.0)		\\ [-0.0ex]
	\T $d_2^{(0)}$		& 0(1)	& -0.22(92)		& -0.32(94)		& -0.02(85)		& -0.21(94)		& 			& $l_2^{(0)}$	& 0(1)	& 0.1(1.0)			& 0.0(1.0)			& 0.0(1.0)			& 0.1(1.0)		\\ [-0.0ex]
	\T $d_2^{(1)}$		& 0(1)	& 0.0(1.0)			& -0.1(1.0)		& 0.0(1.0)			& -0.07(99)		& 			& $l_2^{(1)}$	& 0(1)	& 0.0(1.0)			& 0.0(1.0)			& 0.0(1.0)			& 0.0(1.0)		\\ [-0.0ex]
	\T $d_2^{(2)}$		& 0(1)	& 0.0(1.0)			& -0.1(1.0)		& 0.0(1.0)			& 0.0(1.0)			& 			& $l_2^{(2)}$	& 0(1)	& 0.0(1.0)			& 0.0(1.0)			& 0.0(1.0)			& 0.0(1.0)		\\ [-0.0ex]
	\T $e_1^{(0)}$		& 0(0.3)	& -0.21(17)		& 0.13(24)		& -0.09(16)		& 0.15(23)		& 			& $l_3^{(0)}$	& 0(1)	& -0.1(1.0)		& 0.0(1.0)			& 0.0(1.0)			& 0.1(1.0)		\\ [-0.0ex]
	\T $e_1^{(1)}$		& 0(0.3)	& -0.01(30)		& 0.00(29)		& 0.00(30)		& -0.06(28)		& 			& $l_3^{(1)}$	& 0(1)	& 0.0(1.0)			& 0.0(1.0)			& 0.0(1.0)			& 0.0(1.0)		\\ [-0.0ex]
	\T $e_1^{(2)}$		& 0(0.3)	& 0.00(30)		& -0.02(30)		& 0.00(30)		& -0.02(30)		& 			& $l_3^{(2)}$	& 0(1)	& 0.0(1.0)			& 0.0(1.0)			& 0.0(1.0)			& 0.0(1.0)		\\ [-0.0ex]
	\T $e_2^{(0)}$		& 0(1)	& 0.40(24)		& 0.12(30)		& 0.26(19)		& 0.02(25)		& 			& $l_4^{(0)}$	& 0(1)	& 0.0(1.0)			& 0.0(1.0)			& 0.0(1.0)			& 0.0(1.0)		\\ [-0.0ex]
	\T $e_2^{(1)}$		& 0(1)	& -0.1(1.0)		& 0.25(94)		& 0.0(1.0)			& -0.04(83)		& 			& $l_4^{(1)}$	& 0(1)	& 0.0(1.0)			& 0.0(1.0)			& 0.0(1.0)			& 0.0(1.0)		\\ [-0.0ex]
	\T $e_2^{(2)}$		& 0(1)	& 0.0(1.0)			& 0.0(1.0)			& 0.0(1.0)			& -0.03(99)		& 			& $l_4^{(2)}$	& 0(1)	& 0.0(1.0)			& 0.0(1.0)			& 0.0(1.0)			& 0.0(1.0)		\\   [0.5ex]
\hline\hline
\end{tabular}
\label{tab-modzpriorsII}
\end{table*}

\section{$B_s \to \eta_s$ Form Factors and Ratios}
\label{sec-Ratio}
\begin{figure}[!h]
{\scalebox{1.0}{\includegraphics[angle=270,width=0.5\textwidth]{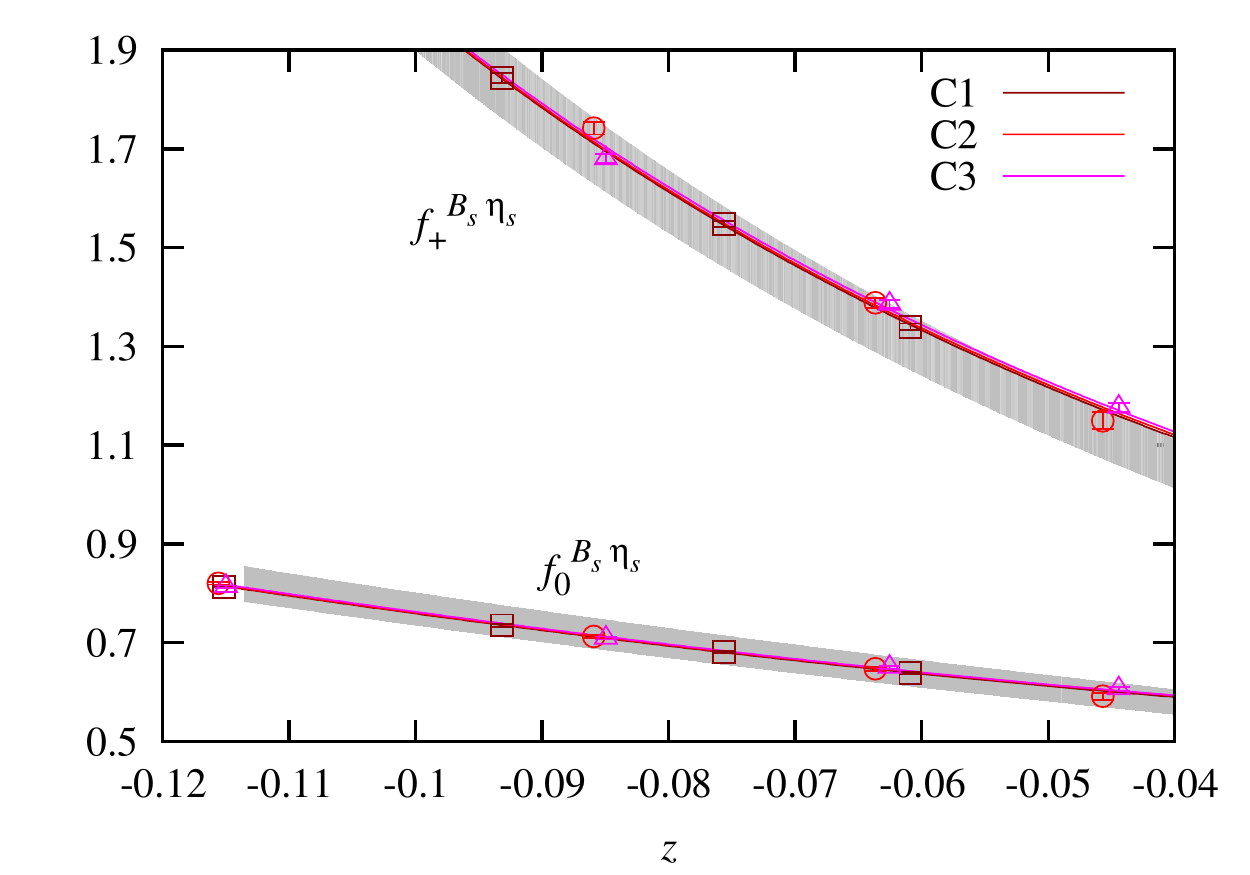}}}
{\scalebox{1.0}{\includegraphics[angle=270,width=0.5\textwidth]{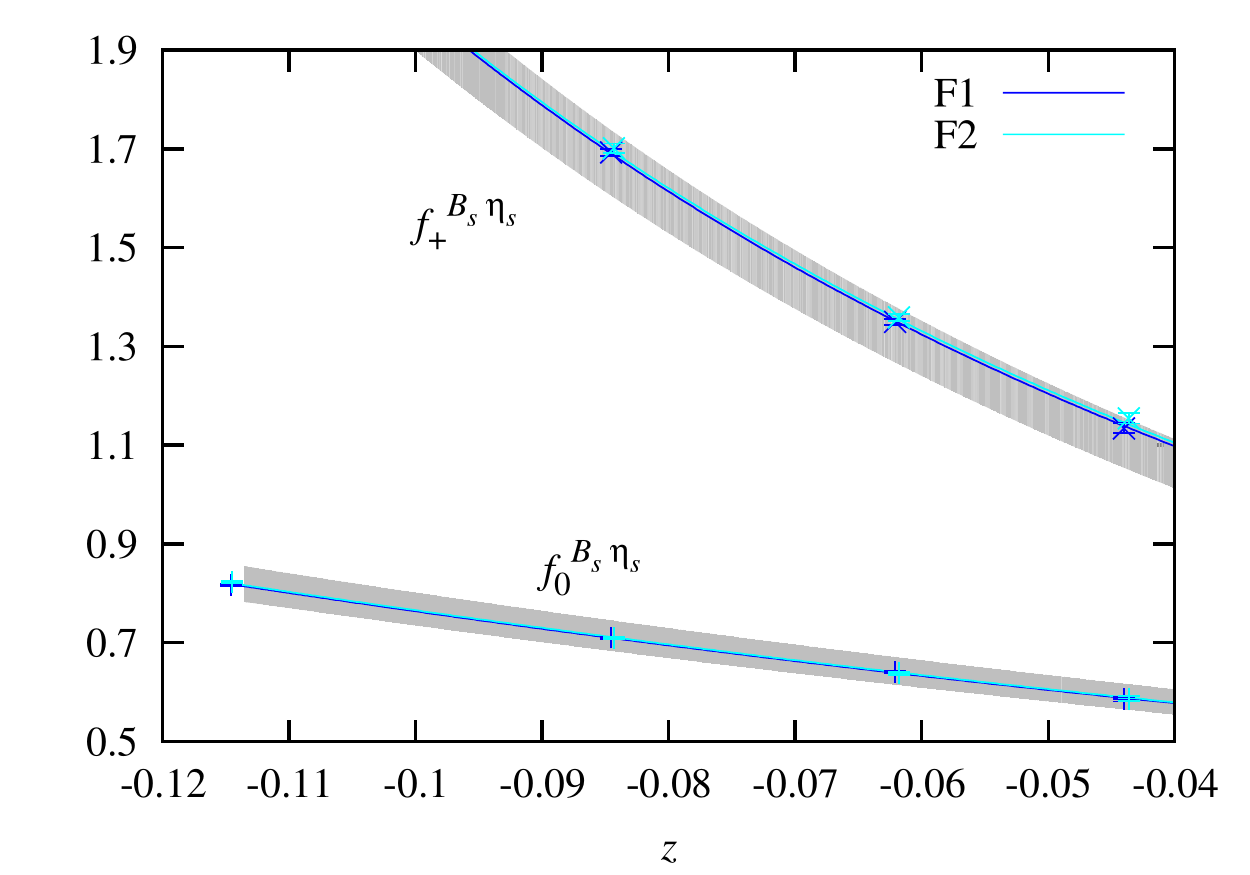}}}
{\scalebox{1.0}{\includegraphics[angle=270,width=0.5\textwidth]{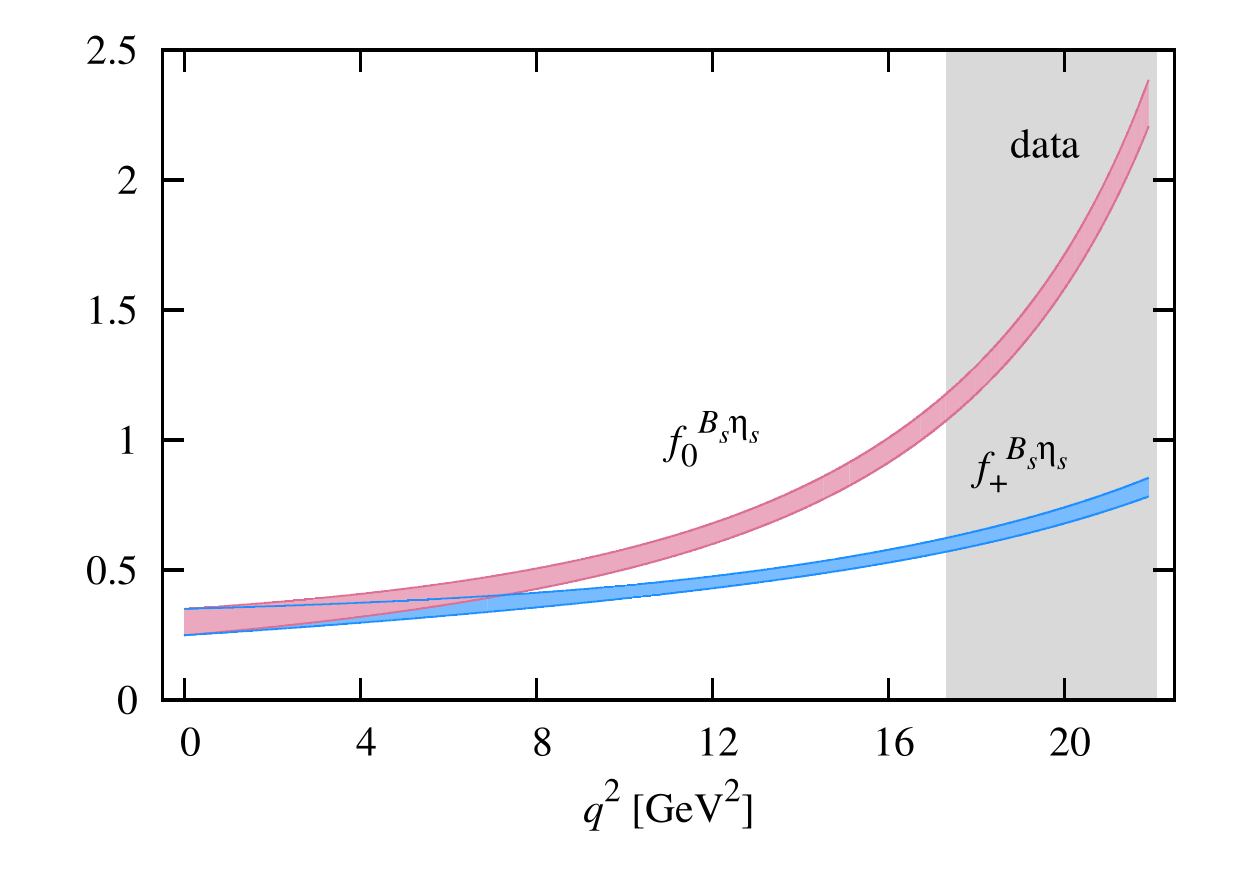}}}
\caption{(color online). $B_s\to \eta_s$ form factor results from a simultaneous HPChPT $z$~expansion are shown ({\it top}) relative to coarse ensemble data (C1, C2, and C3), ({\it middle}) relative to fine ensemble data (F1 and F2), and ({\it bottom}) in the continuum limit with physical masses, extrapolated over the full kinematic range.}
\label{fig-HPChPTfit_BsEtas}
\end{figure}

%%%%%%%%%%%%%%%%%%%%%%%%%%%%%%%%%%%%%%%%%%%%%%%%%%%%%%%
%%%%%%%%%%%%%%%%%%%%%%%%%%%%%%%%%%%%%%%%%%%%%%%%%%%%%%%
%%%%%%%%%%%%%%%%%%%%%%%%%%%%%%%%%%%%%%%%%%%%%%%%%%%%%%%
The results of $B_s\to\eta_s$ correlator fits are tabulated in Table~\ref{tab-BsEtascorrfits} and plotted as data points in the top two panels of Fig.~\ref{fig-HPChPTfit_BsEtas}.  From these plots one sees that simulation data exhibit very small light sea quark mass and lattice spacing dependence.
These fit results are obtained from a single fit to both the $B_s\to K$ and $B_s\to\eta_s$ data described in Sec.~\ref{sec-CorrFits}.  
As a result, the $B_s\to\eta_s$ fit results of Table~\ref{tab-BsEtascorrfits} are correlated with the $B_s\to K$ results of Table~\ref{tab-BsKcorrfits}, as shown in Fig.~\ref{fig-correlations}.
\setlength{\tabcolsep}{0.045in}
\begin{table}[t]
%\vspace{0.1in}
\caption{Fit results for the scalar and vector $B_s\to\eta_s$ form factors on each ensemble and for each simulated momentum.}
\begin{tabular}{ccccc}
\hline\hline
	\T Ensemble   	& $f^{B_s \eta_s}_0(000)$	& $f^{B_s \eta_s}_0(100)$	& $f^{B_s \eta_s}_0(110)$	& $f^{B_s \eta_s}_0(111)$ \\ [0.5ex]
	\hline
	\T C1 	& 0.8135(17)	& 0.7352(22)	& 0.6813(19)	& 0.6381(21)	\\ [-0.2ex]
	\T C2	& 0.8205(21) 	& 0.7127(33)	& 0.6475(39)	& 0.5921(70)	\\ [-0.2ex]
	\T C3	& 0.8140(26) 	& 0.7095(32)	& 0.6504(31)	& 0.6069(39)	\\ [-0.2ex]
	\T F1		& 0.8179(20) 	& 0.7107(23)	& 0.6410(26)	& 0.5862(47)	\\ [-0.2ex]
	\T F2		& 0.8229(24)	& 0.7096(31)	& 0.6383(33)	& 0.5874(51)	\\ [0.5ex]
\\ [-2.5ex]
	\T Ensemble   	&& $f^{B_s \eta_s}_+(100)$& $f^{B_s \eta_s}_+(110)$	& $f^{B_s \eta_s}_+(111)$ \\ [0.5ex]
	\hline
	\T C1 	&& 1.843(10)	& 1.5476(62)	& 1.3400(63)	\\ [-0.2ex]
	\T C2	&& 1.742(13)	& 1.3885(99)	& 1.150(17)	\\ [-0.2ex]
	\T C3	&& 1.6802(95) 	& 1.3855(84)	& 1.1771(85)	\\ [-0.2ex]
	\T F1		&& 1.6928(71)	& 1.3497(55)	& 1.134(10)	\\ [-0.2ex]
	\T F2		&&  1.7012(97)	& 1.3588(72)	& 1.155(11)	\\ [0.5ex]
\hline\hline
\end{tabular}
\label{tab-BsEtascorrfits}
\end{table}

\begin{figure}[t]
%{\scalebox{1.0}{\includegraphics[angle=270,width=0.5\textwidth]{/Users/cmb/fitting/09July2013/fitscripts/modz/BstoK_Etas/plots/stability_v9a2/BsK_Etas_v9a2_stab_chi2-eps-converted-to.pdf}}}
{\scalebox{1.0}{\includegraphics[angle=270,width=0.5\textwidth]{BsK_Etas_v9a2_stab_chi2-eps-converted-to.pdf}}}
\\
%{\scalebox{1.0}{\includegraphics[angle=270,width=0.505\textwidth]{/Users/cmb/fitting/09July2013/fitscripts/modz/BstoK_Etas/plots/stability_v9a2/BsK_Etas_v9a2_stab_BsEtas-eps-converted-to.pdf}}}
{\scalebox{1.0}{\includegraphics[angle=270,width=0.505\textwidth]{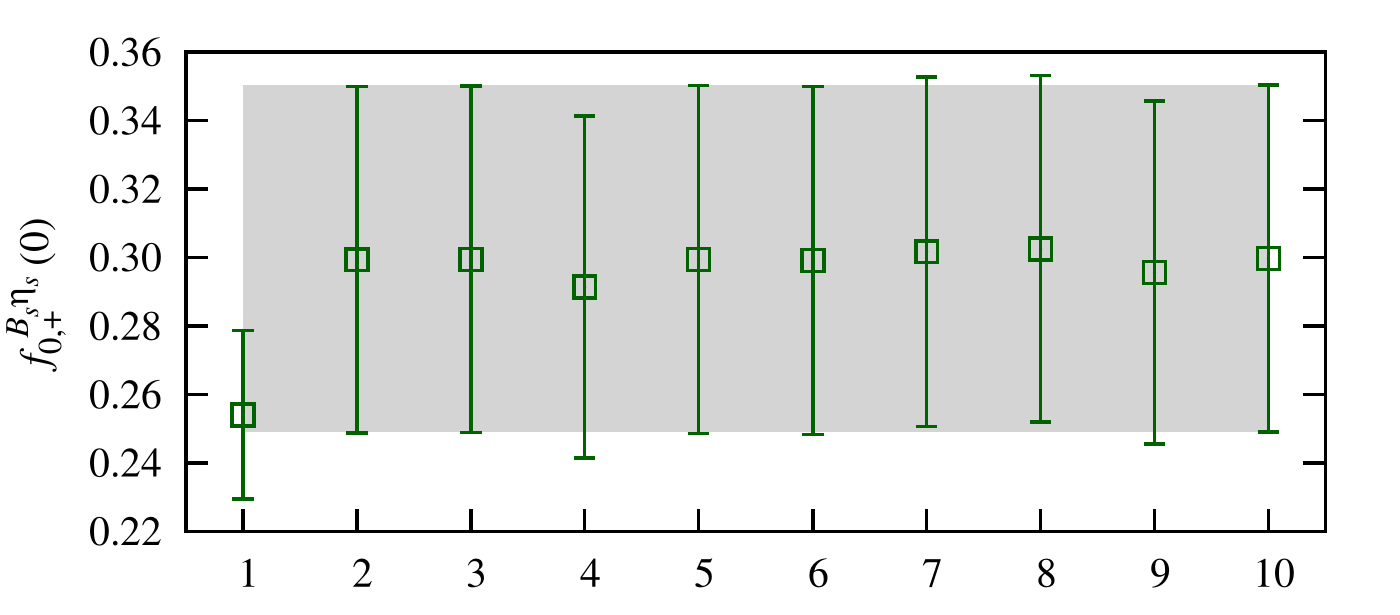}}}
\caption{(color online). The stability of the HPChPT $z$~expansion is demonstrated by studying the fit results under various modifications, discussed in Sec.~\ref{sec-Extrap} of the text.}
\label{fig-BsEtasstability}
\end{figure}

\begin{figure}[t]
%\vspace{-0.03in}
%{\scalebox{1.0}{\includegraphics[angle=270,width=0.5\textwidth]{/Users/cmb/fitting/09July2013/fitscripts/modz/BstoK_Etas/plots/v9a2/BstoK_Etas_f0BsEtas_error_breakdown_v9a2_lines-eps-converted-to.pdf}}}
{\scalebox{1.0}{\includegraphics[angle=270,width=0.5\textwidth]{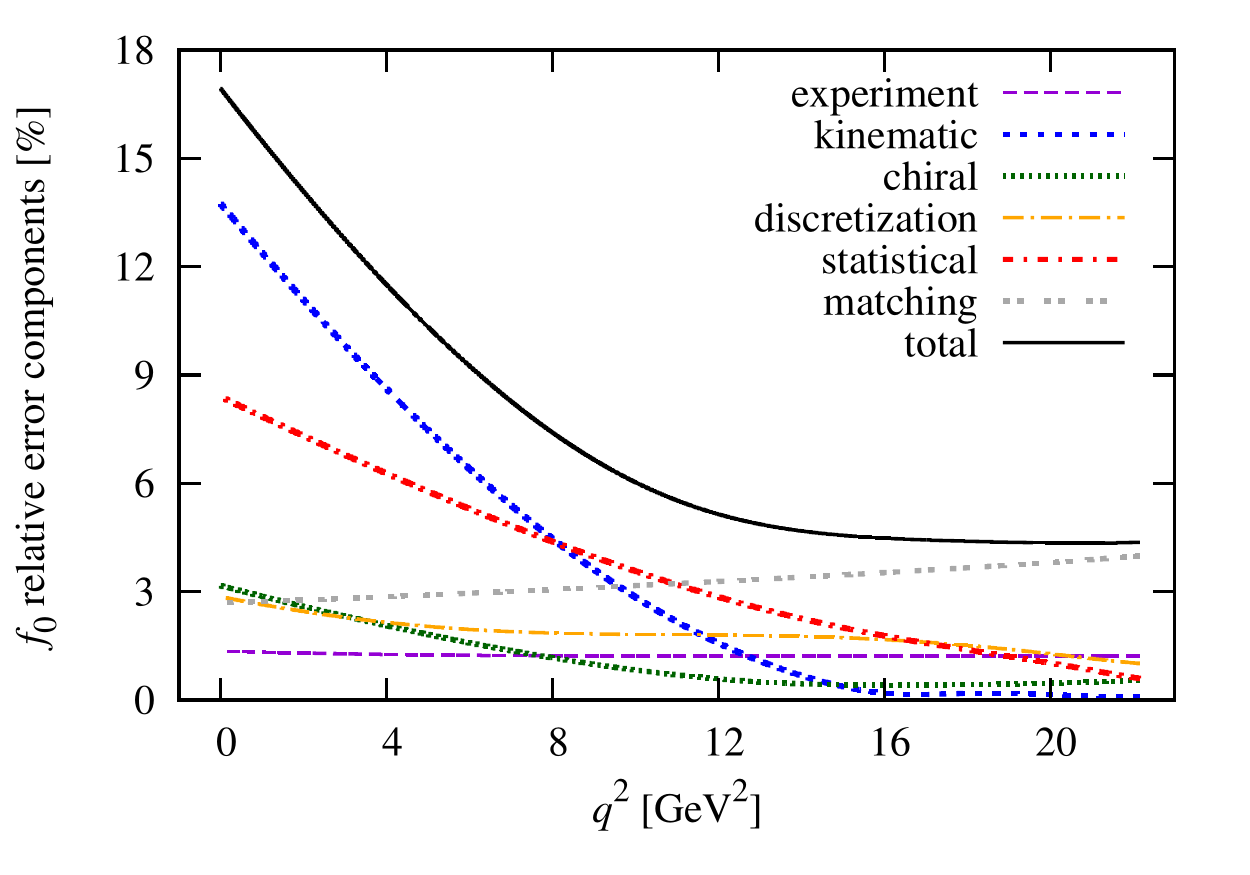}}}
\\
%{\scalebox{1.0}{\includegraphics[angle=270,width=0.5\textwidth]{/Users/cmb/fitting/09July2013/fitscripts/modz/BstoK_Etas/plots/v9a2/BstoK_Etas_fpBsEtas_error_breakdown_v9a2_lines-eps-converted-to.pdf}}}
{\scalebox{1.0}{\includegraphics[angle=270,width=0.5\textwidth]{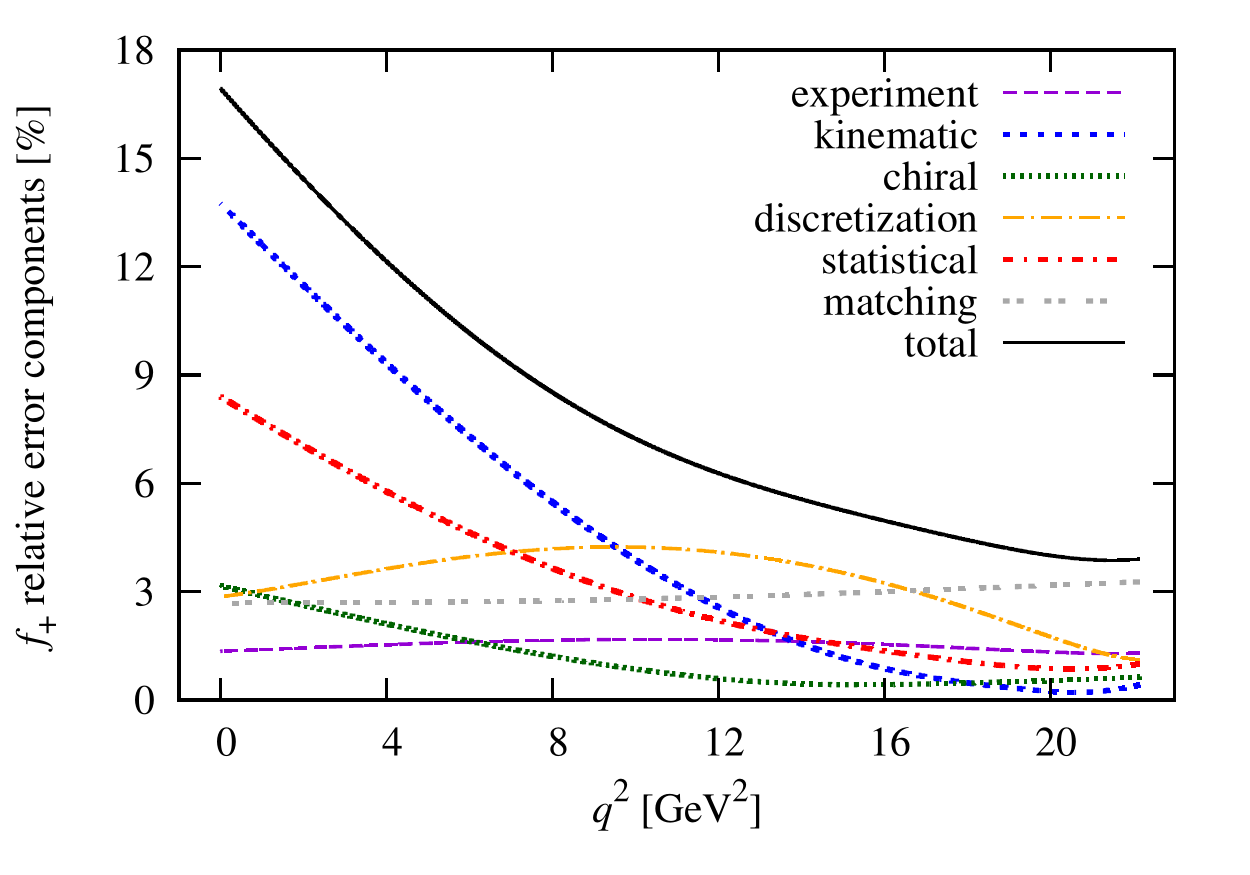}}}
\caption{\label{fig-modzerr_BsEtas}(color online). $B_s\to \eta_s$ ({\it top}) $f_0$ and ({\it bottom}) $f_+$ relative error components.  The total error (solid line) is the sum in quadrature of the components.}
\vspace{-0.03in}
\end{figure}

\begin{figure}[t]
%{\scalebox{1.0}{\includegraphics[angle=270,width=0.5\textwidth]{/Users/cmb/fitting/09July2013/fitscripts/modz/BstoK_Etas/plots/Rpar_v9a2_k3_err_lines-eps-converted-to.pdf}}}
{\scalebox{1.0}{\includegraphics[angle=270,width=0.5\textwidth]{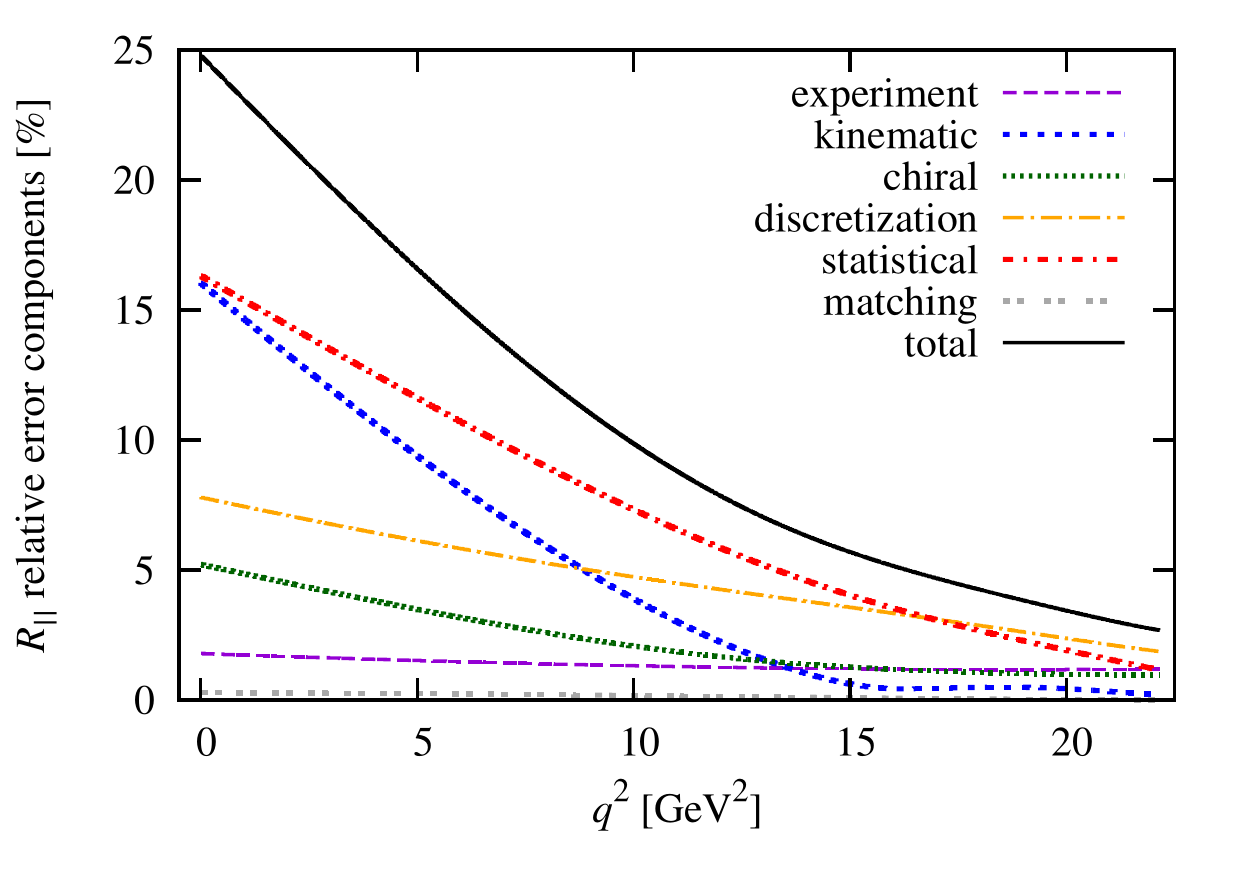}}}
\\
%{\scalebox{1.0}{\includegraphics[angle=270,width=0.5\textwidth]{/Users/cmb/fitting/09July2013/fitscripts/modz/BstoK_Etas/plots/Rperp_v9a2_k3_err_lines-eps-converted-to.pdf}}}
{\scalebox{1.0}{\includegraphics[angle=270,width=0.5\textwidth]{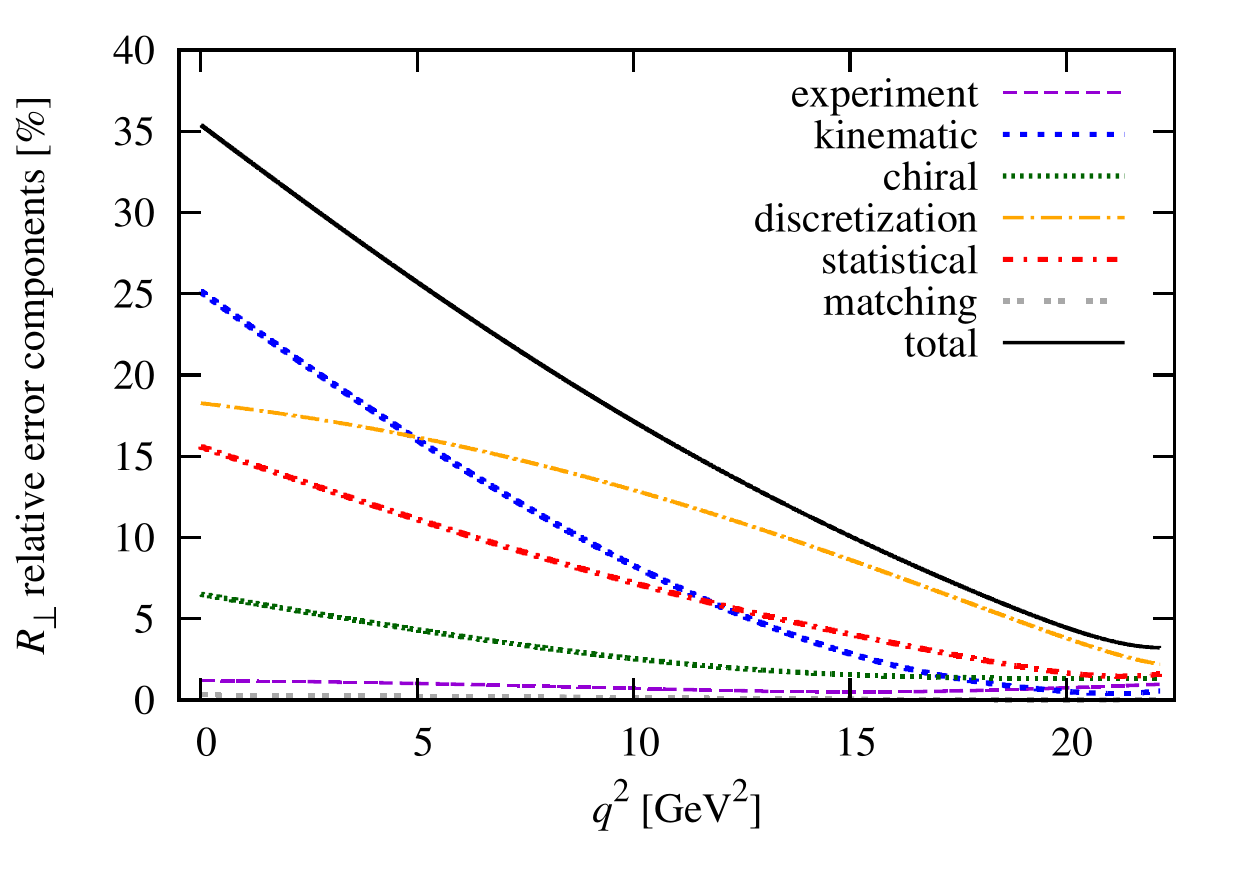}}}
\caption{(color online). Relative error components for ({\it top}) $R_{\parallel}$ and ({\it bottom}) $R_{\perp}$ as a function of $q^2$.  The total error is the sum in quadrature of the components.}
\label{fig-FFratios_err}
\end{figure}

The $B_s\to\eta_s$ form factor data of Table~\ref{tab-BsEtascorrfits} is extrapolated to the physical quark mass, the continuum limit, and over the entire kinematic range using the HPChPT $z$~expansion described in Sec.~\ref{sec-Extrap}.   
This fit is also done simultaneously with the extrapolation of the $B_s\to K$ data.  
The fit functions for the simultaneous chiral, continuum, and kinematic extrapolation of $B_s\to\eta_s$ are equivalent to those of Sec.~\ref{sec-Extrap}, with Eqs.~(\ref{eq-Dk}) and (\ref{eq-logs}) modified as follows:
\begin{eqnarray}
D_k &=& 1 + c^{(k)}_1 x_K + c^{(k)}_2 \delta x_K + c^{(k)}_3 \delta x_{\eta_s}  \nonumber \\
&+& d^{(k)}_1 (a/r_1)^2 + d^{(k)}_2 (a/r_1)^4 \nonumber \\
&+& e^{(k)}_1 (aE_{\eta_s})^2 + e^{(k)}_2 (aE_{\eta_s})^4, \label{eq-DkEtas} \\{}
[\text{logs}] &=& -\frac{1+3g^2}{2} x_K \log x_K - \frac{1+3g^2}{6}x_{\eta} \log x_{\eta}, \nonumber \\
\end{eqnarray}
with implicit indices in Eq.~(\ref{eq-DkEtas}) specifying scalar or vector form factor.  
Results of this fit for the $B_s\to\eta_s$ form factors are shown relative to data, and extrapolated over the full kinematic range of $q^2$, in Fig.~\ref{fig-HPChPTfit_BsEtas}.
The HPChPT $z$~expansion stability analysis outlined in Sec.~\ref{sec-Extrap} involved simultaneous fits to both $B_s\to K$ and $B_s\to \eta_s$ data.  The $B_s\to \eta_s$ fit results for each of the modifications discussed in that analysis are shown in Fig.~\ref{fig-BsEtasstability}.  Because these results are from a simultaneous fit, the values of $\chi^2$ in Fig.~\ref{fig-BsKstability} are applicable here as well and are reproduced for convenience in Fig.~\ref{fig-BsEtasstability}.
Note that the chiral analytic terms for $B_s\to\eta_s$ differ slightly from those for $B_s\to K$, c.f. Eqs.~(\ref{eq-Dk}) and (\ref{eq-DkEtas}).  As a result, the NNLO analytic terms added to the $B_s\to\eta_s$ fit function in modification 7 differ from those listed in Eq.~(\ref{eq-NNLO}).
Error breakdown plots for the $B_s\to\eta_s$ form factors are shown in Fig.~\ref{fig-modzerr_BsEtas}.
\begin{figure}[t]
%{\scalebox{1.0}{\includegraphics[angle=270,width=0.5\textwidth]{/Users/cmb/fitting/09July2013/fitscripts/modz/BstoK_Etas/plots/Rpar_v9a2_k3-eps-converted-to.pdf}}}
{\scalebox{1.0}{\includegraphics[angle=270,width=0.5\textwidth]{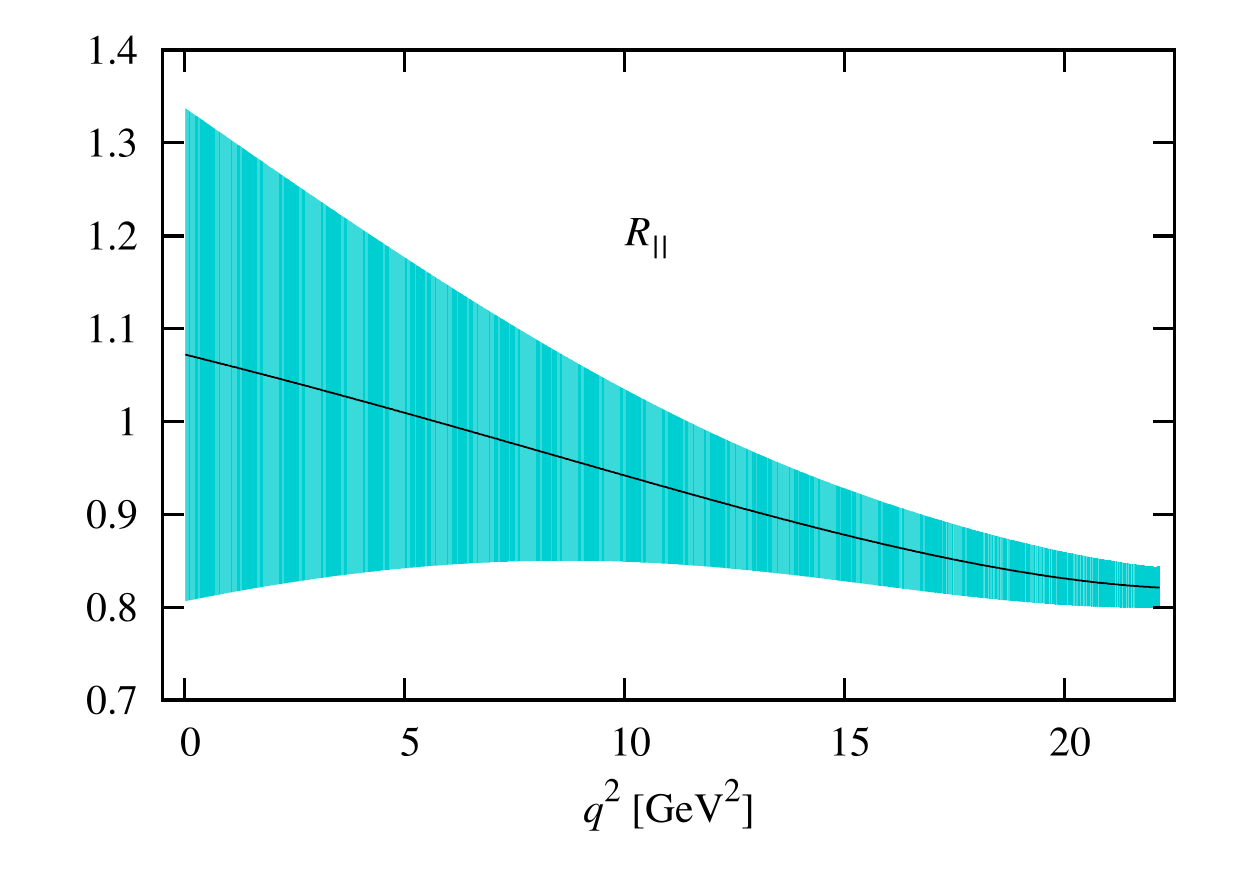}}}
\\
%{\scalebox{1.0}{\includegraphics[angle=270,width=0.5\textwidth]{/Users/cmb/fitting/09July2013/fitscripts/modz/BstoK_Etas/plots/Rperp_v9a2_k3-eps-converted-to.pdf}}}
{\scalebox{1.0}{\includegraphics[angle=270,width=0.5\textwidth]{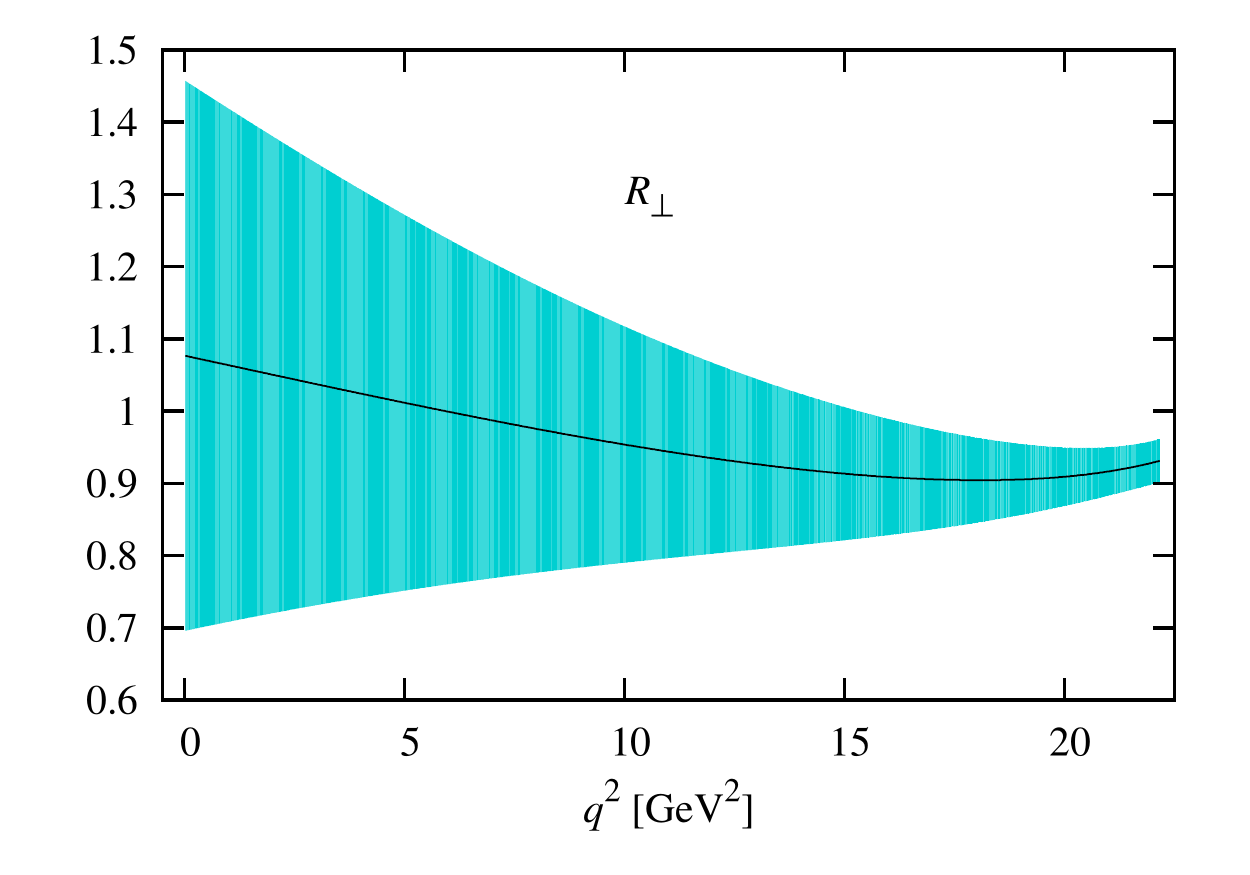}}}
\caption{(color online). Ratios of $B_s\to K$ to $B_s\to\eta_s$ form factors $R_{\parallel, \perp}$ as functions of $q^2$.}
\label{fig-FFratios}
\end{figure}
In the ratios of form factors,
\begin{eqnarray}
R_\parallel(q^2) &=& \frac{f_\parallel^{B_s K}(q^2)}{f_\parallel^{B_s\eta_s}(q^2)}, \\
R_\perp(q^2) &=& \frac{f_\perp^{B_s K}(q^2)}{f_\perp^{B_s\eta_s}(q^2)},
\end{eqnarray}
the leading systematic error, that due to one-loop perturbative matching, largely cancels. 
Fig.~\ref{fig-FFratios} plots the ratios as functions of $q^2$ and shows that they are most precisely determined at $q^2=(M_{B_s}-M_{\eta_s})^2$, where
\begin{eqnarray}
R_\parallel\big((M_{B_s}-M_{\eta_s})^2\big) &=& 0.821(22) , \\
R_\perp\big((M_{B_s}-M_{\eta_s})^2\big) &=& 0.931(30).
\end{eqnarray}
The errors of the ratios are broken down into components in Fig.~\ref{fig-FFratios_err}.
Neglecting correlations among the $B_s\to K$ and $B_s\to\eta_s$ decays yields ratios at this $q^2$ with $\sim\!30\%$ larger errors.
When combined with lattice results for $f_{\parallel, \perp}^{B_s\eta_s}$ using HISQ $b$ quarks, these ratios will provide a nonperturbative determination of the NRQCD $b\to u$ current matching factor, applicable to both $B_s\to K$ and $B\to\pi$.

\clearpage
%================================= BIBLIOGRAPHY =================================

%-----------------------------------o
%  References
%-----------------------------------o
%\clearpage
%\bibliography{mbw}
%\bibliographystyle{elsart-num}
%\bibliographystyle{apsrev}

%======================================================= The End ========

\end{document}